\documentclass[submission,copyright,creativecommons]{eptcs}
\providecommand{\event}{Under review for a Journal since Feb.2018.} 
\pdfoutput=1

\usepackage[margin=20mm,top=20mm,bottom=30mm]{geometry} 

\usepackage{enumitem}
\usepackage{graphicx}
\usepackage{float,csquotes,tikz,pgf-umlsd,graphicx,listings,caption,booktabs,array,multirow,adjustbox,rotating,siunitx,pifont}
\usepackage{subcaption}
\usepackage{tikz}
\usepackage{wrapfig}
\newcommand{\pr}[1]{\begin{tikzpicture}[scale=0.1]%
    \draw (0,0) circle (1);
    \fill[fill,fill=black] (0,0) -- (90:1) arc (90:90-#1*3.6:1) -- cycle;
    \end{tikzpicture}}

\begin{document}

\title{The Snowden Phone: A Comparative Survey of Secure Instant Messaging Mobile Applications} 

\author{Christian Johansen
\institute{Department of Technology Systems, University of Oslo}
\email{christian@johansenresearch.info}
\and Aulon Mujaj 
\institute{Department of Informatics, University of Oslo}
\and Hamed Arshad 
\institute{Department of Informatics, University of Oslo}
\and Josef Noll
\institute{Department of Technology Systems, University of Oslo}
}
\def\titlerunning{The Snowden Phone: A Comparative Survey of Secure Instant Messaging Mobile Applications (authors' version)}
\def\authorrunning{
C. Johansen, A. Mujaj, H. Arshad, J. Noll
}

\maketitle

\begin{abstract}

In recent years, it has come to attention that governments have been doing mass surveillance of personal communications without the consent of the citizens. As a consequence of these revelations, developers have begun releasing new protocols for end-to-end encrypted conversations, extending and making popular the old Off-the-Record protocol. Several new implementations of such end-to-end encrypted messaging protocols have appeared, and commonly used chat applications have been updated with these implementations as well. In this survey, we compare the existing implementations, where most of them implement one of the recent and popular protocols called Signal. 
We conduct a series of experiments on these implementations to identify which types of security and usability properties each application provides. The results of the experiments demonstrate that the applications have variations of usability and security properties, and none of them are infallible. Finally, the paper gives proposals for improving each application w.r.t. security, privacy, and usability.
\end{abstract}

\tableofcontents

\footnotetext{\textbf{Disclaimer:} All the tests reported here were performed in the summer of 2017, and since applications in this area are very dynamic, some of the specific implementation recommendations and observations that we make may have already been treated by the developers. However, this work still should provide guidance for a new user on how to check which desired features are implemented by a specific application; even more so if the application is among the six surveyed here.
More details for this paper can be found in our technical reports \cite{johansen2017SignalRep,aulon2017MSc}.
}

\section{Introduction}\label{intro1}
In recent years, the trend to use mobile applications for communication has grown and become a standard method of communication between people. New messaging applications started to emerge and try to replace traditional SMS, but building them with security and privacy in mind was not important for the developers in the beginning. The popular messaging tools used in recent years did not support end-to-end encryption, only standard client to server encryption, which gives the service providers access to more private information than necessary. When Edward Snowden published the secret papers about NSA,
\footnote{\url{https://en.wikipedia.org/wiki/Edward_Snowden\#Global_surveillance_disclosures} \\
Two feature films on this topic are:
Oliver Stone's \url{https://en.wikipedia.org/wiki/Snowden_(film)}
and 
Laura Poitras' \url{https://en.wikipedia.org/wiki/Citizenfour}
}
people finally understood that mass surveillance was an issue, and secure mobile messengers became more critical and popular. 

People are more prone to understand the privacy implications of mass surveillance \cite{schneier2015Book}. Edward Snowden has sparked a heated debate throughout the world about individual privacy which is undermined by the mass surveillance that multiple countries have been doing for  decades.\footnote{\url{https://en.wikipedia.org/wiki/List_of_government_mass_surveillance_projects}}
No need to look further than the first quarter of 2017, when WikiLeaks\footnote{\url{https://wikileaks.org/}} leaked documents from the U.S.~Central Intelligence Agency (CIA). The leak, codenamed ``Vault 7'' by WikiLeaks is the largest ever publication of confidential documents from the agency.\footnote{	``Wikileaks Unveils 'Vault 7': The Largest Ever Publication Of Confidential CIA Documents; Another Snowden Emerges'', authored by Tyler Durden in the Zerohedge, March 2017. Available at \url{http://www.zerohedge.com/news/2017-03-07/wikileaks-hold-press-conference-vault-7-release-8am-eastern}}
The documents that leaked have information on how to get access to mobile phones or personal computers without the user's knowledge, and how the CIA did  mass surveillance.

Several companies started implementing secure messaging protocols and applications to counter the mass surveillance and offer an end-to-end encrypted messaging system which does not leak any information about the user's message content. 
However, the problem with new and bleeding-edge applications is their adoption. After a while, companies such as Google, Facebook, and Open Whisper Systems joined forces to implement protocols into already widely adopted applications such as WhatsApp, which has over one billion monthly active users.\footnote{The Statistics Portal: \textit{``Number of monthly active Whatsapp users worldwide from April 2013 to December 2017 (in millions)''} \url{https://www.statista.com/statistics/260819/number-of-monthly-active-whatsapp-users/}}

Instant Messaging clients that did not provide asynchronous communication became uninteresting because of the rise of smart-phones and applications that were not always online. The most mature secure messaging protocol, Off-the-Record, did not support asynchronous messaging, which motivated the development of new protocols with asynchronous communication built-in. 
The most notable is the Signal application with their protocol also called Signal. After a while, the new protocol became quite popular among developers and researchers \cite{sok,frosch2016secure,formal-signal,signal-audit-inria,kristoffer2017HotSpot}.
Subsequently, the Signal protocol started to be implemented in other applications, which were supporting only client-to-server encryption until then.



Quite a number of new secure messaging applications exist (in 2017 we counted six, which we survey in this work) that offer end-to-end encrypted message conversations over mobile phones and computers, but these often sacrifice usability aspects for security. In the light of the above motivations one would probably prioritise privacy and security, but in order to attract most normal users it should be possible to have the best of both worlds. Applications should give enough information for the users to know when or if a conversation is not secure anymore and the options to secure it once again. Moreover, the security controls should be intuitive and usable enough to be handled by a majority of people, not only for the technology inclined ones.


\subsection*{The goals of this study} 

The area of end-to-end encryption in instant messaging applications has become rather broad recently.
It is difficult for a user to find digestible 
information sources, and even less when it comes to comparative integrated studies. Therefore, our first goal is:

\begin{itemize}
\item[G1:] Provide comprehensible and comparative study of relevant approaches to end-to-end encrypted messaging applications.
\end{itemize}

A detailed analysis of the security and privacy properties provided by secure messaging protocols is not easy and there are very few such studies (which we build upon). End-to-end encrypted messaging technologies should be both usable so to allow a wide adoption, but also have rather strong security requirements. These two, i.e., usability and security, are usually conflicting, and a good balance is difficult to find. This leads to our second goal:

\begin{itemize}
\item[G2:] Overview the security and privacy properties provided by current end-to-end messaging technologies, and to what extent existing applications achieve these properties.
\end{itemize}

The Signal application (and protocol) is one of the most used end-to-end messaging technologies currently available for smart phones and desktop PCs. Moreover, the Signal protocol is employing state of the art encryption and key establishment techniques. 

\begin{itemize}
\item[G3:] Describe for non-experts the security mechanisms behind the Signal protocol.
\end{itemize}

There is little research in the area of usability vs.\ security in secure messaging applications. In the last couple of years, it has become increasingly important for researchers to look at the usability and not purely to the technical issues surrounding secure messengers. Schroder et al. \cite{signal-fan} were the first to look at the usability issues for end-to-end encrypted messengers, doing a comprehensive user study of the usability of Signal's security features and proposing fixed to the issues they found with users failing to detect and deter man-in-the-middle attacks. 
In this paper we look at the same types of potential attack spots as \cite{signal-fan} did, looking also at the application interface, and extending to five more applications than Signal.

Unger et al. \cite{sok} did a comprehensive study of secure messaging protocols, looking at the security properties around trust establishment, conversation security, and transport privacy. Their survey shows that protocols that specialise in encryption do not achieve every important security and privacy property. 
In this paper, our properties of interest are based on those from \cite{sok}, focusing on the conversation security for the three major end-to-end encrypted messaging protocols that we found.

\subsection*{Main Contributions}

\begin{itemize}
\item We make a comprehensive analysis of applications that implement secure messaging, by performing five testing scenarios to study their essential security and usability properties.
We also provide suggestions for improvements.
\item We provide an (updated) overview of conversation security in secure messaging protocols, following \cite{sok}. Subsequently, we describe the inner security workings of the latest versions of two major protocols (OTR and Signal), striving to make these understandable for a general audience.
\end{itemize}

The rest of the paper is organised as follows. 
Section~\ref{review-protocols} presents a systematisation of knowledge about three secure end-to-end encrypted messaging protocols, with a discussion of their security properties. 
Section~\ref{review-applications} presents the study testing six mobile phone applications that support either the secure messaging protocols presented before or their own variants which are not open source applications. 
Section~\ref{results} summarises the results from the test scenarios in a unified and comparative manner. 
Section~\ref{discussion} discusses the applications as a whole also providing recommendations for improvements. 
Finally, Section~\ref{conclusion} concludes the paper.

\section{Background on Secure Messaging Protocols}
\label{review-protocols}

This section provides background on secure messaging protocols that are implemented by the applications analysed in this paper. 
First we present the attacker models that we consider and assumptions that we make about the user applications, then we review basic properties relevant for end-to-end encrypted messaging.
Section~\ref{subsec-OTR} surveys Off-the-Record (OTR), which is the baseline for the two other protocols, Signal (surveyed in Section~\ref{subsec_signalNutShell}) and Matrix, which are new protocols that are actually implemented (or copied) by the current popular secure messaging applications.

This section builds on the comprehensive survey \cite{sok}, as well as on various other resources regarding these protocols. Most of the resources for the two new protocols Signal and Matrix are online, since these protocols have not come out of academia. However, both are built on the good foundation laid by the OTR protocol, which has been well studied in academia \cite{otr-com,user-study-otr,secureOTR,user-auth-otr,group-otr,impro-gotr,group-otr2}, and also complemented by significant online resources.

\subsection{Relevant Threat Models}

We assume the following adversaries:

\begin{itemize}
    \item \textbf{Active adversaries:} Man-in-The-Middle attacks are possible on both local and global networks by adding a proxy between the applications and servers handling the messages. These are under the usual assumptions of a Dolev-Yao model \cite{dolev1983security}.
    \item \textbf{Passive adversaries:} These adversaries log everything that is sent to and from a user and could potentially use that information to keep track of who users talk to and when. Passive adversaries could also log information such as messages and keys, even though the contents of the messages are encrypted.
    \item \textbf{Service providers:} The messaging systems that require centralized infrastructure (such as Signal and Matrix) need to keep the information about users secure. The service operators could at any time become a potential adversary.
\end{itemize}

We assume the endpoints of the messaging applications, e.g., an app on a smartphone, are secure and that the devices do not have malware that could exploit the messaging application.

\subsection{Security principles relevant for end-to-end encrypted messaging}

Different secure messaging protocols capture different security principles in various degrees, and are important when comparing specifications of applications implementing them. Most security and privacy features that we review here are also being found in \cite{sok}.

\begin{description}
 \item[End-to-End Encryption:]
%
Communication encryption protocols like Transport Layer Security (TLS) \cite{tls-ietf} are designed to secure communications between a client and server.
Messaging applications that allow two parties to communicate to each other through a server, can use TLS to secure their communication against network attackers.
Messages sent to the server are decrypted by the server, which means that it can read, store or edit the message before encrypting again and sending it to the other user. 
Often servers cannot be trusted, as they can be hacked by an adversary, or may be contacted by law enforcement to give information sent by clients through the server \cite{ncc-group}.
End-to-end encryption ensures that the endpoints do the encryption while the servers only transmit the messages without network attackers nor a corrupted server being able to see the content.

 \item[Confidentiality:]
%
Confidentiality ensures that the necessary level of secrecy is enforced at each junction of data processing, preventing unauthorized disclosure \cite{harris2012cissp}. 
Confidentiality can be provided by encrypting data while it is stored and transmitted. 
In cryptographic protocols confidentiality is essential to ensure that keys and other data are available only as intended \cite{boyd2003protocols}.
Attackers try to break confidentiality by stealing password files, breaking encryption schemes, etc. Users, on the other hand, can intentionally or accidentally disclose sensitive information by not encrypting it before sending it to another person, or by falling prey to a social engineering attack \cite{harris2012cissp}.

 \item[Integrity:]
%
Integrity ensures that no one throughout the transmission modifies the messages. Hardware, software, and communication mechanisms must work in concert to maintain, process, and move data to intended destinations, without unexpected alterations.
Systems that enforce and provide this security property ensure that attackers, or mistakes by users, do not compromise the integrity of systems or data \cite{harris2012cissp}. This can be achieved through the use of hash functions in combination with encryption, or by use of a message authentication code (MAC) to create a separate check field. Data integrity is a form of integrity that is essential for most cryptographic protocols to protect elements such as identity fields or nonces \cite{boyd2003protocols}.

 \item[Authentication:]
%
Authentication is meant to identify the parties in a conversation. Message authentication is also called \textit{data-origin authentication}, and protects the integrity of the sender of the message \cite{black1999umac,black2000cbc}.
Message authentication codes can provide assurance about the source and integrity of a message. A message authentication code is computed by using the message and a shared secret between the two parties \cite{gollmann}. If an adversary changes the message, then the computed MAC would be different as well, and moreover, an adversary cannot produce a valid MAC because only the sender and receiver have the shared secret.

 \item[Perfect Forward Secrecy:]
%
A key establishment protocol provides forward secrecy if a compromise of long-term keys of a set of principals does not compromise the session keys established in previous protocol runs involving those principals.
Typical examples of protocols which provide forward secrecy are key agreement protocols where the long-term key is only used to authenticate the exchange. Key transport protocols in which the long-term key is used to encrypt the session key cannot provide forward secrecy \cite{boyd2003protocols}.

 \item[Future Secrecy:]
%
Future secrecy, as it is called by Open Whisper Systems \cite{adv-ratch}, (sometimes also called backward secrecy) is the guarantee that the compromise of long-term keys does not allow subsequent ciphertexts to be decrypted by passive adversaries \cite{sok}. A protocol supports future secrecy when it can provide the ``self-healing'' aspect of the Diffie-Hellman ratchet, which will be described in section \ref{subsec-OTR}, because if any ephemeral key is compromised or found to be weak at any time, the ratchet will heal itself and compute new ephemeral keys for the future messages sent during the conversation \cite{adv-ratch}.

\item[Deniability:]
%
Deniability is a property common to new secure messaging protocols, where it is not possible for others to prove that the data was sent by some particular conversation party. If Bob receives a message from Alice, he can be sure it was Alice that sent it, but cannot prove to anyone else that. To provide deniability, usually secure messaging protocols have a mechanism to allow anyone to forge messages, after a conversation, to make them look like coming from someone in the conversation. Deniability also includes authenticity during the conversation so that the participants are assured that the messages they see are authentic and are not modified by anyone \cite{def-den}.
Deniability can be divided into three different parts:
\begin{enumerate}
\item \textbf{Message Unlinkability:} If a judge is convinced that a participant authored one message in the conversation, it does not provide evidence that they authored other messages.
\item \textbf{Message Repudiation:} Given a conversation transcript and all cryptographic keys, there is no evidence that a given message was authored by any particular user. We assume the accuser has access to the session keys, but not the other participants' long-term secret keys.
\item \textbf{Participation Repudiation:} Given a conversation transcript and all cryptographic key material for all but one accused (honest) participant, there is no evidence that the honest participant was in a conversation with any of the other participants.
\end{enumerate}

\item[Synchronicity:]
There are two types of communication, synchronous and asynchronous. Synchronous protocols require all participants to be online for them to receive or send messages. Chat applications are traditionally synchronous communications. Alternatively, asynchronous messaging means that the participants do not need to be online to receive messages, such as SMS text messaging or emails, since there is a third party, like a server, to save the information until the recipient gets online again.
Modern chat protocols do not use synchronous protocols, usually because of social or technical constraints, such as device battery, limited reception or other social happenings which do not allow people to be constantly online to receive messages. That is why the majority of Instant Messaging (IM) solutions provide an asynchronous environment by having a third party server to store the messages until the other participant gets online to receive it.

\item[Group Chat Properties:]
Group conversations are popular nowadays, e.g., using Facebook Messenger\footnote{\url{https://www.messenger.com}}, Slack\footnote{\url{https://slack.com}} or other popular messaging applications\footnote{\url{https://www.engadget.com/2016/09/30/12-most-used-messaging-apps/}}. 
Security properties in the context of group chats include:

\begin{enumerate}
\item \textbf{Computational Equality:} Do the participants share an equal computational load when talking to each other.

\item \textbf{Trust Equality:} No single participant has more trust or responsibility, within the group, than any other.

\item \textbf{Subgroup Messaging:} Participants can send messages to only a subgroup without generating a new conversation.

\item \textbf{Contractible Membership:} No need to restart the security protocol when a member leaves the conversation.

\item \textbf{Expandable Membership:} There is no need to restart the security protocol when adding a new member after the group has been generated.
\end{enumerate}

It is important to be able to change the cryptographic keys when a new user joins the secure group conversation, since then the new users will not have the ability to decrypt previously exchanged messages. New cryptographic keys should also be exchanged when a user leaves the conversation. Changing the keys can easily be done by restarting the protocol, but this is often computationally expensive. Protocols which offer contractible and expandable memberships usually achieve these features without restarting the protocol.

\item[Other Security Properties:]
A protocol or application for end-to-end secure IM may implement any (if not all) of the following.

\begin{enumerate}
	\item \textbf{Participant Consistency:} At any point when a message is accepted by an honest party, all honest parties are guaranteed to have the same view of the participant list.
	\item \textbf{Destination Validation:} When a message is accepted by an honest party, they can verify that they were included in the set of intended recipients for the message.
	\item \textbf{Anonymity Preserving:} Any anonymity features provided by the underlying transport privacy architecture (like the Tor\footnote{https://www.torproject.org/} network \cite{dingledine2004tor,tor2008}) are not undermined (e.g., if the transport privacy system provides anonymity, the conversation security level does not deanonymize users by linking key identifies).
	\item \textbf{Speaker Consistency:} All participants agree on the sequence of messages sent by each participant. A protocol might perform consistency checks on blocks of messages during the protocol, or after every message is sent.
	\item \textbf{Causality Preserving:} Implementations can avoid displaying a message before messages that causally precede it.
	\item \textbf{Global Transcript:} All participants see all messages in the same order. When this security feature is assured, it implies both speaker consistency and causality preserving are assured.
\end{enumerate}

\end{description}

\subsection{Properties for Usability and Adoption}  \label{techback:secprin-usabAdop}

Various aspects need to be taken into account when looking at usability and adoption of a secure IM application.

\begin{enumerate}
	\item \textbf{Out-of-order Resilience:} If a message is delayed in transit, but eventually arrives, its contents are accessible upon arrival.
	\item \textbf{Dropped Message Resilient:} Messages can be decrypted without receipt of all previous messages. This is desirable for asynchronous and unreliable network services.
	\item \textbf{Asynchronous:} Messages can be sent securely to devices which are not connected to the Internet at the time of sending.
	\item \textbf{Multi-Device Support:} A user can connect to the conversation from multiple devices at the same time, and have the same view of the conversation as the others. 
	\item \textbf{No Additional Service:} The protocol does not require any infrastructure other than the protocol participants. Specifically, the protocol must not require additional servers for relaying messages or storing any kind of key material.
\end{enumerate}

\subsection{Off-the-Record in a nut shell}\label{subsec-OTR}

Intuitively, the Off-the-Record (OTR) protocol \cite{otr-com,user-study-otr,secureOTR,user-auth-otr,group-otr,impro-gotr,group-otr2} wants to provide for online conversations the same features that reporters want when talking with a news source. Take a scenario where Alice and Bob are alone in a room. Nobody can hear what they are saying to each other unless someone records them. No one knows what they talk about, unless Alice and Bob tell them, and no one can prove that what they said is true, not even themselves. A good thing about an Off-the-Record conversation (in reality) is the legal support behind it since it is illegal to record conversations without participants knowing. It also applies to conversations over the phone, since by law, it is illegal to tap phone lines. There are however no similar laws for communications over the web. OTR-like protocols aim to provide this using cryptography techniques, and thus need to provide at least perfect forward secrecy and deniability/repudiation.
%
%
Full details can be found in \cite[Sec.2.4]{aulon2017MSc}.

\paragraph{Step 1: Authenticated Key Exchange} \label{otr:step1}
The latest version of OTR \cite{otr} use a variation of Diffie-Hellman Key Exchanged called SIGMA \cite{sigma}. 
Alice and Bob also have long-term authentication public keys $pub_A$ and $pub_B$, respectively. The point is to do an unauthenticated Diffie-Hellman key exchange to set up an encrypted channel, and inside that channel do mutual authentication.
%
%
The plain Diffie-Hellman key exchange is vulnerable to man-in-the-middle attacks which would break the authentication that OTR needs \cite{secureOTR}. Therefore, OTR implements a signature-based authenticated DH exchange, named SIGMA, which solves this weakness \cite{sigma}. 

The SIGMA acronym is short for ``SIGn-and-MAc,'' because SIGMA decouples the authentication of the DH exponentials from the binding of key and identities. The former authentication task is performed using digital signatures while the latter is done by computing a MAC function keyed via 
the common DH secret
and applied to the sender's identity \cite{sigma}.  
OTR uses a four message variant known as SIGMA-R, since it provides defence both against active attacks on the responder's identity and passive attacks on the initiator's identity. 

\paragraph{Step 2: Message Transmission} \label{otr:step2}
The message transmission step, before sending, performs encryption and authentication of messages using AES \cite{aes-book} in counter mode \cite{aes-ctr} using  message authentication codes (HMAC) \cite{hmac} for authentication.
Using AES in counter mode provides a malleable encryption scheme which allows deniability. Malleability allows to transform a ciphertext into another ciphertext which then decrypts to a related plaintext \cite{secureOTR}. This means that a valid ciphertext cannot be connected with neither Alice nor Bob since anyone can create a ciphertext that can be decrypted correctly and then compute a valid MAC from the ciphertext, because old MAC keys are published (more about this in step 4). 
Entire new messages, or full transcripts, can thus be forged.

For sending messages Alice needs to compute an \textit{Encryption Key} and a \textit{MAC Key}. The encryption key is used to encrypt the message while the MAC key is used to ensure the authenticity of the message. This method is called encrypt-then-mac, where usually the encryption key is a hash of the shared secret, $EK = Hash(SS)$, and then the encryption key is hashed a second time to compute the MAC key ($MK = Hash(EK)$).
After the encryption and MAC key are computed, Alice encrypts first the message, $Enc_{EK}(M)$ and then MACs the encrypted message, $MAC(Enc_{EK}(M),MK)$.  
Bob computes the same $EK$ and $MK$ from the common secret, used to verify the MAC and decrypt the message.

\paragraph{Step 3: Re-key}  \label{otr:step3}
Off-the-Record changes the keys every time the conversation changes directions, to make the duration of vulnerability to attacks as short as
possible. 
Once the new key is established it will be used to encrypt and authenticate new messages, while the previous ones are erased \cite{secureOTR}. After establishing new secrets and keys, the partners erase the old secret $SS$ and encryption key $EK$, so that no one can forge or decrypt the messages that have been sent. The reason to securely erase this information is to get perfect forward secrecy. The $MK$ key is not erased, but published in the next step.

\paragraph{Step 4: Publish MK} \label{otr:step4}
The next step of OTR is to publish the old MAC keys by adding them to the next message that Alice or Bob sends to each other, in plaintext. Alice and Bob both know that they have moved over to $MK'$, hence if one of them receives a message with the old $MK$, they will know that the message has been forged.
Publishing $MK$ allows others to forge transcripts of conversations between Alice and Bob. This is useful since it provides extra deniability to both parties \cite{signal-deniability}. 
%
In short, Alice's secrecy relies on Bob deleting the encryption keys, whereas Alice's deniability relies only on Alice publishing her $MK$.

\paragraph{Socialist Millionaire Protocol} \label{otr:smp}
The problem with secure instant messaging is that there is no way to tell if there has been a Man-In-The-Middle attack. Therefore, the parties need to make sure they have the \textit{same secret} which is done using the Socialist Millionaire Protocol (SMP) \cite{1996Millionairs,yao1982protocols}.
Intuitively, SMP allows two millionaires who want to exchange information to see whether they are equally rich, without revealing anything about the fortunes themselves.
Between Alice and Bob the SMP allows to know whether $ss^{A}
=ss^{B}$, i.e., the respective computed secrets, without revealing these secrets to anyone \cite{user-auth-otr}. 

\subsection{Signal in a nut shell}\label{subsec_signalNutShell}

Signal is a new end-to-end encryption protocol which has recently seen larger adoption than the Off-the-Record protocol. OTR had an original feature, i.e., refresh the message encryption keys often, which has became known as ratcheting and adopted in Signal as well \cite{formal-signal}.
The Signal Protocol is designed by Moxie Marlinspike and Trevor Perrin from Open Whisper Systems\footnote{\url{https://whispersystems.org/}}  to work in both synchronous and asynchronous messaging environments.\footnote{Advanced Ratcheting, by Moxie Marlinspike at Open Whisper Systems, November 26 2013, \url{https://whispersystems.org/blog/advanced-ratcheting/}}
The goals of Signal include end-to-end encryption and advanced security properties such as forward secrecy and future secrecy \cite{formal-signal}. Initially, Signal was divided into two different application, TextSecure\footnote{\url{https://whispersystems.org/blog/the-new-textsecure/}} and RedPhone\footnote{\url{https://whispersystems.org/blog/low-latency-switching/}}. The former was for SMS and instant messaging, while the latter was an encrypted VoIP\footnote{\url{https://www.voip-info.org/wiki/view/What+is+VOIP}} application. TextSecure was based on the OTR protocol, extending the ratcheting into a Double Ratchet, combining OTR's asymmetric ratchet with a symmetric ratchet \cite{formal-signal}, and naming it \textit{Axolotl Ratchet}. Open Whisper Systems later combined TextSecure and RedPhone to form the new Signal application that implements the protocol with the same name.

In recent years, the Signal Protocol has been adopted by numerous companies, such as WhatsApp\footnote{\url{https://whatsapp.com}} by Facebook, the Messenger\footnote{\url{https://messenger.com}} also by Facebook, and Google's new messaging app, Allo\footnote{\url{https://allo.google.com}}.

Signal uses the Double Ratchet algorithm \cite{double-ratchet} to exchange encrypted messages based on a shared secret key that the two parties have. To agree on the shared secret key Signal uses the X3DH Key Agreement \cite{x3dh} Protocol (standing for \textit{Extended Triple Diffie-Hellman}\footnote{\url{https://whispersystems.org/docs/specifications/x3dh/}}) which we describe later in section
\ref{signal:x3dh-intro}.
Full details can be found in \cite[Chap.3]{aulon2017MSc}.

\subsubsection{The Double Ratchet Algorithm}\label{signal:dr-intro}

The Double Ratchet Algorithm uses Key Derivation Function Chains (KDF Chains) \cite{double-ratchet} to constantly derive keys for encrypting each message, and it combines two different ratchet algorithms, a symmetric-key ratchet for deriving keys for sending and receiving messages, and a Diffie-Hellman ratchet used to provide secure inputs to the symmetric-key ratchet.

A KDF Chain is a sequencing of applications of a Key Derivation Function which returns one key used as a new KDF key for the next chain cycle as well
as an output key for messages. 
The KDF Chain has the following important properties \cite{double-ratchet}:

\begin{itemize}
	\item \textbf{Resilience:} The output keys appear random to an adversary without knowledge of the KDF keys, even if the adversary has control of the KDF inputs.
	\item \textbf{Forward security:} Output keys from the past appear random to an adversary who learns the KDF key at some future point.
	\item \textbf{Break-in recovery:} Future output keys appear random to an adversary who learns the KDF key at some point in time, provided the future inputs have added sufficient entropy.
\end{itemize}

\begin{figure}[t]
\centering
	\begin{subfigure}{.42\textwidth}
  		\centering
\includegraphics[width=.9\linewidth]{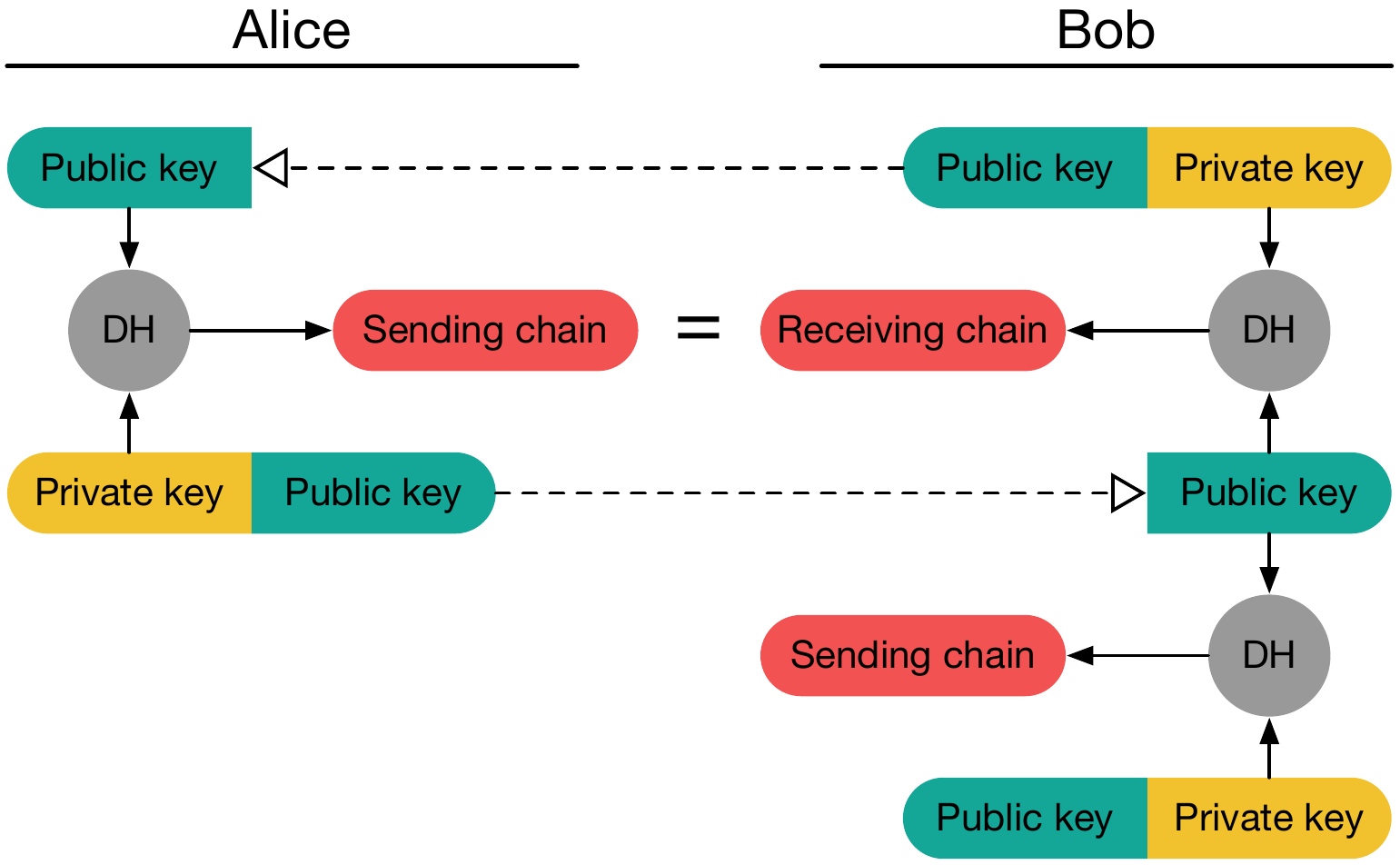}
\caption{Sending and receiving chains.}
\label{fig:dhratchet3}
	\end{subfigure}%
	\begin{subfigure}{.56\textwidth}
  		\centering
  \includegraphics[width=.9\linewidth]{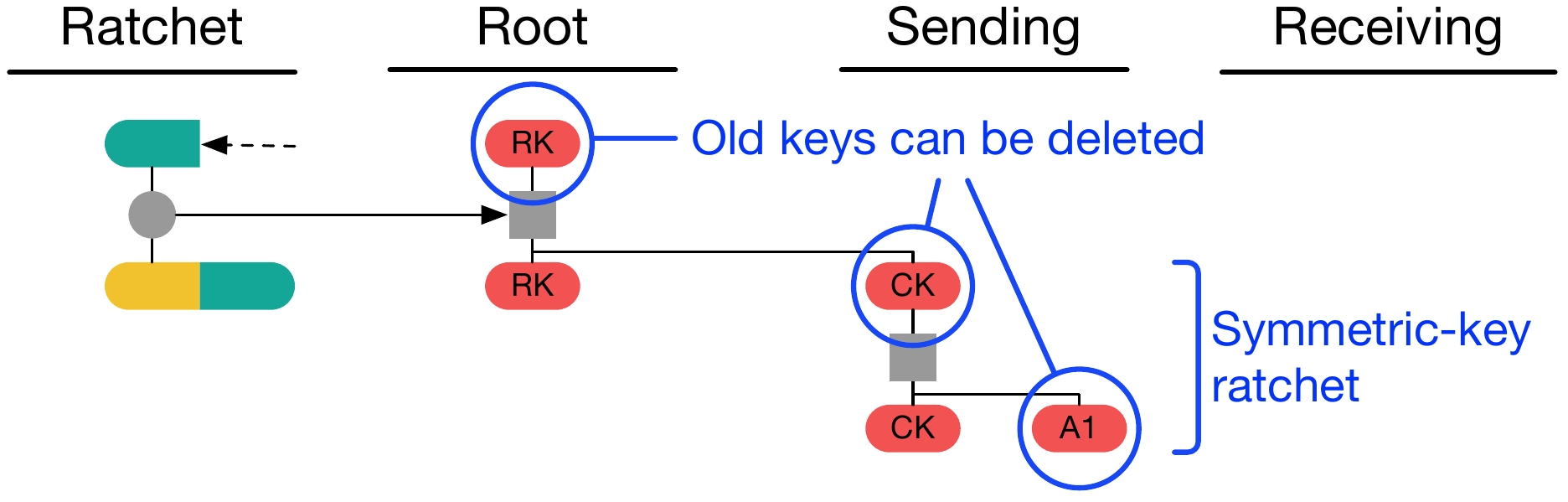}\\ \ \\ \ \\
\caption{First double ratchet message key A1 for Alice.}
\label{fig:doubleratchet2}
	\end{subfigure}
	\caption{Signal double ratchet Key-Derivation Function (KDF) Chains; (reproduced from \cite{double-ratchet}).}
\end{figure}

The Double Ratchet generates and maintains keys of each party for three chains: a \textit{root chain}, a \textit{sending chain}, and a \textit{receiving chain} (Alice's sending chain matches Bob's receiving chain, and vice versa; see Figure\ref{fig:dhratchet3}).
While the parties exchange messages, they also exchange new Diffie-Hellman public keys. The secrets from the \textit{Diffie-Hellman ratchet} output become the inputs to the root chain of the KDF chain, and then the output keys from the root chain become new KDF keys for the sending and receiving chains, which need to advance for each sending and receiving message. This is called the \textit{symmetric-key ratchet}.

The output keys from the symmetric-key ratchet are unique message keys which are used to either encrypt or decrypt messages. The sending and receiving chains ensure that each message is encrypted or decrypted with a unique key which can be deleted after use. 
The message keys are not used to derive new message keys or chaining keys. Because of this, it is possible to store the message key without affecting the security of other keys, only the message that belongs to the particular message key may be read if this key is compromised. This is \textit{useful for handling out-of-order messages} because a participant can store the message key and decrypt the message later when they receive the respective message. 
See more about out-of-order messages in Section~\ref{signal:ooomes}.

The Double Ratchet is formed by combining the symmetric-key ratchet and the Diffie-Hellman ratchet. If the Double Ratchet did not use the Diffie-Hellman ratchet to compute new chain keys for the sending and receiving chain keys, an attacker that can steal one of the chain keys can then compute all future message keys and decrypt all future messages \cite{double-ratchet}.

Each party generates a DH key pair, a public and a private key, which will be their first ratchet key pair. When a message is sent, the header must contain the current public key. When a message is received, the receiver checks the public keys that are given with the message and do a DH ratchet step to replace the receiver's existing ratchet key pair with a new one \cite{double-ratchet}.

The result is a kind of ``ping-pong'' behaviour as the two parties take turns replacing their key pairs. An attacker that compromises one message key has little use of it since this will soon be replaced with a new, unrelated key \cite{double-ratchet}.
The DH ratchet produces as output the sending  and the receiving chain keys, which have to correspond, i.e., the sending chain of one party is the same as the receiving chain of the other party (see Figure\ref{fig:dhratchet3}). 
Using a KDF chain here improves the resilience and break-in recovery. 

%
%

%

Combining the symmetric-key ratchet and Diffie-Hellman ratchet is as follows.
(A) When a message is sent or received, a symmetric-key  ratchet step is applied to the sending or receiving chain to derive the message key.
(B) When a new ratchet public key is received, a DH ratchet step is performed prior to the symmetric-key ratchet to replace the chain keys.
%

Fig.~\ref{fig:doubleratchet2} shows the information used by Alice to send her first message to Bob. The sending chain key (CK) is used on a symmetric-key ratchet step to derive a new CK and a message key, A1, to encrypt her message. The new CK is stored for later use, while the old CK and the message key can be deleted.
To ensure that the secrecy throughout the Double Ratchet is upheld, the old root key (RK) is deleted after it has been used to derive a new RK.
%

\begin{figure}[t]
\centering
	\begin{subfigure}{.50\textwidth}
  		\centering
\includegraphics[width=.95\linewidth]{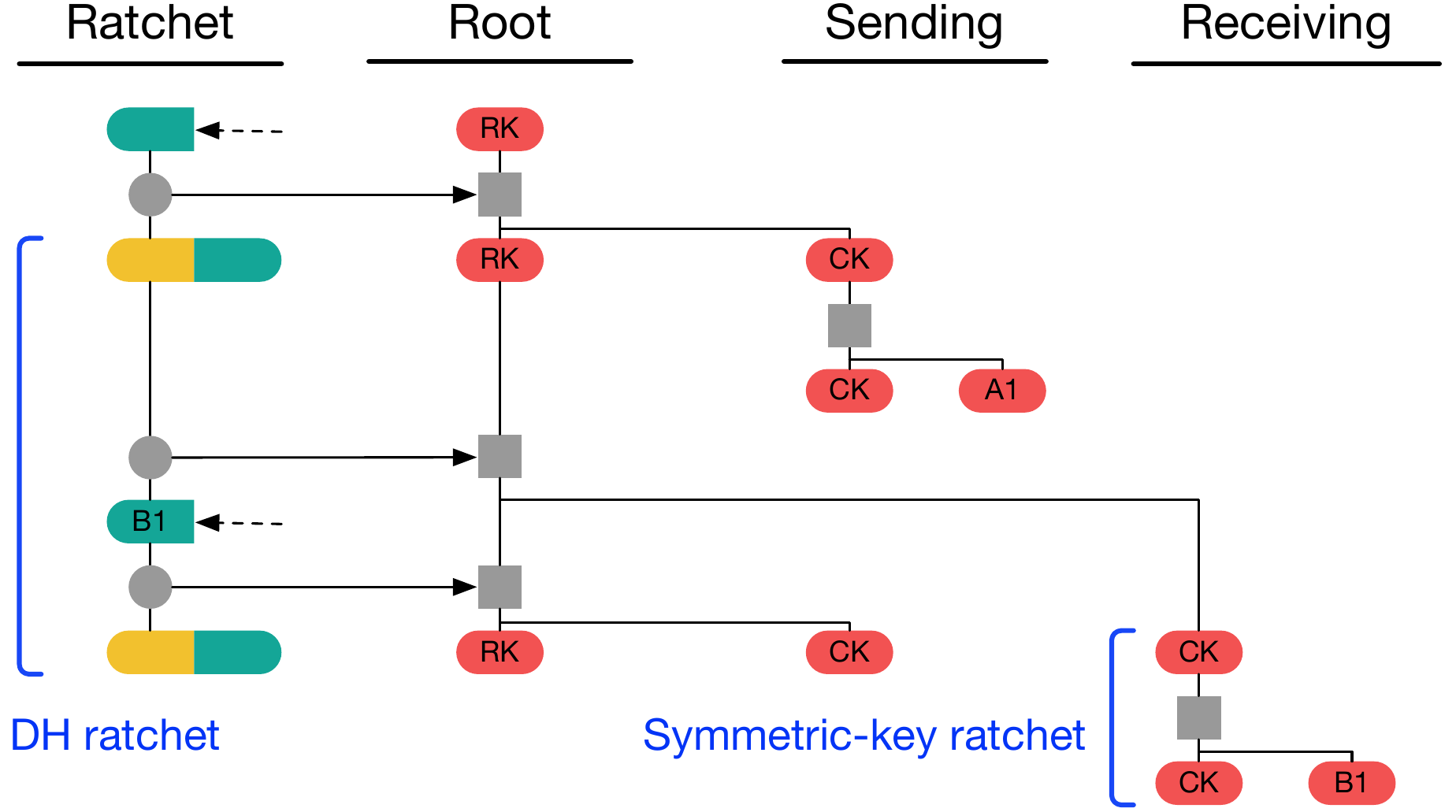}\\ \ \\ \ \\ \ \\
\caption{First receiving double ratchet message key B1 computed by Alice.}
\label{fig:doubleratchet3}
	\end{subfigure}%
	\begin{subfigure}{.52\textwidth}
  		\centering
  \includegraphics[width=.9\linewidth]{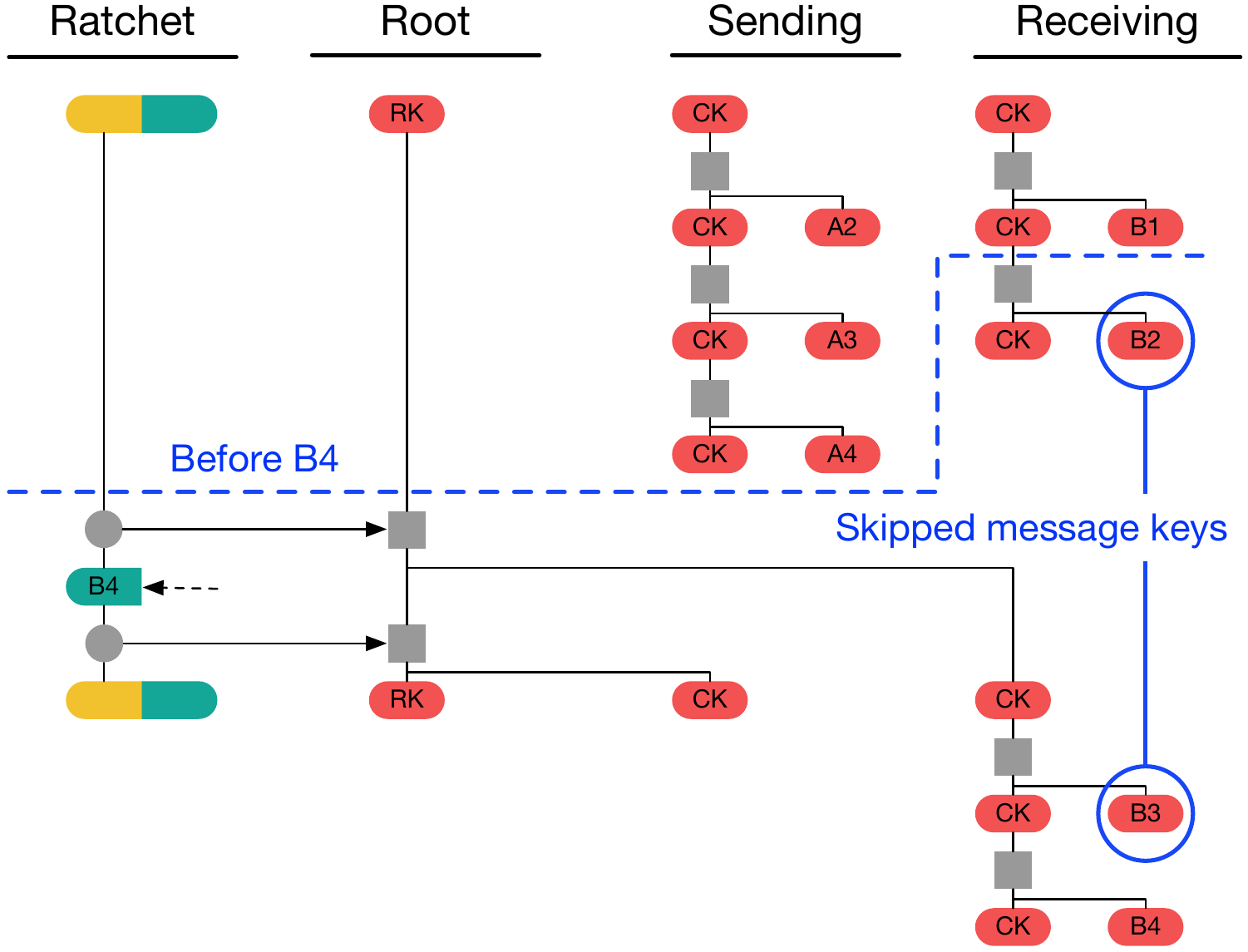}
\caption{Handling of Out-of-Order Messages.}
\label{fig:outoforderdubleratchet}
	\end{subfigure}
	\caption{Signal more advanced aspects of the Double Ratchet protocol; (reproduced from \cite{double-ratchet}).}
\end{figure}

Fig.~\ref{fig:doubleratchet3} shows the computations that Alice does when receiving a response from Bob, which includes his new ratchet public key. Alice applies a new DH ratchet step to derive a new receiving and sending chain keys. The receiving CK is used to derive a new receiving CK and a message key, B1, to decrypt Bob's message. Then she derives a new DH output for the next root KDF chain with her new ratchet private key to derive a new RK and a sending CK.

%

\subsubsection{Out-of-Order Messages} \label{signal:ooomes}

The Double Ratchet handles lost or out-of-order messages by including in each message header the message's number in the sending chain (N) and the length (number of message keys) in the previous sending chain (PN) \cite{double-ratchet}. This allows the receiver to advance the keys to the relevant message key, while still storing the skipped message keys in case they receive an older message at a later time.
%
%
Consider the example from Fig.~\ref{fig:outoforderdubleratchet} where we assume that Alice has already received message B1, and now 
she receives message B4 from Bob, with the PN = 2 and N = 1. Alice sees that she would need to do a DH ratchet step, but first, she calculates how many message keys she needs to store from her current receiving chain (Bob's previous sending chain). Since PN = 2 and her current receiving chain length is 1, the number of stored keys from the current receiving chain is 1 message key (i.e., B2). Then she does a DH ratchet step where a new receiving chain is derived. Because the length of her new receiving chain is 0, she needs to store a message key from her new receiving chain (i.e., B3). After Alice has stored B2 and B3, she can derive the last message key to decrypt message B4.


\subsubsection{The X3DH Key Agreement Protocol} \label{signal:x3dh-intro}

For Signal the X3DH is designed for asynchronous settings where one user, Bob, is offline but has published information to a server, and another user, Alice, wants to use that information to send encrypted data to Bob \cite{x3dh}. The Extended Triple Diffie-Hellman key agreement protocol (X3DH) thus establishes a shared secret between two parties who mutually authenticate each other based on public keys, and at the same time provides both forward secrecy and cryptographic deniability.

To provide asynchrony, a server is used to store messages from Alice and Bob which later can be retrieved; and the same server keeps the sets of keys for Alice and Bob to retrieve when needed \cite{x3dh}. 
%
%
The X3DH Protocol has three different phases:

\begin{enumerate}
\item Bob publishes his elliptic curve public keys to the server:
(i) Bob's identity key $IK_B$,
(ii) Bob's signed prekey $SPK_B$,
(iii) a set of Bob's one-time prekeys ($OBK_B^1$, $OBK_B^2$, $OBK_B^3$, \dots).
Identity keys need to be uploaded to the server once, while the other keys, such as new one-time prekeys can be uploaded again later if the server is getting low.
The server will delete a one-time prekey each time it sends it to another user. 

\item Alice fetches from the server Bob's identity key, signed prekey, prekey signature, and optionally a single one-time prekey. 
If the verification of the prekey signature fails the protocol is aborted.
Otherwise, Alice generates an ephemeral key pair with her public key $EK_A$, and will use the prekey to calculate several DH keys with the purpose to provide mutual authentication and forward secrecy (see details in \cite{x3dh}) used to generate the secret key for encryption (SK). After calculating the SK, Alice will delete her ephemeral private key and the DH
outputs to preserve secrecy.

Alice uses the key to send an initial message to Bob containing:
%
(i) Alice's identity key $IK_A$;
(ii) Alice's ephemeral key $EK_A$;
(iii) Identifiers stating which of Bob's prekeys Alice used;
(iv) An initial ciphertext encrypted with some AEAD encryption scheme \cite{AEAD} using AD as associated data and using an encryption key which is either SK or the output from some cryptographic pseudo-random function keyed by SK.
Alice's initial ciphertext is typically used as the first message in a post-X3DH communication protocol, such as the Double Ratchet protocol in the case of Signal. 

\item Bob receives and processes Alice's initial message. 
Bob will load his identity private key and the private key(s) corresponding to the signed prekey and one-time prekey that Alice used \cite{x3dh}.
Bob repeats the same steps with DH and KDF calculations to derive his own SK and then deletes the DH values, the same as Alice did. 
Afterwards, he tries to decrypt the initial ciphertext. 
The decryption is the only difference between what Bob does and what Alice did on her side. 
If the decryption fails, Bob will delete the SK and the protocol aborts, and the participants need to restart the protocol \cite{x3dh}.
If the decryption is successful, he gets the information that Alice had encrypted, and the protocol is complete for Bob. He deletes any one-time prekey private key that was used during the protocol in order to uphold the forward secrecy.
\end{enumerate}

\subsection{Security properties of protocols for end-to-end encrypted instant messaging}

\begin{table}[t]
\centering
\caption[Comparison of secure messaging protocols.]{Comparison of secure messaging protocols (reproduced from \cite{sok}).
\\
\pr{100} : Provides the property; \ \ \pr{50} : Partially provides the property; \ \ \pr{0} : Does not provide the property.
}
\label{tbl-properties}
\begin{sideways}
\begin{tabular}{@{\hspace{0ex}}ll@{\hspace{1ex}}l@{\hspace{1ex}}l@{\hspace{0ex}}}
\hline
Properties  & \multicolumn{3}{l}{\textbf{Protocol} / Client}                     \\ \cline{2-4}
                             &\textbf{OTR} & \textbf{Signal}  & \textbf{Matrix}  \\ 
                             &Pidgin &Signal & Riot \\ \hline
 \textbf{Security and Privacy}           &                 &         &       \\\vspace{1mm}
 Confidentiality           & \pr{100}                & \pr{100}        & \pr{100}      \\\vspace{1mm}
                      Integrity                 & \pr{100}                & \pr{100}        & \pr{100}      \\\vspace{1mm}
                     Authentication            & \pr{100}                & \pr{100}        & \pr{100}      \\\vspace{1mm}
                      Participant Consistency   & \pr{100}                & \pr{100}        & \pr{100}      \\\vspace{1mm}
                      Destination Validation    & \pr{100}                & \pr{100}        & \pr{100}      \\\vspace{1mm}
                      Forward Secrecy           & \pr{50}                 & \pr{100}        & \pr{50}       \\\vspace{1mm}
                      Backward Secrecy          & \pr{100}                & \pr{100}        & \pr{50}       \\\vspace{1mm}
                      Anonymity Preserving      & \pr{100}                & \pr{0}               & \pr{0}             \\\vspace{1mm}
                      Speaker Consistency       & \pr{50}                 & \pr{100}        & \pr{100}      \\\vspace{1mm}
                      Causality Preserving      & \pr{50}                 & \pr{100}        & \pr{100}      \\\vspace{1mm}
                      Global Transcript         & \pr{0}                       & \pr{0}               & \pr{0}             \\\vspace{1mm}
                      Message Unlinkability     & \pr{100}                & \pr{100}        & \pr{100}      \\\vspace{1mm}
                      Message Repudiation       & \pr{100}                & \pr{100}        & \pr{100}      \\\vspace{3.5mm}
                      Participation Repudiation & \pr{50}                 & \pr{100}        & \pr{100}      \\\vspace{1mm}
 \textbf{Usability and Adoption }    &                  &         &       \\\vspace{1mm}
 Out-of-Order Resilient    & \pr{50}                 & \pr{100}        & \pr{100}      \\\vspace{1mm}
                      Dropped Message Resilient & \pr{50}                 & \pr{100}        & \pr{100}      \\\vspace{1mm}
                      Asynchronicity            & \pr{0}                       & \pr{100}        & \pr{100}      \\\vspace{1mm}
                      Multi-Device Support      & \pr{0}                       & \pr{50}         & \pr{100}      \\\vspace{3.5mm}
                      No Additional Service     & \pr{100}                & \pr{0}               & \pr{0}             \\\vspace{1mm}
            \textbf{Group Chat}    &                         &        &    \\\vspace{1mm}
 Computational Equality    &  \pr{0}                & \pr{100}        & \pr{100}      \\\vspace{1mm}
                      Trust Equality            &     \pr{0}                    & \pr{100}        & \pr{100}      \\\vspace{1mm}
                      Subgroup Messaging        &     \pr{0}                    & \pr{100}        & \pr{100}      \\\vspace{1mm}
                      Contractable membership   &     \pr{0}                    & \pr{100}        & \pr{100}      \\
                      Expandable membership     &     \pr{0}                    & \pr{100}        & \pr{100}      \\ \hline
\end{tabular}

\end{sideways}

\end{table}

Table \ref{tbl-properties} shows that none of the secure messaging protocols we have gone through in this section can give the users every security property (for more information about the definition of the properties used in Table~\ref{tbl-properties} please refer to \cite{sok}). 
In this section we briefly comment on each of these, including for completeness also the Matrix\footnote{Matrix protocol having several IM implementations; at \url{https://matrix.org}} as a major protocol.

While the Off-the-Record protocol does not need any additional services or servers, it cannot provide group conversation (in the current version and implementations). There have been research works investigating group conversations on top of OTR \cite{group-otr,impro-gotr,group-otr2}, but they have not received enough attention from the developers mainly because these do not support asynchronous chat conversations.
While Signal supports desktops through the Chrome Extensions, it does not support native desktop application. Moreover, it only allows for one device to be used, i.e., multiple mobile phones cannot be added to a user's account. 
This could be achieved using the same functionality for group conversations, but efficiency could be a problem.

The Matrix protocols and application (see Section~\ref{subsubsec_matrix} for details) supports multiple devices, without affecting the efficiency of the conversations. However, it does not achieve full forward and backward secrecy in the protocols, but the implementation does.
%
The Signal protocol has been audited by two research groups in 2017 \cite{formal-signal,signal-audit-inria} and since it is open source, the community can improve it. 
The Matrix protocol has also been audited  \cite{ncc-group-matrix}.
This indicates that researchers are taking these protocols seriously and want to strengthen their credibility.

\subsubsection{Off-the-Record}

OTR uses an encrypt-then-MAC approach to protect messages (see Section~\ref{subsec-OTR} for details) which provides \textit{confidentiality, integrity, and authentication}. 
The SIGMA protocol (a variant of authenticated Diffie-Hellman key exchange) ensures \textit{participation consistency} for the key exchange \cite{sigma}.
\textit{Forward secrecy} is ensured by the fact that message keys are regularly replaced with new key material during the conversation. 
\textit{Backward secrecy} is ensured by the fact that message keys are computed by new DH values which are advertised by the sender with each sent message. 
\textit{Anonymity preservation} is ensured by the fact that the long-term public keys are never observed, neither during the key exchange nor during the conversation. 
\textit{Causality preservation} is only partially achieved, as messages implicitly reference their causal predecessors based on which keys they use \cite{sok}. 
\textit{Speaker consistency} is only partially achieved since an adversary cannot drop messages without also dropping all future messages, for otherwise the recipients would not be able to decrypt subsequent messages \cite{sok}. The aftermath of the speaker consistency is that the recipient needs to save out-of-order messages because if they do not come in order the message will be encrypted with an unexpected key, and at the same time the window of compromises enlarges, and the OTR would end up only partially providing the forward secrecy. 
\textit{Out-of-order and dropped messages} are only partially provided because if a message is out-of-order or dropped during the transmission, the protocol can store the decryption key until the participant receives that message. The problem of storing the decryption key is that it raises the possibility of successful attacks on the client side.

The OTR protocol signs the messages with the shared MAC keys and not the long-term keys. To strengthen the \textit{message unlinkability} and \textit{message repudiation} features, OTR uses malleable encryption and the MAC keys are published after each message exchange \cite{otr-com}. OTR only signs the ephemeral keys and not every parameter during the key exchange, which provides only \textit{partial participation repudiation} since the conversation partners can use the signed ephemeral keys to forge transcripts.
The OTR protocol is intended for instant messaging, and thus does not provide asynchronous messaging. However, the synchronous only requirement allows OTR to not rely on additional services for establishing a connection between two participants.

\subsubsection{Signal}

The design of the Signal protocol extends OTR, thus maintaining the same security features, while in some cases adding stronger or new features as well. 
\textit{Forward secrecy} is provided because of the use of thee KDF ratchets, whereas \textit{backward secrecy} is provided because even when KDF keys are comprised, they are soon replaced by new keys.
The X3DH handshake (Section~\ref{signal:x3dh-intro}) provides the same level of \textit{authentication} as the SIGMA from OTR, but X3DH achieves full \textit{participation repudiation} since anybody can forge a transcript between two parties \cite{sok}. However, Signal \textit{fails to provide anonymity preserving} because X3DH uses the long-term public keys during the initial key agreement.
The prekeys are used to provide an \textit{asynchronous messaging} system by sending a set of prekeys to a central server, and then a sender can request the next prekey for the receiver to compute encryption keys. By using a central server to keep the prekeys, the Signal protocol \textit{loses the no additional service property}.
\textit{Out-of-order and dropped messages} are fully supported on one-to-one conversations asynchronously by the use of prekeys.

\textit{Group conversation} is achieved by using multicast encryption, which when sending a single encrypted message to the group, it is sent to a server and then relays it to the other participants while the decryption key is sent as a standalone message to each member of the group conversation. The group conversation provides asynchronous messaging, speaker consistency, and causality preservation, by attaching message identifiers, of the messages before, to the new message \cite{sok}, but it cannot guarantee participant consistency. Multi-device is partly provided, in the sense that only an extra computer can join in a conversation by using the Signal Desktop application\footnote{\url{https://whispersystems.org/blog/signal-desktop/}}, which is only a Chrome Extension\footnote{\url{https://en.wikipedia.org/wiki/Browser_extension}} and not an own application.

The Signal protocol provides computational and trust equality, subgroup messaging, contractible and expandable membership properties. By using pairwise group messaging and multicast encryption, Signal has the ability to push group management into the client apps, which makes it easier for the users to change the group, expand it or shrink it in size, without having to restart the whole group conversation and protocol. When users want to send a group message they send a message to each of the users that are participating and adding a parameter to the header marking that it is meant for the specific group chat. The Signal server does not know about the group conversation, since the messages are encrypted using their normal public key. The pairwise group messaging also makes the computation of new cryptographic keys and trust equality as computationally demanding as if there was only a one-to-one conversation.

\subsubsection{Matrix}\label{subsubsec_matrix}

The Matrix protocol consists of two different algorithms, the Olm\footnote{\url{https://matrix.org/docs/spec/olm.html}} for one-to-one conversations and 
Megolm\footnote{\url{https://matrix.org/docs/spec/megolm.html}} for group conversations between multiple devices. The Olm algorithm is based on the Signal protocol, which means they achieve the same security properties as Signal does, while the Megolm algorithm is a new AES-based cryptographic ratchet developed for group conversations.
%
Multiple devices are possible with Matrix because Megolm implements a separate ratchet per sending device that is participating in a group conversation.\footnote{Matrix.org Launches Cross-platform Beta of End-to-End Encryption Following Security Assessment by NCC Group \url{https://pr.blonde20.com/matrix-e2e/}}
The protocol does not restart when the ratchet is replaced with a new one, which provides computational and trust equality, subgroup messaging, contractible and expandable membership properties.

The NCC Group has audited both algorithms \cite{ncc-group-matrix} and found that Megolm has some security flaws about forward and future secrecy. If an attacker manages to compromise the key to Megolm sessions, then it can decrypt any future messages sent to the participants in a group conversation. The Matrix SDK, which is used in the applications that have implemented the Matrix protocol such as Riot\footnote{\url{https://about.riot.im/}}; ensures that the Megolm keys get refreshed after a certain amount of messages. Forward secrecy is only partially provided since the Megolm maintains a record of the ratchet value which allows them to decrypt any messages sent in the session after the corresponding point in the conversation. The Matrix developers have stated that this is intentionally designed \cite{matrix-conf}, but also said that it is up to the application to offer the user the option to discard old conversations.

\section{Analysis of Applications Implementing Secure Messaging}
\label{review-applications}

This section surveys applications (mostly smartphone apps) that advertise secure messaging conversation capabilities between one-to-one and/or many-to-many users. We investigate a set of usability properties relevant for secure messaging. 
\subsection{Test Scenarios}
We first explain in this subsection the test scenarios that we carried and the security and usability properties we are looking for.
The test scenarios are the same for each application, and screenshots were taken during the testing phases to gather enough information for later analysis.
We are going to study a set of properties 
in 
each test scenario, 
chosen to discover how secure and usable each application is. 
The test scenarios that the applications are going through are described in the following subsections.

\subsubsection{Initial Set Up}
This test scenario includes two stages: ``Setup and Registration''  and ``Initial Contact''. 
Stage one is the first process a user needs to go through after installing an application. Here we test how the applications handle the registration process, what the user needs to do to register a new account, and whether there are multiple ways to register.
The properties of interest are:
\begin{itemize}
	\item \textbf{Phone registration:} User can register an account with a phone number.
	\item \textbf{E-mail registration:} User can register an account with an e-mail address.
	\item \textbf{Verification by SMS:} Receive verification code through SMS.
	\item \textbf{Verification by Phone Call:} Receive verification code through a phone call.
	\item \textbf{Access SMS inbox:} The app requires access to the SMS Inbox in order to read the verification code automatically.
	\item \textbf{Contact list upload:} The app requires to upload contacts to a server in order to see if others are using the same application.
\end{itemize}


Stage two examines how applications handle the first message sent from one participant to another, whether the participants are informed of the secure messaging capabilities or whether the app shows how the cryptographic keys are used. 
Properties:
\begin{itemize}
	\item \textbf{Trust-On-First-Use:} Automatically verify each other on initiation.
	\item \textbf{Notification About E2E Encryption:} The app presents notifications to explain to the user that messages are encrypted.
\end{itemize}

\subsubsection{Message After a Key Change}

This scenario tests how the application handles changes of cryptographic keys after Bob deletes the application in the middle of a conversation with Alice. After Bob has reinstalled his application, Alice sends him a new message and examines if the application gives Alice any information about the key changes.
When a user deletes a secure messaging application, the cryptographic keys are normally deleted from the device to strengthen the security of the messages the participant has already sent. When a participant then reinstalls the application, a new set of cryptographic keys are generated.  
%
Properties of interest are:
\begin{itemize}
	\item \textbf{Notification about key changes:} Notifies Alice that Bob has changed cryptographic keys.
	\item \textbf{Blocking message:} Blocks new messages from being sent until Alice and Bob verify each other.
\end{itemize}

\subsubsection{Key Change While a Message is in Transit}

Cryptographic key changes while a message is in transit is similar to the test scenario before; however, we are interested in what happens when a message is lost before new keys are generated. Bob deletes his application without telling Alice; she then sends Bob a message, but the message is lost in transit. 
%
Properties of interest are:
\begin{itemize}
	\item \textbf{Re-encrypt and send message:} Does the application re-encrypt the message and sends it again after the receiver has generated new cryptographic keys or is the message lost forever?
	\item \textbf{Details about transmission of message:} Users can see the difference between sent and delivered messages. 
\end{itemize}

\subsubsection{Verification Process Between Participants}

In a conversation, Alice and Bob want to verify each other to ensure that they are having a conversation with honest participants. This test scenario looks at how the verification process works and if it is a secure and usable method of doing it.
%
\begin{itemize}
	\item \textbf{QR-code:} Verify each other through a QR-code.
	\item \textbf{Verify by Phone call:} Call each other with E2E-encrypted phone call and read keys out loud.
	\item \textbf{Share keys through 3rd party:} Share the keys through other applications.
	\item \textbf{Verified check:} Users can check later if a specific user is already verified.
\end{itemize}

\subsubsection{Other Security Implementations}

Each application may have additional security and privacy features meant to protect from various intrusions or attacks.
%
\begin{itemize}
	\item \textbf{Passphrase/code:} Possibility of a passphrase/code that only the user knows and enters to access the application.
	\item \textbf{Two-step verification:} When registering after a reinstall or new device, then a second passphrase/code is needed which only the specific user knows. 
	\item \textbf{Screen security:} The user is not allowed to screenshot within the application.
	\item \textbf{Clear trusted contacts:} Can all verified contacts be cleared, which means the user needs to verify each contact again?
	\item \textbf{Delete devices from account:} If the application allows multiple devices, is there an option to delete devices which are not in use anymore.
\end{itemize}


\subsection{Running The Different Test Cases}

\subsubsection{Case 1: Signal} 
\label{impl:signal}

Signal is an instant messaging application as well as a voice calling application, for both Android and iOS. What sets the Signal application apart from the other applications, except for Riot, is the fact that it is completely open source. This reassures people that it does what the developers claim, since anyone can audit the source code as well as the employed cryptographic protocols.


\paragraph{Initial Set Up:}\label{impl:signal-setup}

\begin{wrapfigure}[19]{r}{.49\textwidth}
\centering
	\begin{subfigure}{.24\textwidth}
  		\centering
  		\includegraphics[width=.95\linewidth]{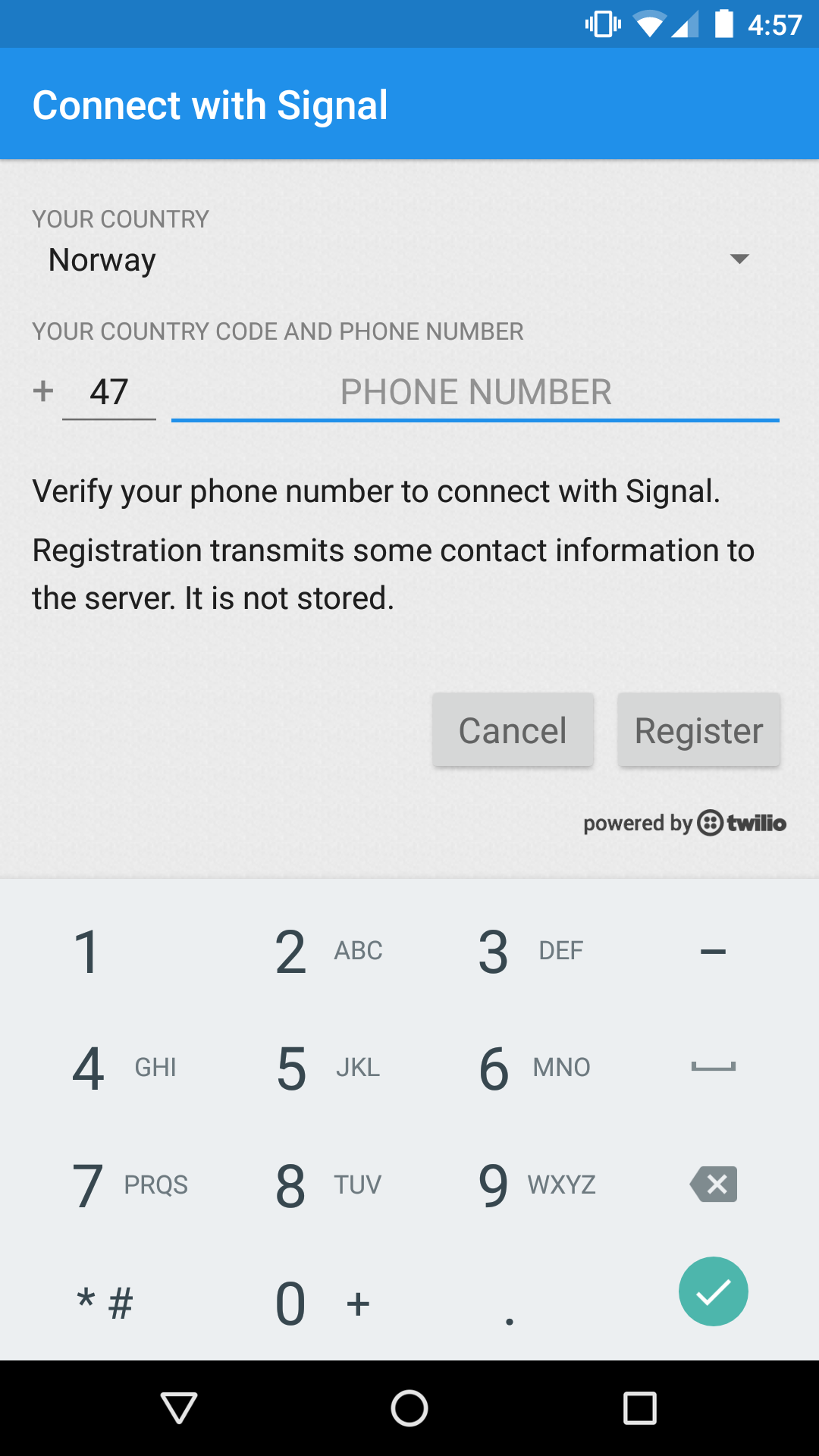}
  		\caption{phone number registration}
		\label{fig:impl-signal-reg-1}
	\end{subfigure}%
	\begin{subfigure}{.24\textwidth}
  		\centering
  		\includegraphics[width=.95\linewidth]{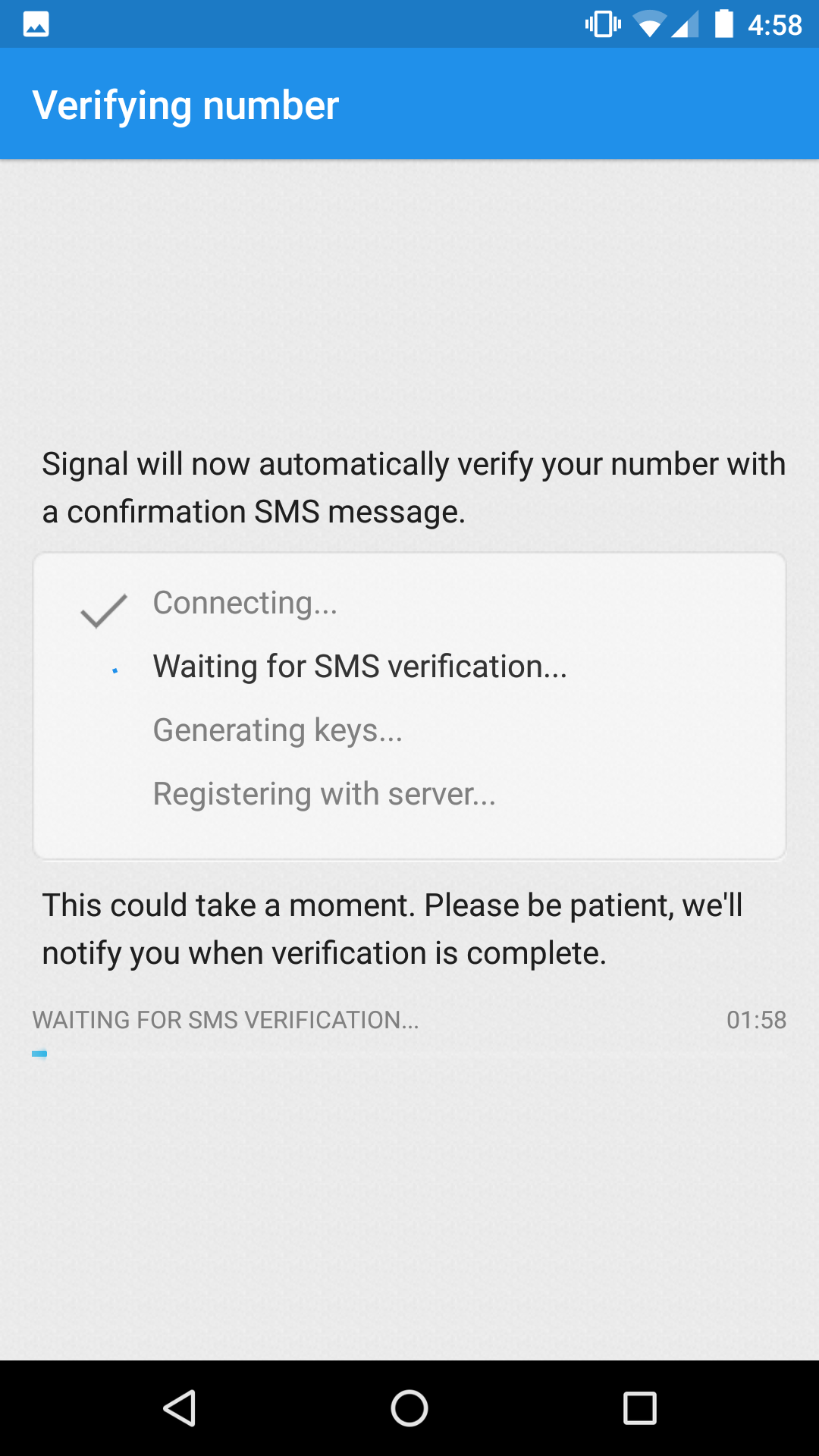}
  		\caption{verifying the phone number}
  		\label{fig:impl-signal-reg-2}
	\end{subfigure}
\caption{Signal: registration process}
\label{fig:impl-signal-1}
\end{wrapfigure}

An account can be registered to one device at a time, which means that using the same number on a second device, will automatically deactivate the first device, to strengthen the security and to keep the private cryptographic keys on one device only.
Fig.~\ref{fig:impl-signal-1}\subref{fig:impl-signal-reg-1} shows the first view a user sees when opening the app for the first time. Twilio
is used for handling the SMS verification process with the Signal server when registering an account. Contact information are transmitted to the server, but are not stored.

Fig.~\ref{fig:impl-signal-1}\subref{fig:impl-signal-reg-2} explains the different steps the Signal app goes through to register and verify a new user account. The verification code is sent as an SMS, and the app reads the SMS automatically to verify the new user. After the verification, the app generates new device cryptographic keys.
At the end, the app registers the account within the Signal server.
If the user does not give the application access to their SMS inbox, then it has to wait for the SMS verification timer to time out, as shown at the bottom of Fig.~\ref{fig:impl-signal-1}\subref{fig:impl-signal-reg-2}, after which the  Signal application calls the user and gives out a verification number to be typed in manually.


\paragraph{Message After a Key Change:}\label{impl:signal-keychanges}


\begin{figure}[h]
\centering
	\begin{subfigure}{.24\textwidth}
  		\centering
  		\includegraphics[width=.95\linewidth]{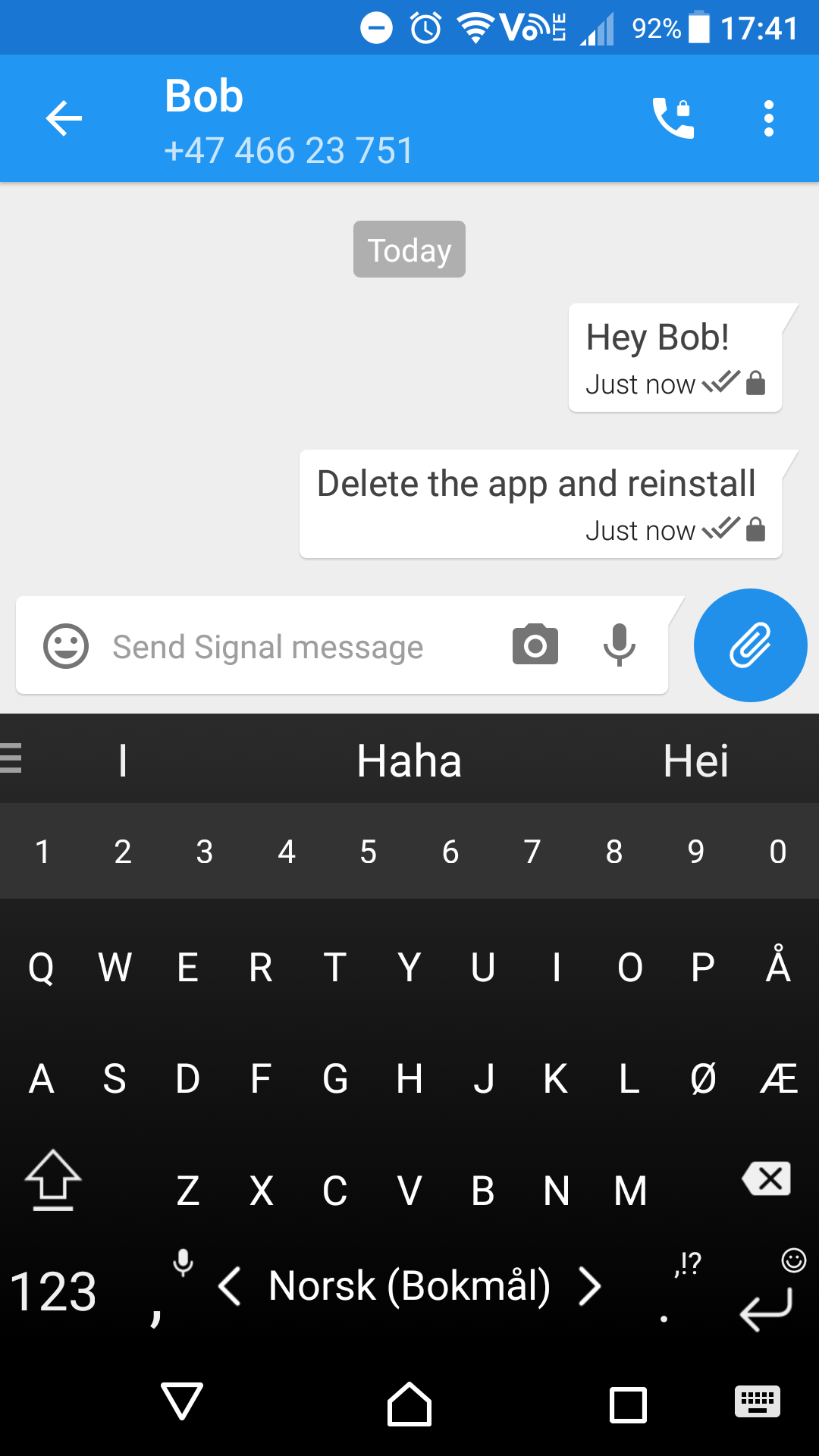}
  		\caption{Alice's first message\\ \ }
  		\label{fig:impl-signal-kc-1}
	\end{subfigure}
	\begin{subfigure}{.24\textwidth}
  		\centering
  		\includegraphics[width=.95\linewidth]{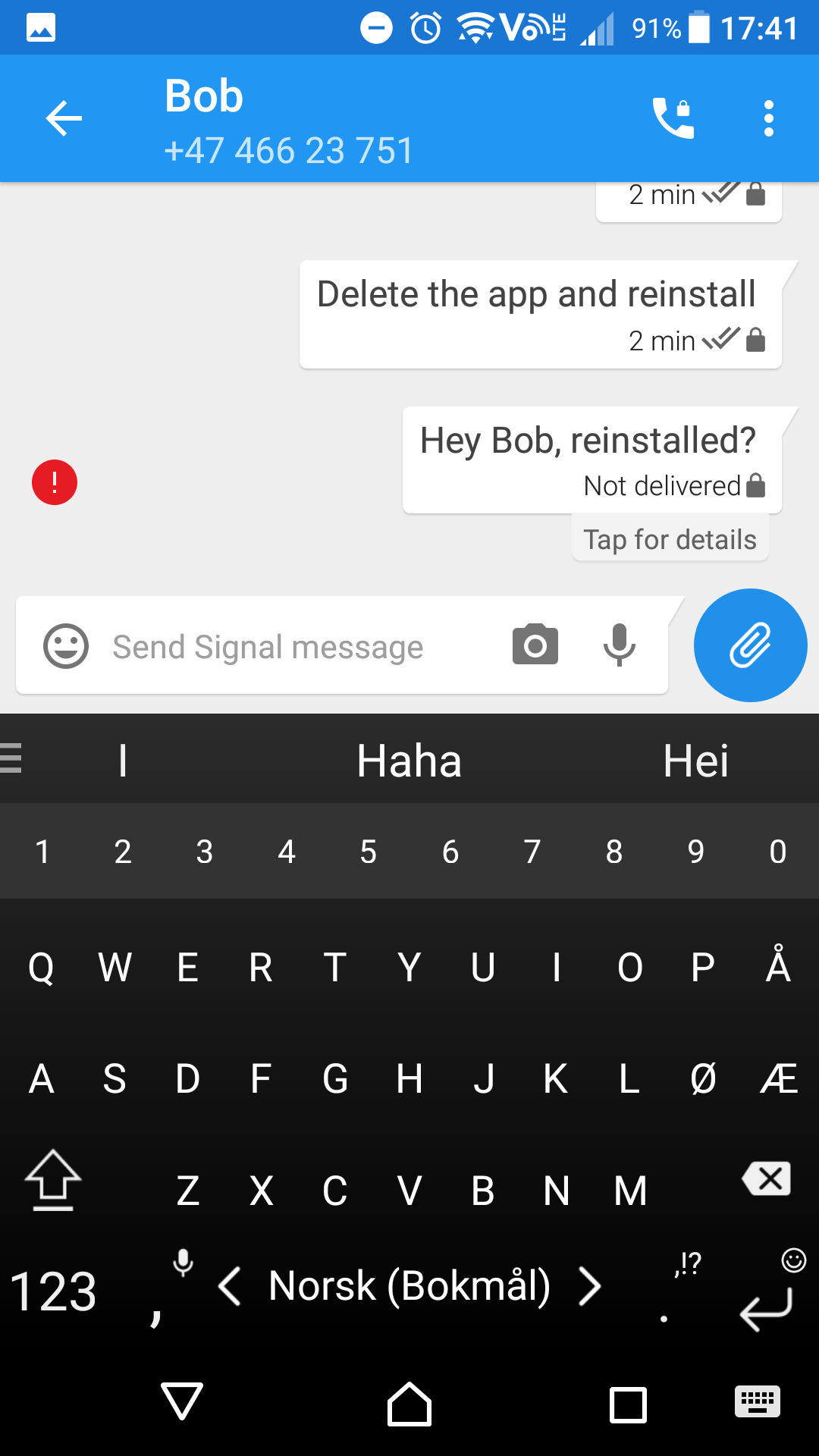}
  		\caption{Message after reinstall\\ \ }
  		\label{fig:impl-signal-kc-2}
	\end{subfigure}
	\begin{subfigure}{.24\textwidth}
  		\centering
  		\includegraphics[width=.95\linewidth]{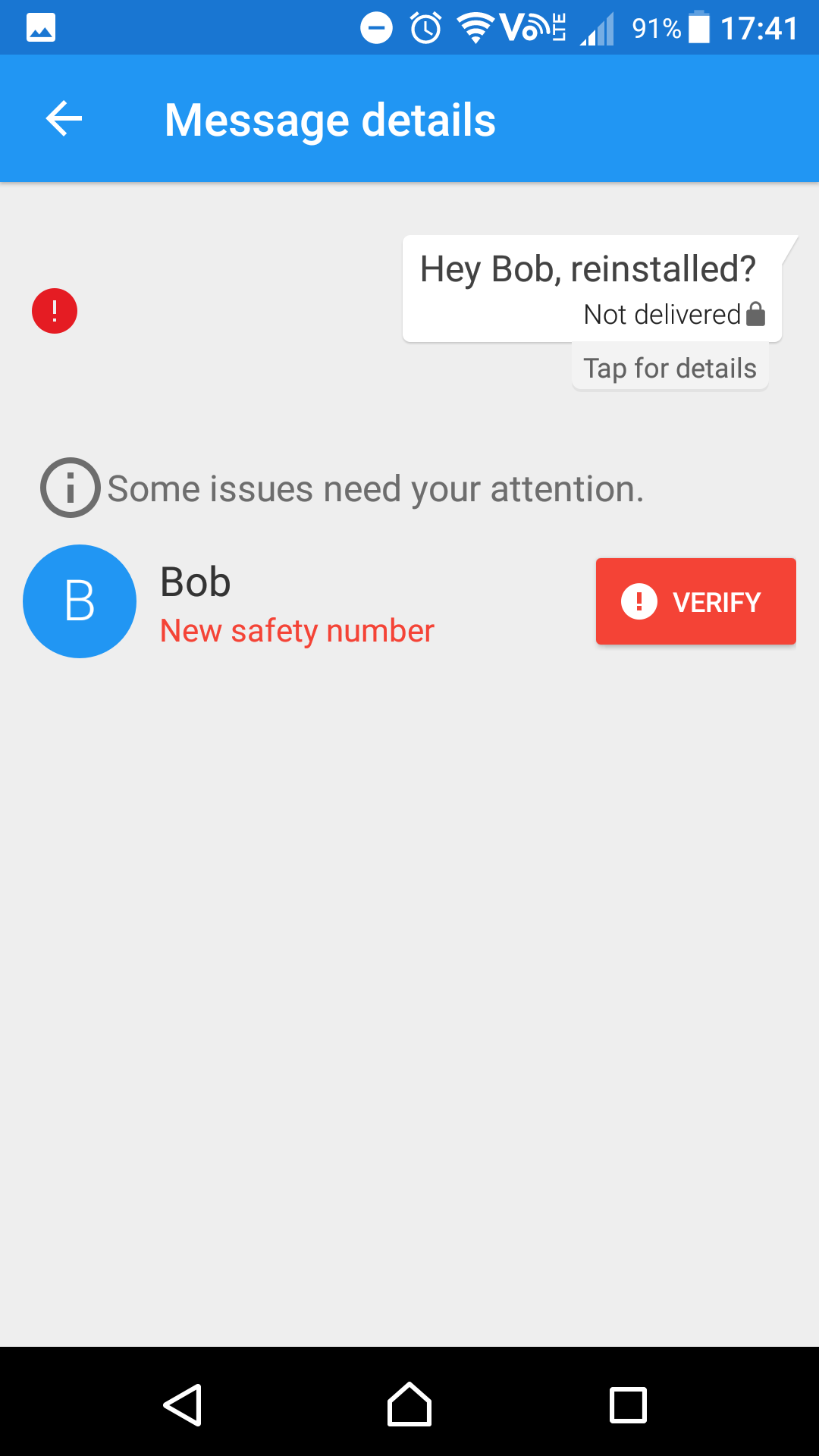}
  		\caption{Verifying Bob again\\ \ }
  		\label{fig:impl-signal-kc-3}
	\end{subfigure}
	\begin{subfigure}{.24\textwidth}
  		\centering
  		\includegraphics[width=.95\linewidth]{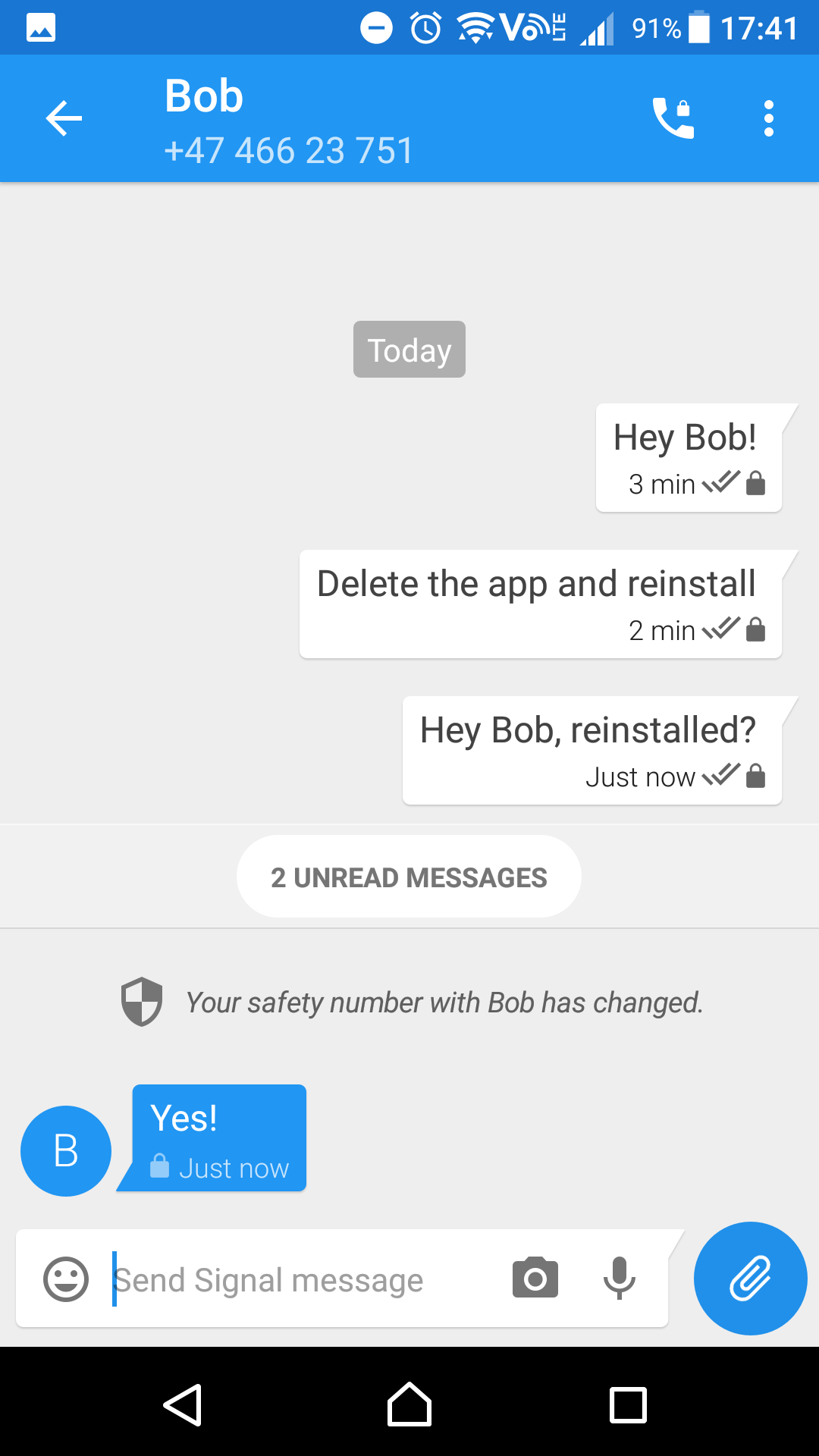}
  		\caption{Message after verification}
  		\label{fig:impl-signal-kc-4}
	\end{subfigure}
	\caption{Signal: Message after key change}
	\label{fig:impl-signal-kc}
\end{figure}

This test scenario analyses what happens when cryptographic keys change, e.g., when a user in a conversation deletes and then reinstalls the Signal app. Fig.~\ref{fig:impl-signal-kc}\subref{fig:impl-signal-kc-1} shows the first two messages that Alice sends to Bob.
The \textit{double checkmark} shown on each message indicates that it has been received and read by Bob. The lock indicates that the message is encrypted from one end to the other, and nobody in between can read it.
Fig.~\ref{fig:impl-signal-kc}\subref{fig:impl-signal-kc-2} shows when Alice sends Bob another message after he has deleted and reinstalled his Signal app. The application notifies Alice that the message has not been delivered with a red notification icon on the left. It also gives information that by pressing on the message the user can get more details about the notification.

Fig.~\ref{fig:impl-signal-kc}\subref{fig:impl-signal-kc-3} is the view the user sees when pressing the message that was not delivered. Alice is presented with information that Bob has a new security number (cryptographic keys), and she needs to verify the new keys to get the ability to send any new message to Bob. 
%
After the verification process between Alice and Bob is done, they can continue the conversation, and a notification is posted in the conversation that Bob has changed his security number, as shown in Fig.~\ref{fig:impl-signal-kc}\subref{fig:impl-signal-kc-4}.

\begin{wrapfigure}[19]{r}{.49\textwidth}
\vspace{-3ex}\centering
	\begin{subfigure}{.24\textwidth}
  		\centering
  		\includegraphics[width=.95\linewidth]{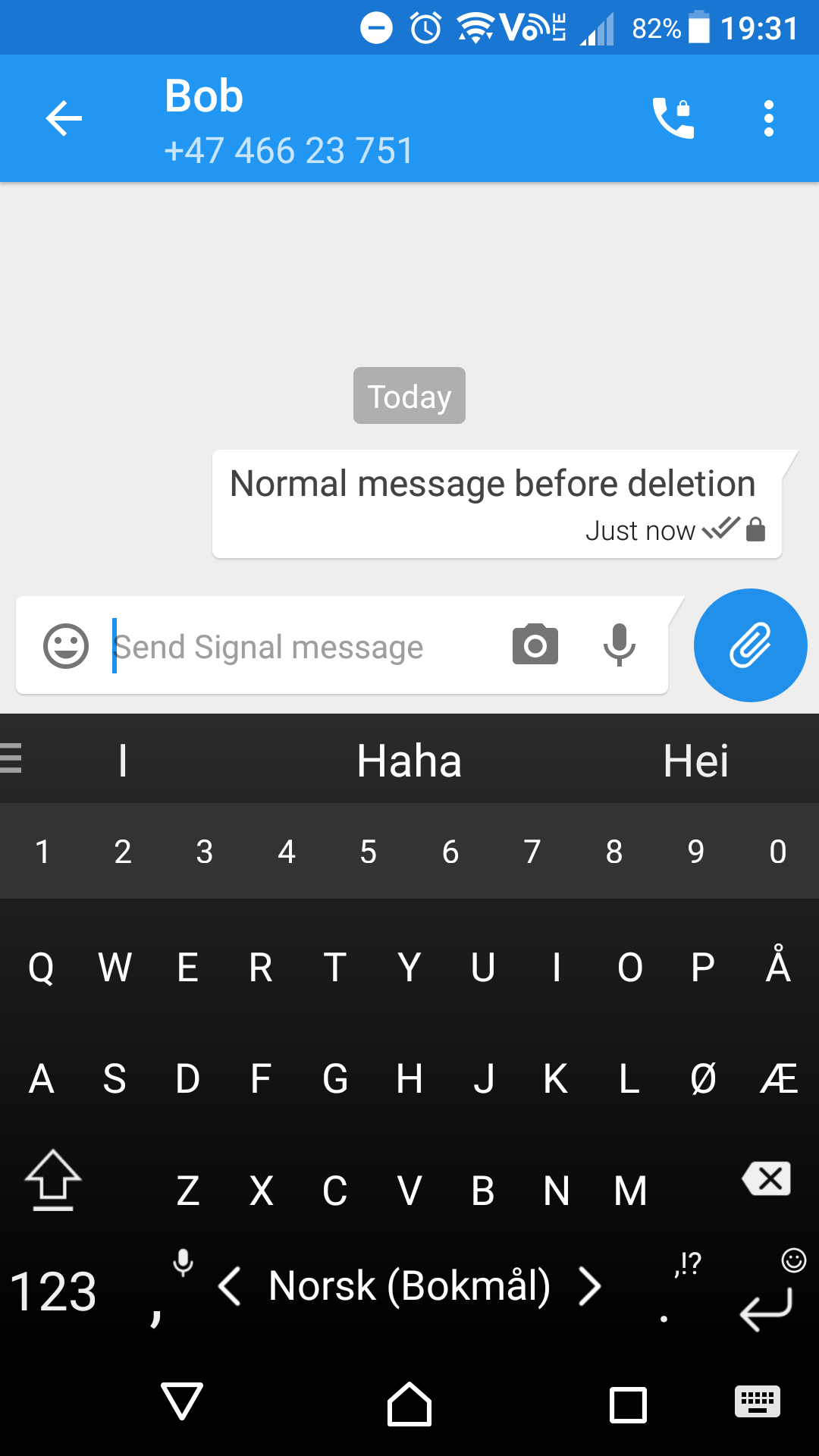}
  		\caption{Message before key change}
  		\label{fig:impl-signal-kc-transit-1}
	\end{subfigure}
	\begin{subfigure}{.24\textwidth}
  		\centering
  		\includegraphics[width=.95\linewidth]{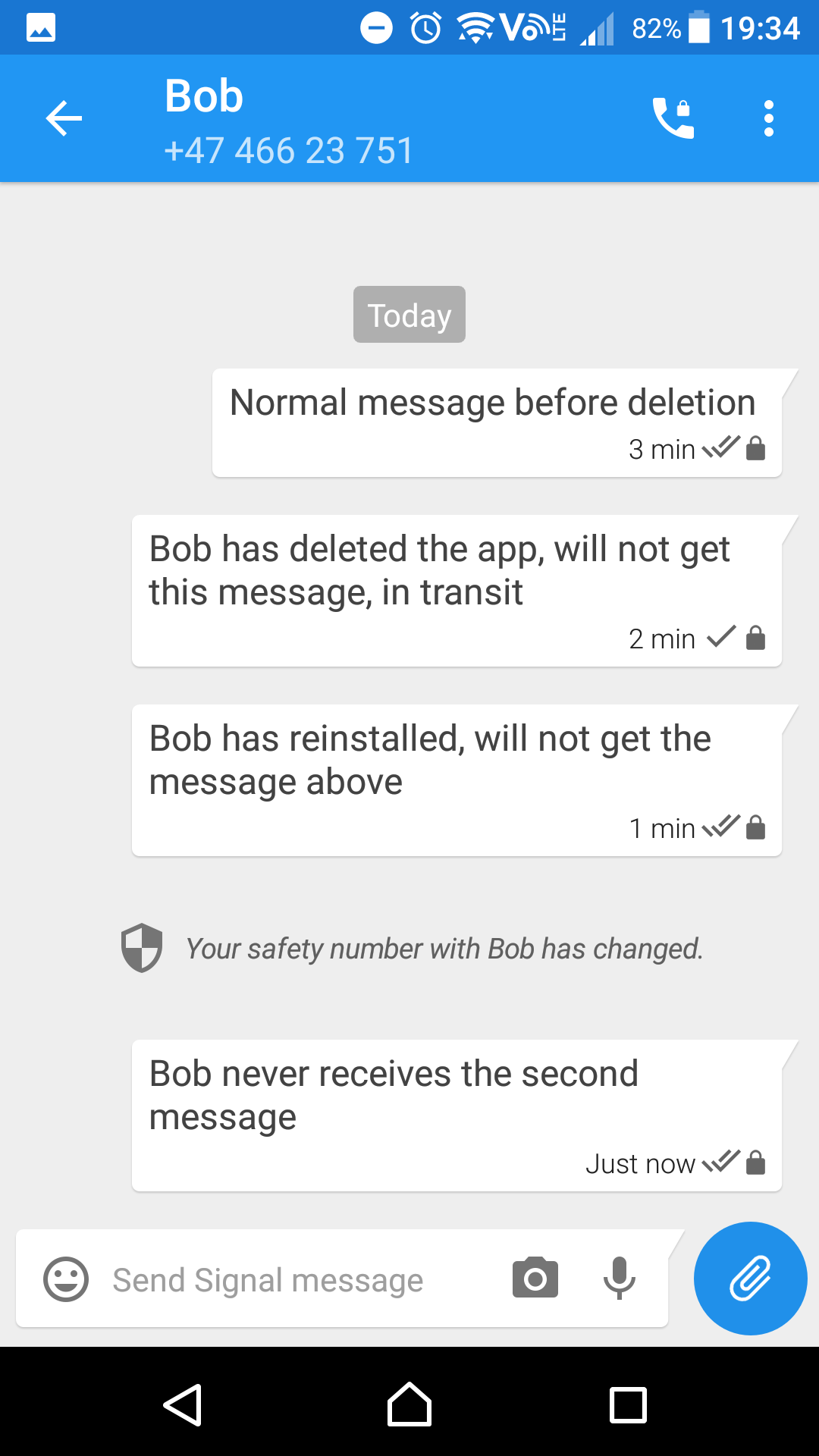}
  		\caption{Message after key change}
  		\label{fig:impl-signal-kc-transit-2}
	\end{subfigure}
	\caption{Signal: Key change while a message is in transit}
	\label{fig:impl-signal-kc-transit}
\end{wrapfigure}
%
\paragraph{Key Change While a Message is in Transit:}

This test scenario is mostly the same as the previous one, with the difference that here we want to check what does the Signal app do when a message is sent before Bob has managed to reinstall his Signal application, i.e., handling of messages lost in transit.
 
Fig.~\ref{fig:impl-signal-kc-transit}\subref{fig:impl-signal-kc-transit-1} shows the initialization of the conversation.
%
Fig.~\ref{fig:impl-signal-kc-transit}\subref{fig:impl-signal-kc-transit-2} shows the conversation after a couple of messages from Alice to Bob. The second message is sent after Bob has deleted his application, and there is only a single checkmark on that message, which means the message has been sent, but not received.
The icons on the third message indicate that Bob has finally reinstalled, but he never received the second message which was sent before he reinstalled. After Alice and Bob verify their new security number, all new messages are received and encrypted by both sides, but the second message is never received.
The reason for never receiving the second message in Fig.~\ref{fig:impl-signal-kc-transit}\subref{fig:impl-signal-kc-transit-2} is because the Signal app never stores messages that are encrypted after they are sent to the server, and the messages are never re-encrypted by Alice when Bob has changed his cryptographic keys.


%
\paragraph{Verification Process Between Participants:}
\begin{wrapfigure}[19]{r}{.49\textwidth}
\vspace{-3ex}\centering
	\begin{subfigure}{.24\textwidth}
  		\centering
  		\includegraphics[width=.95\linewidth]{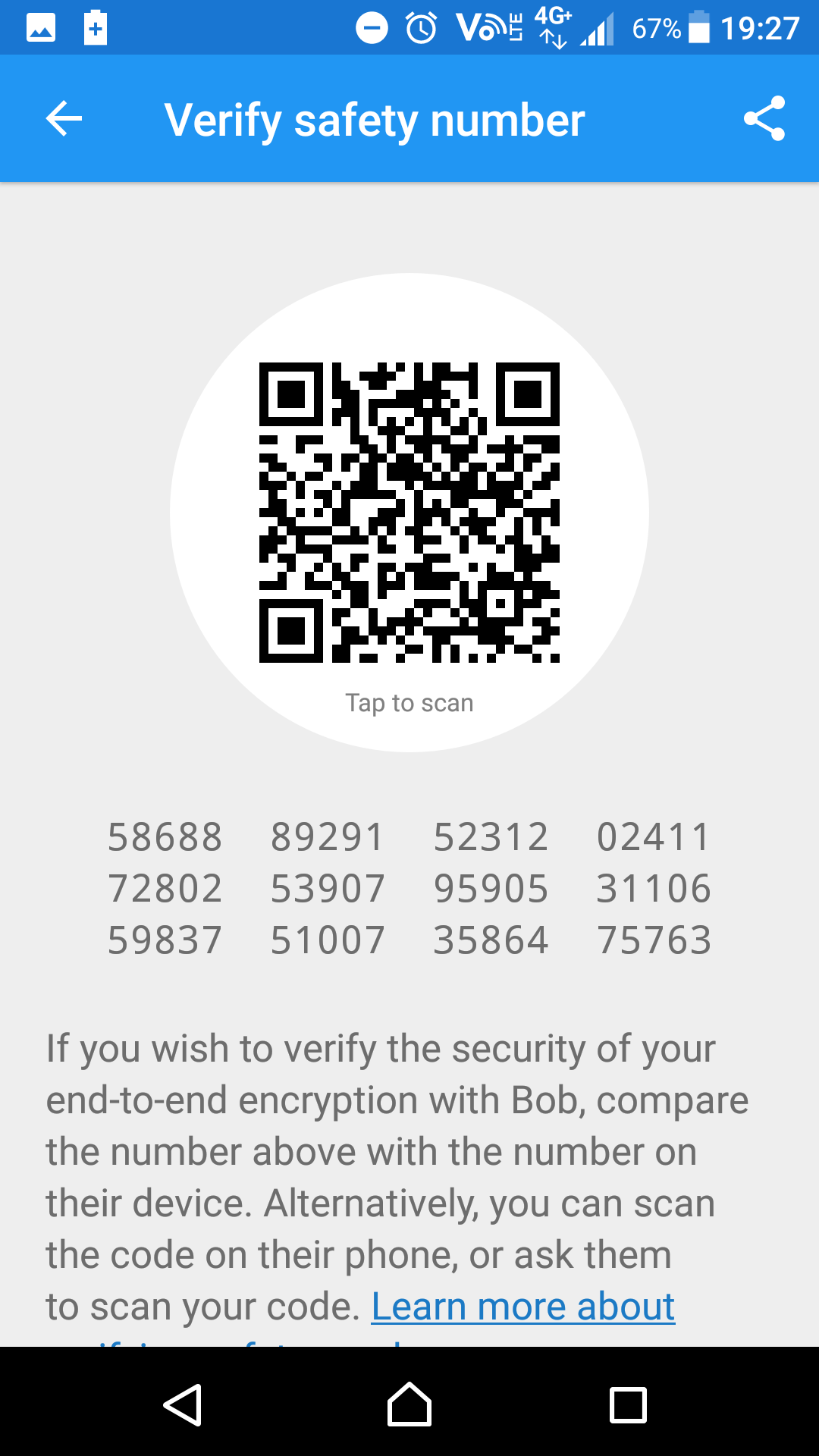}
  		\caption{Verification page}
  	\label{fig:impl-signal-verify-1}
	\end{subfigure}%
	\begin{subfigure}{.24\textwidth}
  		\centering
  		\includegraphics[width=.95\linewidth]{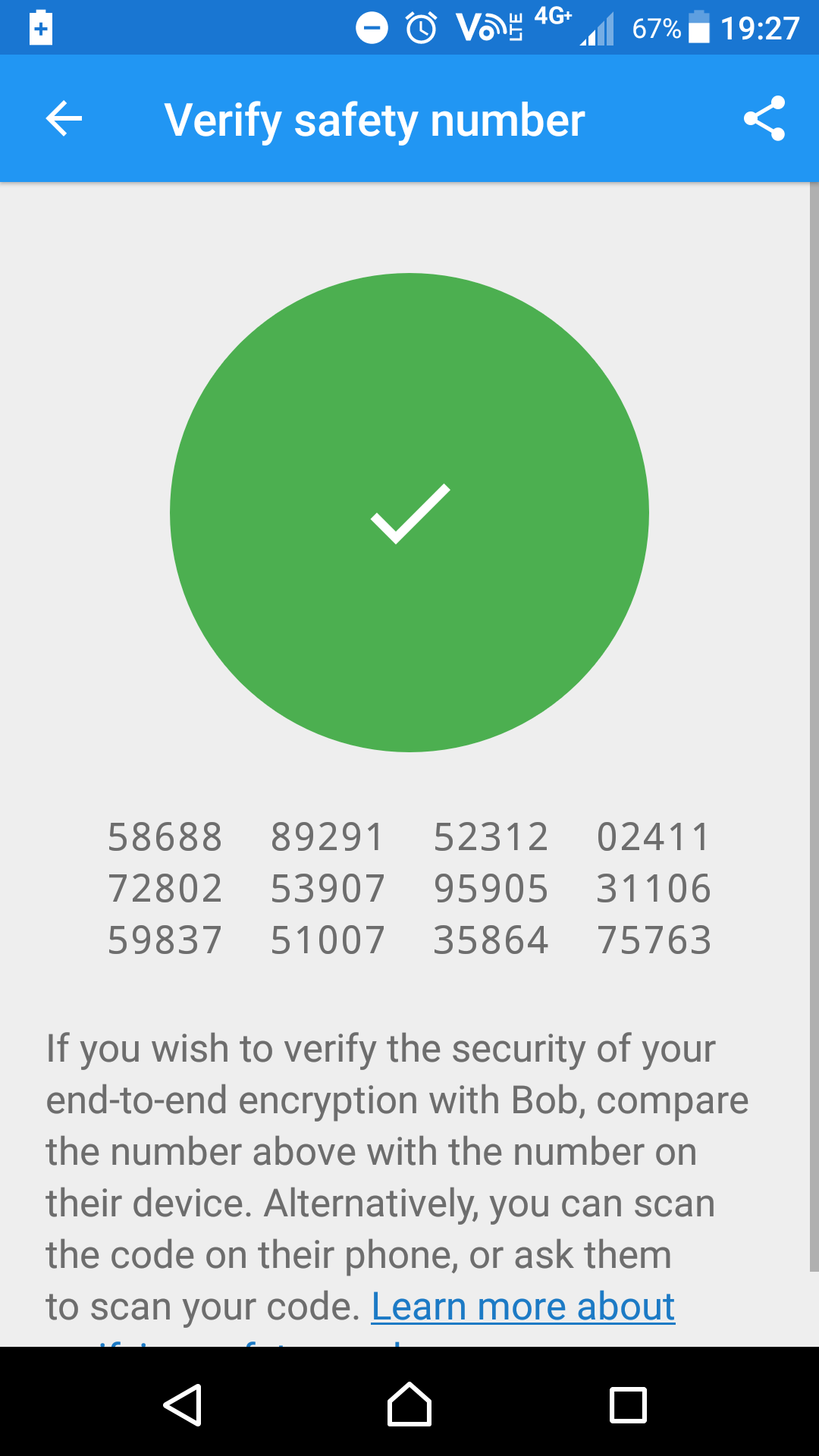}
  		\caption{Verified}
  		\label{fig:impl-signal-verify-2}
	\end{subfigure}
	\caption{Signal: verification process}
	\label{fig:impl-signal-verify}
\end{wrapfigure}

Signal supports three different methods for the users to verify each other. The first verification process uses the built-in calling option of Signal which is end-to-end encrypted and then read out loud to the other participant the security numbers that are shown in Fig.~\ref{fig:impl-signal-verify}\subref{fig:impl-signal-verify-1}. 
If the Signal calling is not regarded as secure enough, users can meet in person and show the numbers to each other.

The second method,
shown in Fig.~\ref{fig:impl-signal-verify}\subref{fig:impl-signal-verify-1}, uses the built-in QR-code scanner to scan the other participants QR-code to verify it is the same person in the chat.
%
The third option to verify the other user is meant to be used when the users do not trust the Signal application for handling the verification process. It is possible to share the security numbers to other applications on the user's phone. The user may have PGP\footnote{\url{https://en.wikipedia.org/wiki/Pretty_Good_Privacy}} 
\cite{Zimmermann1995pgpGuide,garfinkel1995pgp} enabled e-mail on their phone, and they trust it more than the Signal application, then this method is a better way of verifying the other user.


\paragraph{Other Security Implementations:}

\begin{wrapfigure}[16]{r}{.49\textwidth}
\vspace{-7ex}\centering
	\begin{subfigure}{.24\textwidth}
  		\centering
  		\includegraphics[width=.95\linewidth]{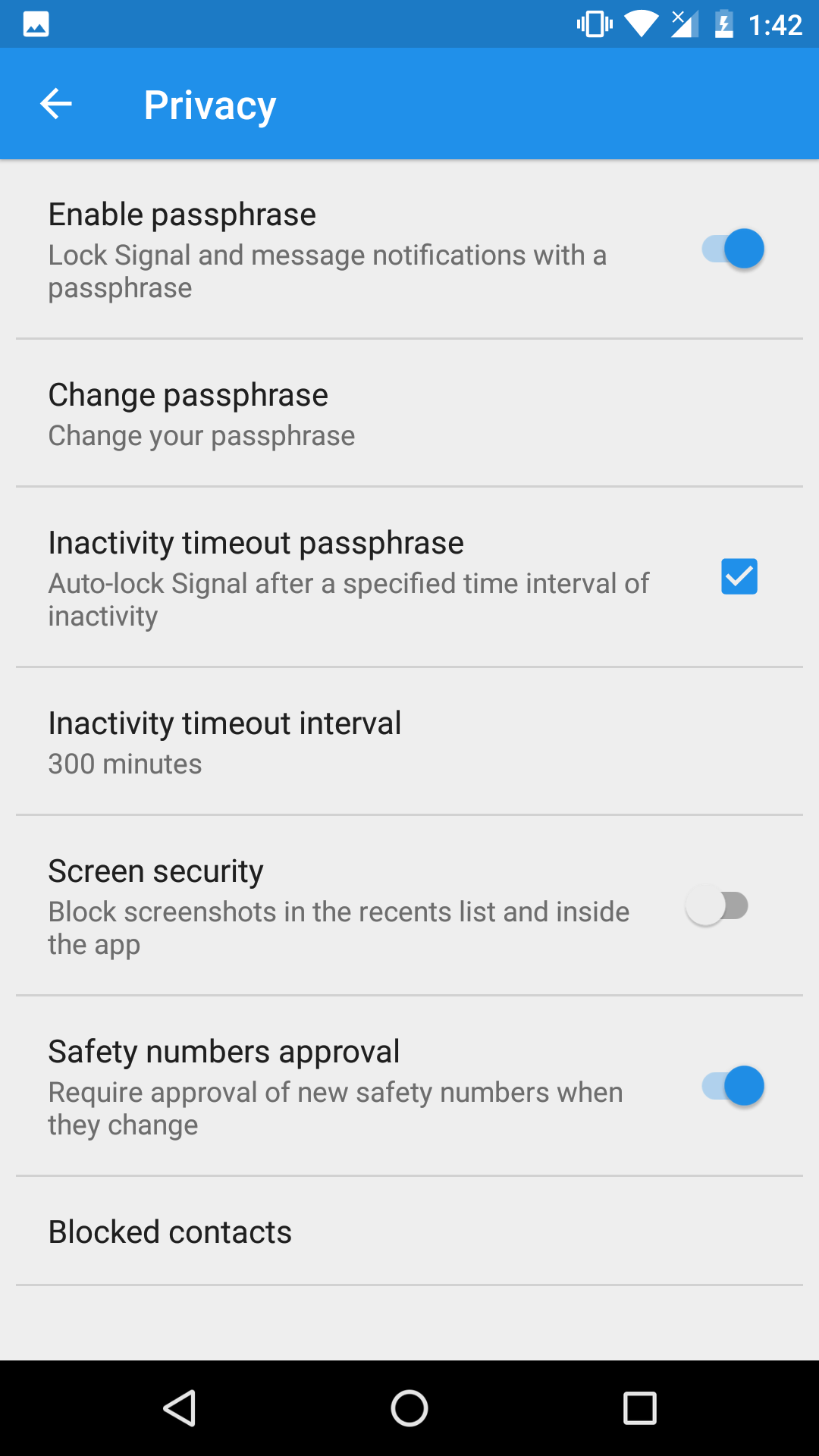}
  		\caption{Privacy settings}
  		\label{fig:impl-signal-priv-1}
	\end{subfigure}
	\begin{subfigure}{.24\textwidth}
  		\centering
  		\includegraphics[width=.95\linewidth]{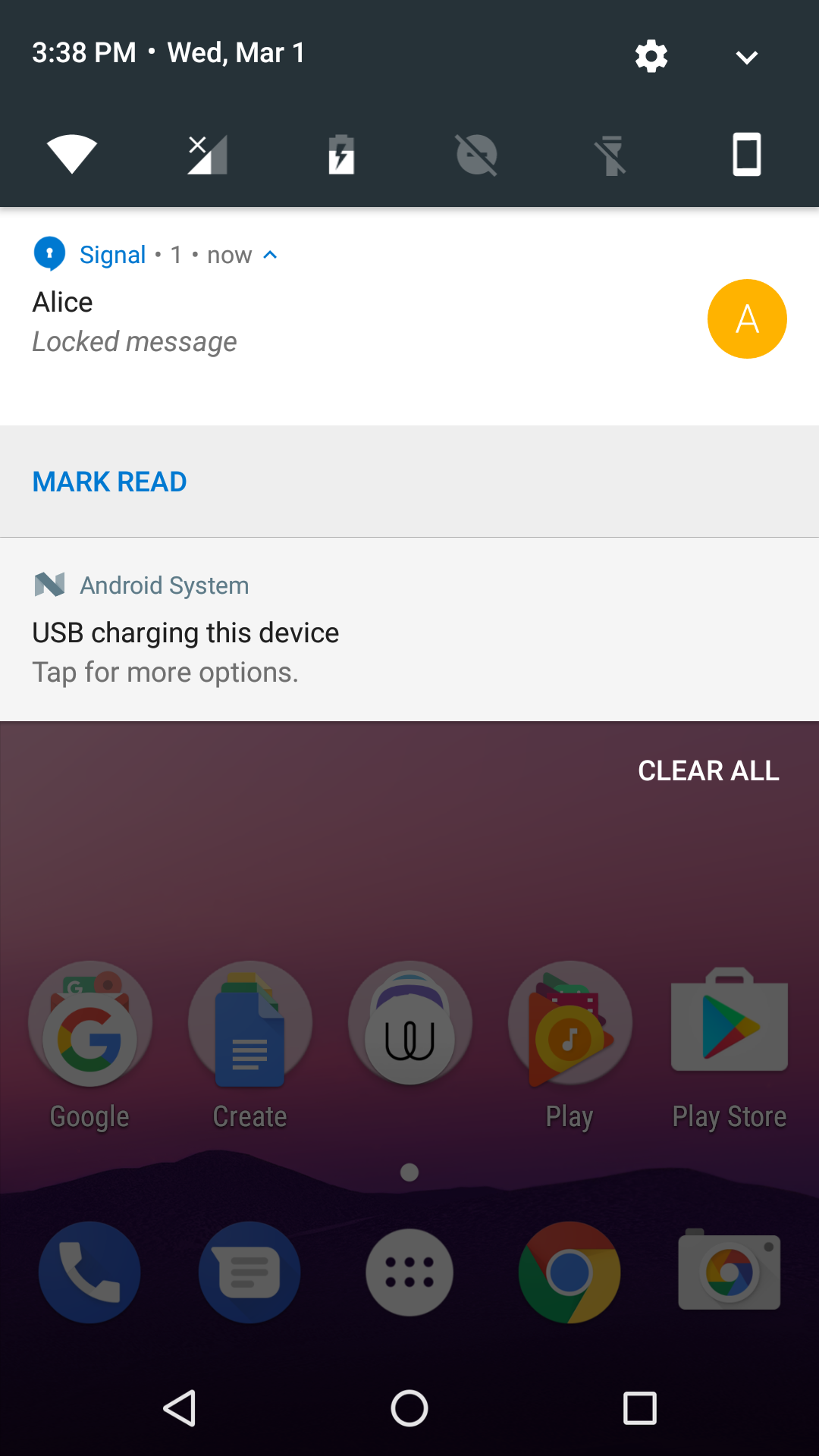}
  		\caption{Notification locked}
  		\label{fig:impl-signal-priv-2}
	\end{subfigure}
	\caption{Signal: other security implementations}
	\label{fig:impl-signal-priv}
\end{wrapfigure}

The Signal application has extra privacy settings. 
The first is the ``Safety numbers approval'' as shown in Fig.~\ref{fig:impl-signal-priv}\subref{fig:impl-signal-priv-1}. The setting is activated by default, which is important. When a user changes the safety numbers (cryptographic keys) by deleting and reinstalling the app, the device keys will also change. When the device keys change, the messages will not be shown to the receiver on a new device, until the new safety numbers are approved.
The second privacy setting is ``Screen security'', which does not allow the user to take screenshots as long as they are inside the Signal application.

The last privacy setting is the ability to enable a passphrase. The passphrase locks the Signal application and all message notifications.
It is possible to add an inactivity timeout passphrase which locks the application after some given time. Fig.~\ref{fig:impl-signal-priv}\subref{fig:impl-signal-priv-2} shows the notification which is locked and when the user tries to open the application, 
she needs to enter their passphrase they chose when the setting was activated.
%

\subsubsection{Case 2: WhatsApp} \label{impl:whatsapp-intro}

WhatsApp started as a small company in 2009, bought by Facebook in 2014 when it had 465 million monthly active users, and in 2017 that number has grown to 1.5 billion.\footnote{The Statistics Portal: \textit{``Number of monthly active Whatsapp users worldwide from April 2013 to December 2017 (in millions)''} \url{https://www.statista.com/statistics/260819/number-of-monthly-active-whatsapp-users/}}
WhatsApp initially was only a cross-platform non-secure instant messaging, but by the end of 2014 they announced that every user was going to start sending end-to-end encrypted messages using the Signal protocol.\footnote{``Open Whisper Systems partners with WhatsApp to provide end-to-end encryption'', announced by Moxie Marlinspike from Open Whisper Systems on November 18 2014, at \url{https://whispersystems.org/blog/whatsapp/}}
This was an important step for the Signal protocol and Open Whisper Systems since now the most popular instant messaging application would use their protocol. In April 2016 a complete transition was made from non-secure messaging to fully end-to-end encryption.\footnote{``WhatsApps's Signal Protocol integration is now complete'', announced by Moxie Marlinspike from Open Whisper Systems on April 05 2016, at \url{https://whispersystems.org/blog/whatsapp-complete/}}

\paragraph{Initial Set Up:} \label{impl:whatsapp-setup}
\begin{wrapfigure}[24]{r}{.49\textwidth}
\vspace{9ex}\centering
	\begin{subfigure}{.24\textwidth}
  		\centering
  		\includegraphics[width=.95\linewidth]{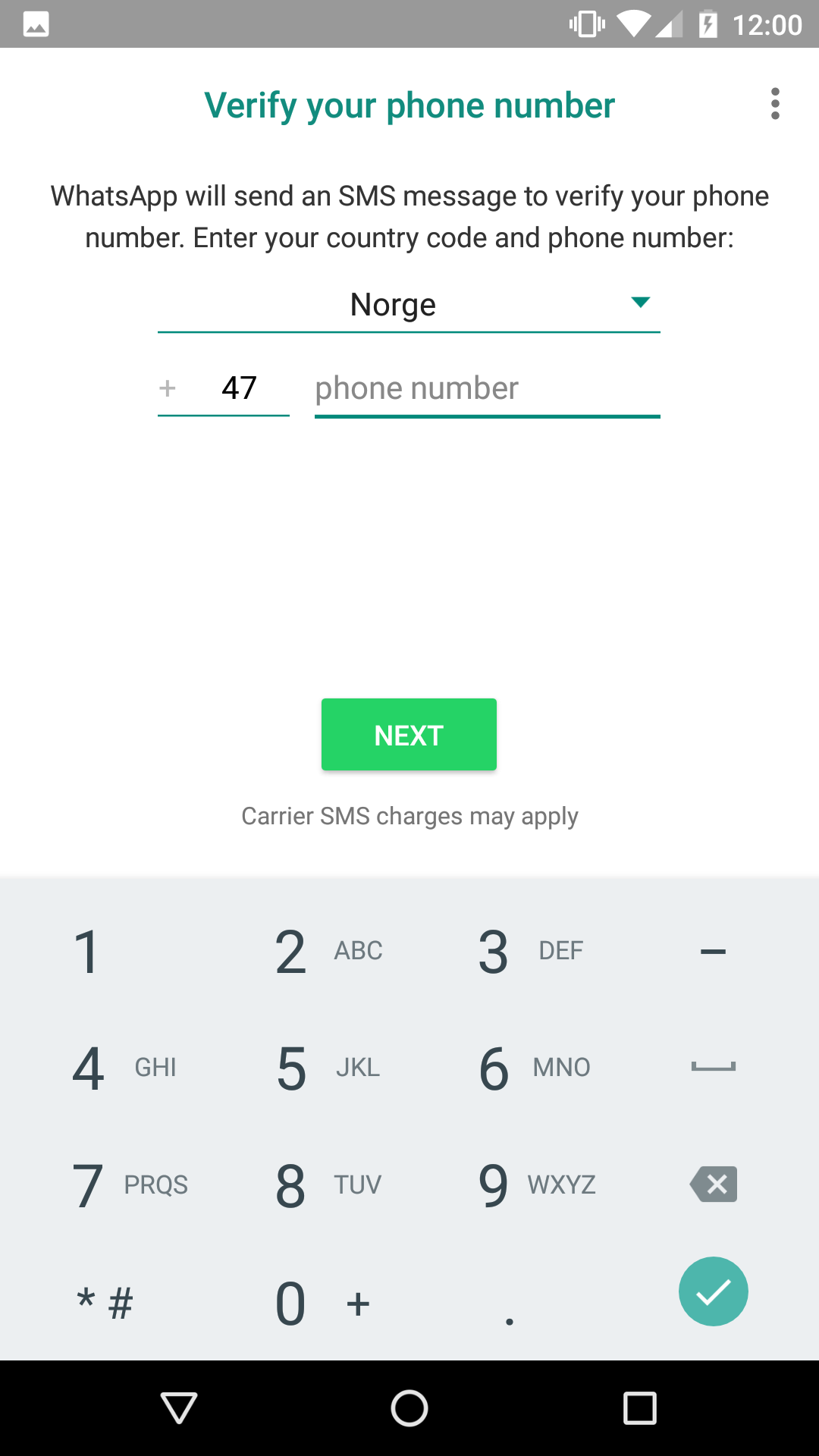}
  		\caption{Phone number registration}
  		\label{fig:impl-whatsapp-reg-1}
	\end{subfigure}%
	\begin{subfigure}{.24\textwidth}
  		\centering
  		\includegraphics[width=.95\linewidth]{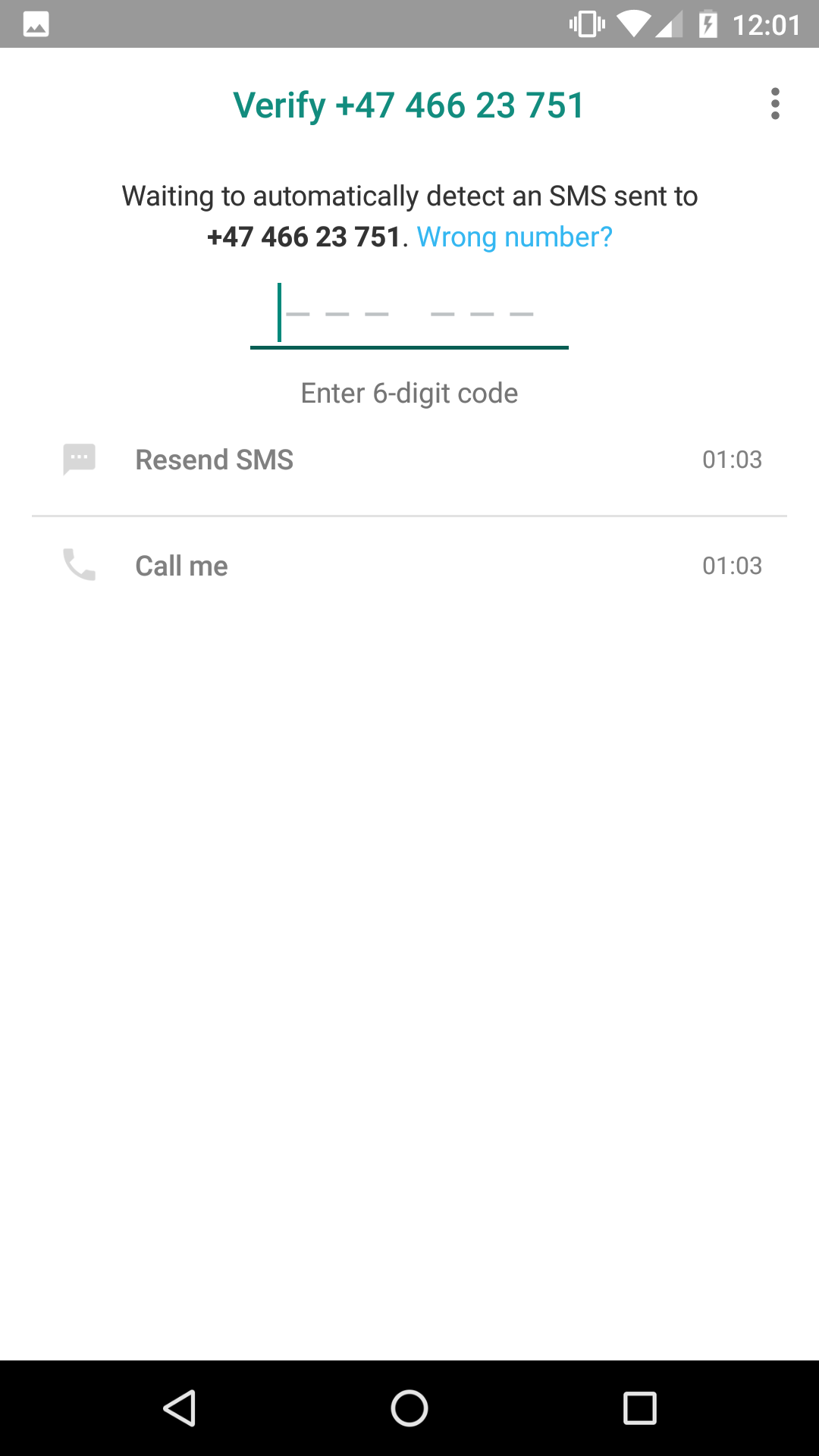}
  		\caption{Verifying the phone number}
  		\label{fig:impl-whatsapp-reg-2}
	\end{subfigure}
	\caption{WhatsApp: registration process}
	\label{fig:impl-whats-1}
\end{wrapfigure}

Establishing an account on WhatsApp is done in the same way as for the Signal application, where the account only works on one device at a time.
Fig.~\ref{fig:impl-whats-1}\subref{fig:impl-whatsapp-reg-1} shows the first page a user sees when starting the application for the first time. WhatsApp uses its own infrastructure to handle the SMS verification process instead of a third party.
Fig.~\ref{fig:impl-whats-1}\subref{fig:impl-whatsapp-reg-2} shows the verification page after the user has entered her phone number. WhatsApp automatically enters the verification code that is sent to the user's SMS inbox, but if the user has not given the app access to the inbox, she can enter the verification code manually. If for some reason the verification code does not arrive, the user has the options to either resend the SMS or ask WhatsApp to call the user to receive the verification code through voice.

\paragraph{Message After a Key Change:}

\begin{figure}[t]
\centering
	\begin{subfigure}{.24\textwidth}
  		\centering
  		\includegraphics[width=.95\linewidth]{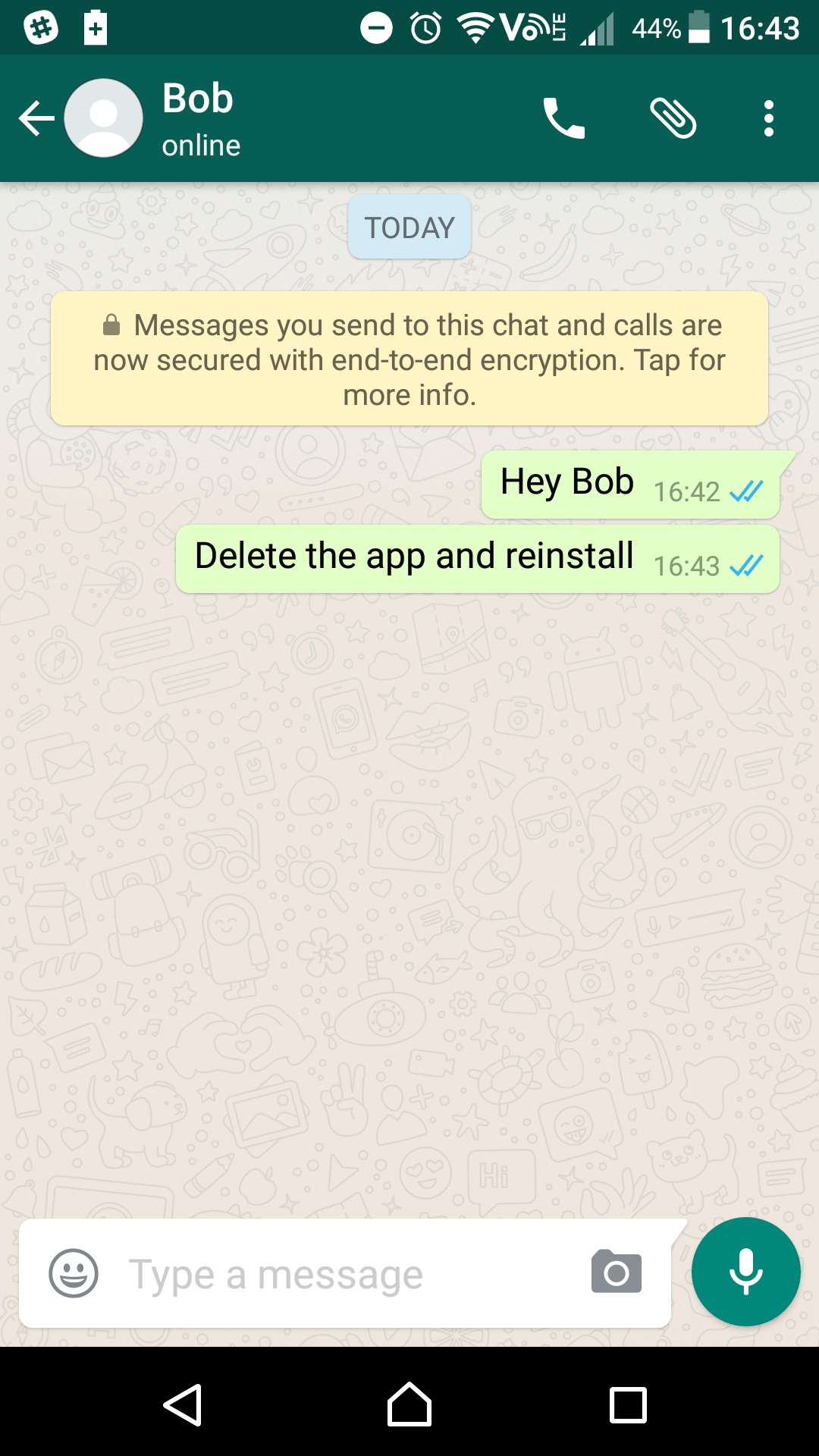}
  		\caption{Alice's first message}
  		\label{fig:impl-whatsapp-kc-1}
	\end{subfigure}
	\begin{subfigure}{.24\textwidth}
  		\centering
  		\includegraphics[width=.95\linewidth]{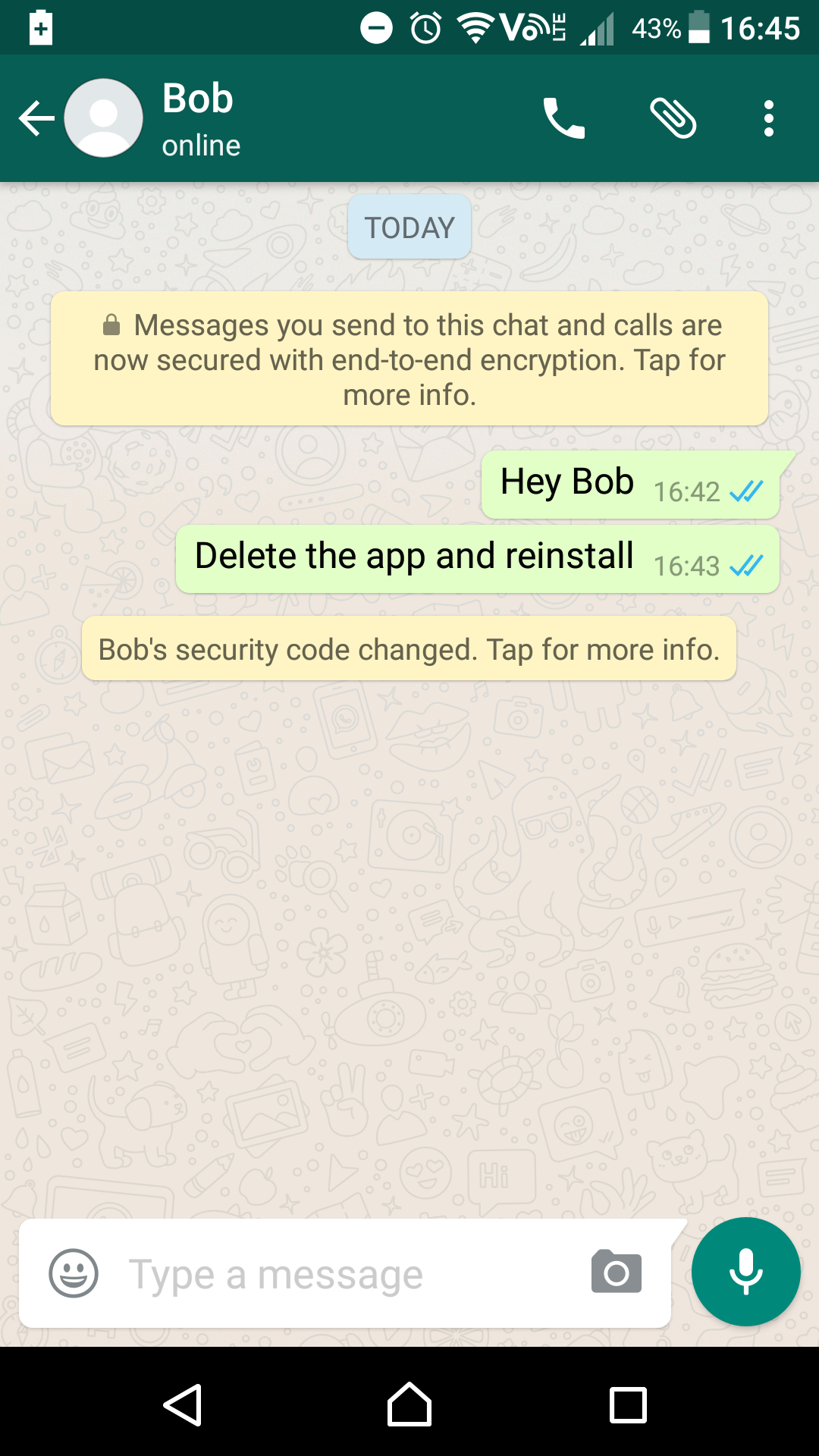}
  		\caption{After Bob has reinstalled}
  		\label{fig:impl-whatsapp-kc-2}
	\end{subfigure}
	\begin{subfigure}{.24\textwidth}
  		\centering
  		\includegraphics[width=.95\linewidth]{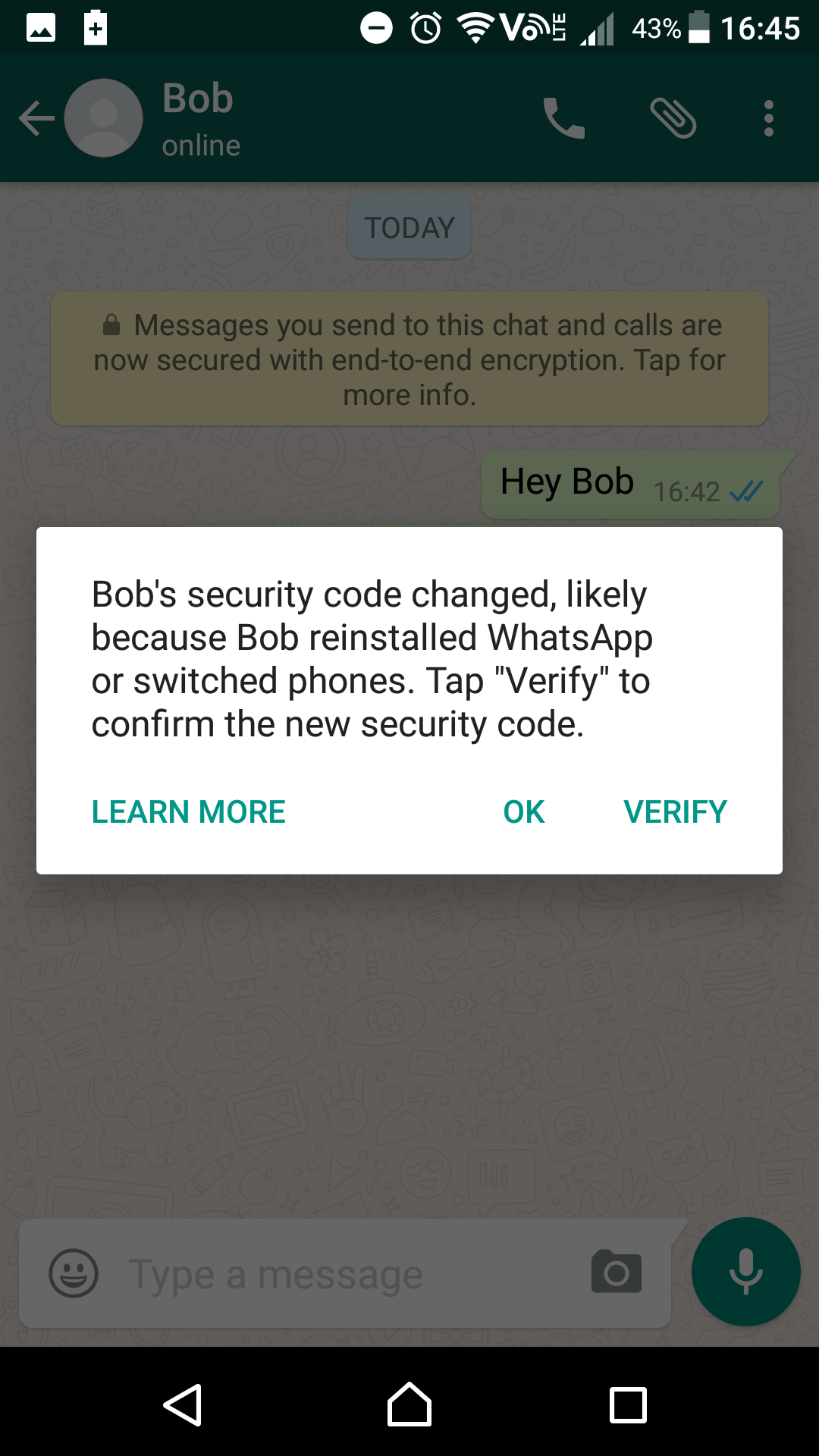}
  		\caption{Info about Bob's new keys}
  		\label{fig:impl-whatsapp-kc-3}
	\end{subfigure}
	\begin{subfigure}{.24\textwidth}
  		\centering
  		\includegraphics[width=.95\linewidth]{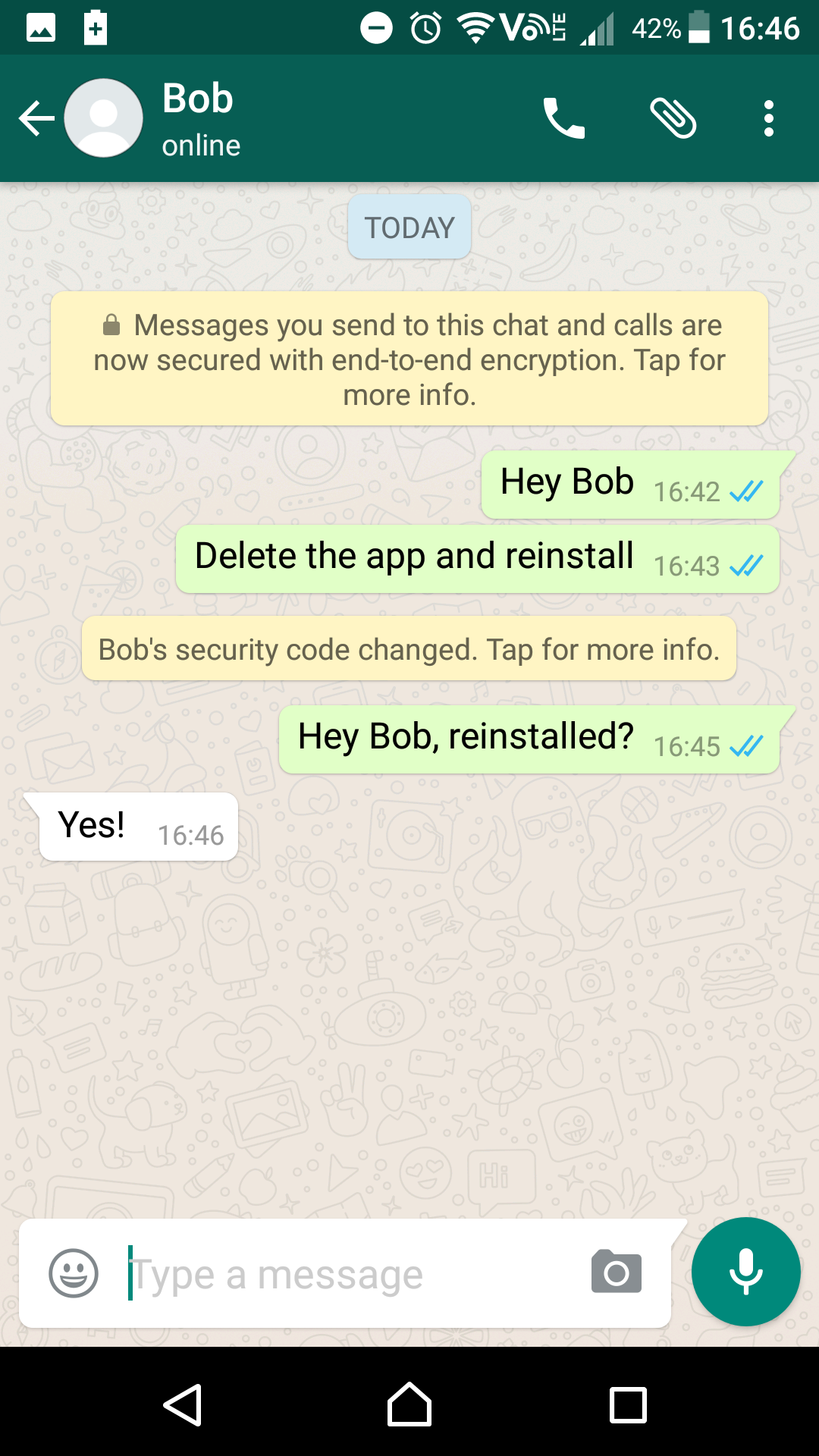}
  		\caption{Message after verification}
  		\label{fig:impl-whatsapp-kc-4}
	\end{subfigure}
	\caption{WhatsApp: message after a key change}
	\label{fig:impl-whatsapp-kc}
\end{figure}

Fig.~\ref{fig:impl-whatsapp-kc}\subref{fig:impl-whatsapp-kc-1} shows when Alice sends her first and second messages to Bob in order to initiate a conversation. The yellow notification box at the top of the conversation is shown to both participants stating that the conversation is end-to-end encrypted and more  information can be obtained by pressing the box. Each message is shown with a double checkmark, as in Signal.

Fig.~\ref{fig:impl-whatsapp-kc}\subref{fig:impl-whatsapp-kc-2} shows a new notification box appearing on Alice's conversation page after Bob has reinstalled his application. WhatsApp automatically checks if new cryptographic keys (security code) are changed even though she has not sent him any message.
When Alice taps the notification box,
a popup (Fig.~\ref{fig:impl-whatsapp-kc}\subref{fig:impl-whatsapp-kc-3}) informs Alice why Bob's cryptographic keys have changed and the option to verify him before she sends new messages.
%
Alice sends a new message to Bob after verifying new cryptographic keys of Bob and the message is labeled (Fig.~\ref{fig:impl-whatsapp-kc}\subref{fig:impl-whatsapp-kc-4}) with a double checkmark meaning that everything went well.

\paragraph{Key Change While a Message is in Transit:}

\begin{wrapfigure}[17]{r}{.49\textwidth}
\vspace{-10ex}\centering
	\begin{subfigure}{.24\textwidth}
  		\centering
  		\includegraphics[width=.95\linewidth]{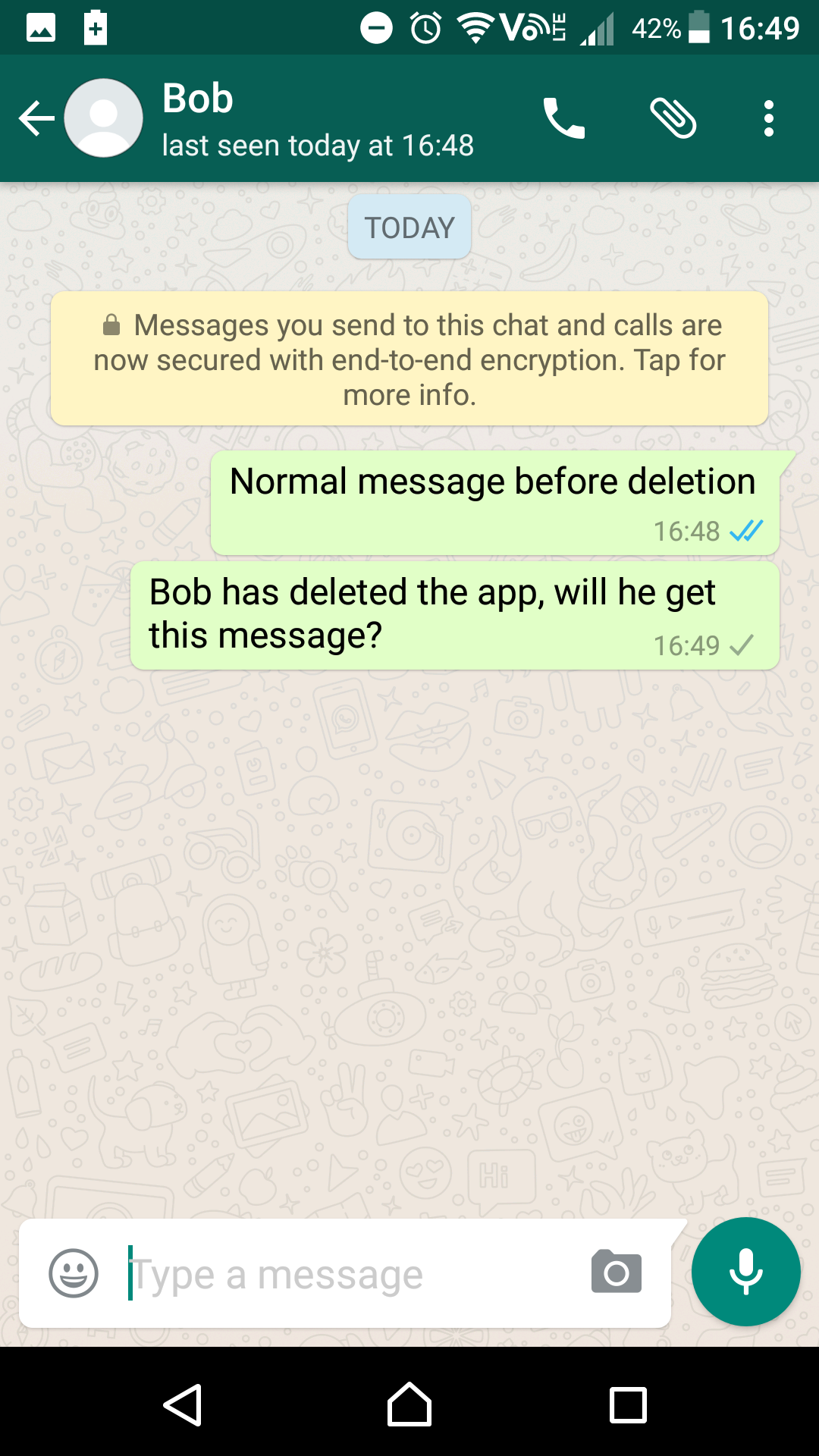}
  		\caption{Bob deletes his app}
  		\label{fig:impl-whatsapp-kc-transit-2}
	\end{subfigure}
	\begin{subfigure}{.24\textwidth}
  		\centering
		\includegraphics[width=.95\linewidth]{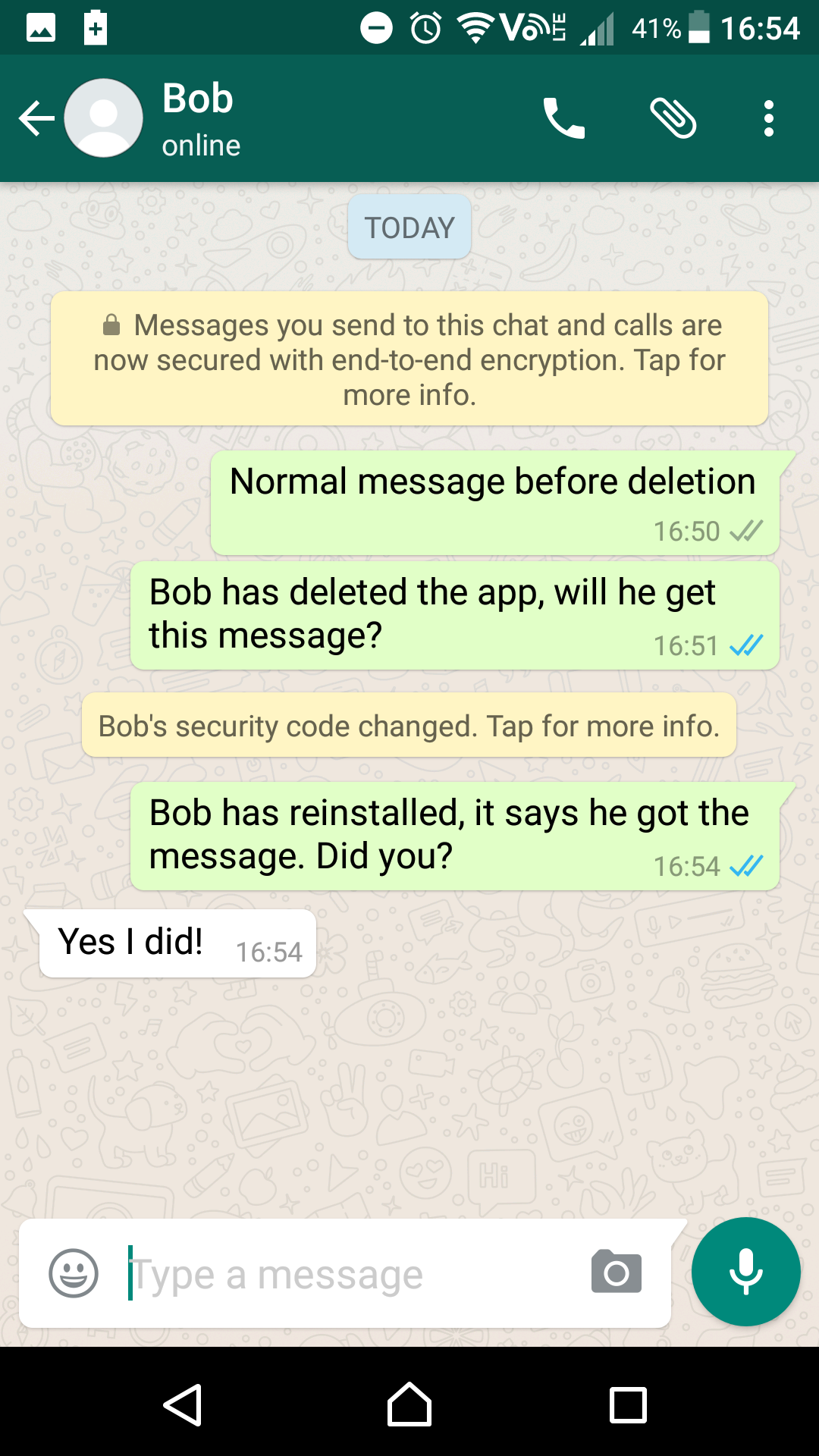}
		\caption{Bob reinstalled}
		\label{fig:impl-whatsapp-kc-transit-3}
	\end{subfigure}
	\caption{WhatsApp: key change while msg.\ in transit}
	\label{fig:impl-whatsapp-kc-transit}
\end{wrapfigure}

This test scenario starts as the previous one, but here we look at how WhatsApp handles messages sent before Bob finishes to reinstall.
%
%
Fig.~\ref{fig:impl-whatsapp-kc-transit}\subref{fig:impl-whatsapp-kc-transit-2} shows Alice sending a second message to Bob after he has deleted his application. The single checkmark on the message means that it has been sent but not received and read by Bob.
When Bob finishes the re-installation of the application, both the second message Alice sent and the same yellow notification box are added to the conversation. Fig.~\ref{fig:impl-whatsapp-kc-transit}\subref{fig:impl-whatsapp-kc-transit-3} displays the conversation after Alice sends a third message, showing that Bob receives the second message that was sent before he reinstalled. This means that WhatsApp re-encrypts messages when the receiver generates new cryptographic keys and the sender does not verify the new keys.

\paragraph{Verification Process Between Participants:}

WhatsApp has implemented the same verification process as Signal. It uses the Signal numerical format for verification, a QR-code for scanning with the built-in scanner, and the user can choose if they want to copy the security numbers outside of the WhatsApp application. The reason for this may be that when they decided to implement the Signal end-to-end security protocol, they implemented every single step of the Signal implementation to uphold the specifications. WhatsApp also features end-to-end encrypted calling, which enables users to call each other and verify the security code.


\paragraph{Other Security Implementations:}

\begin{figure}[t]
\centering
	\begin{subfigure}{.24\textwidth}
  		\centering
		\includegraphics[width=.95\linewidth]{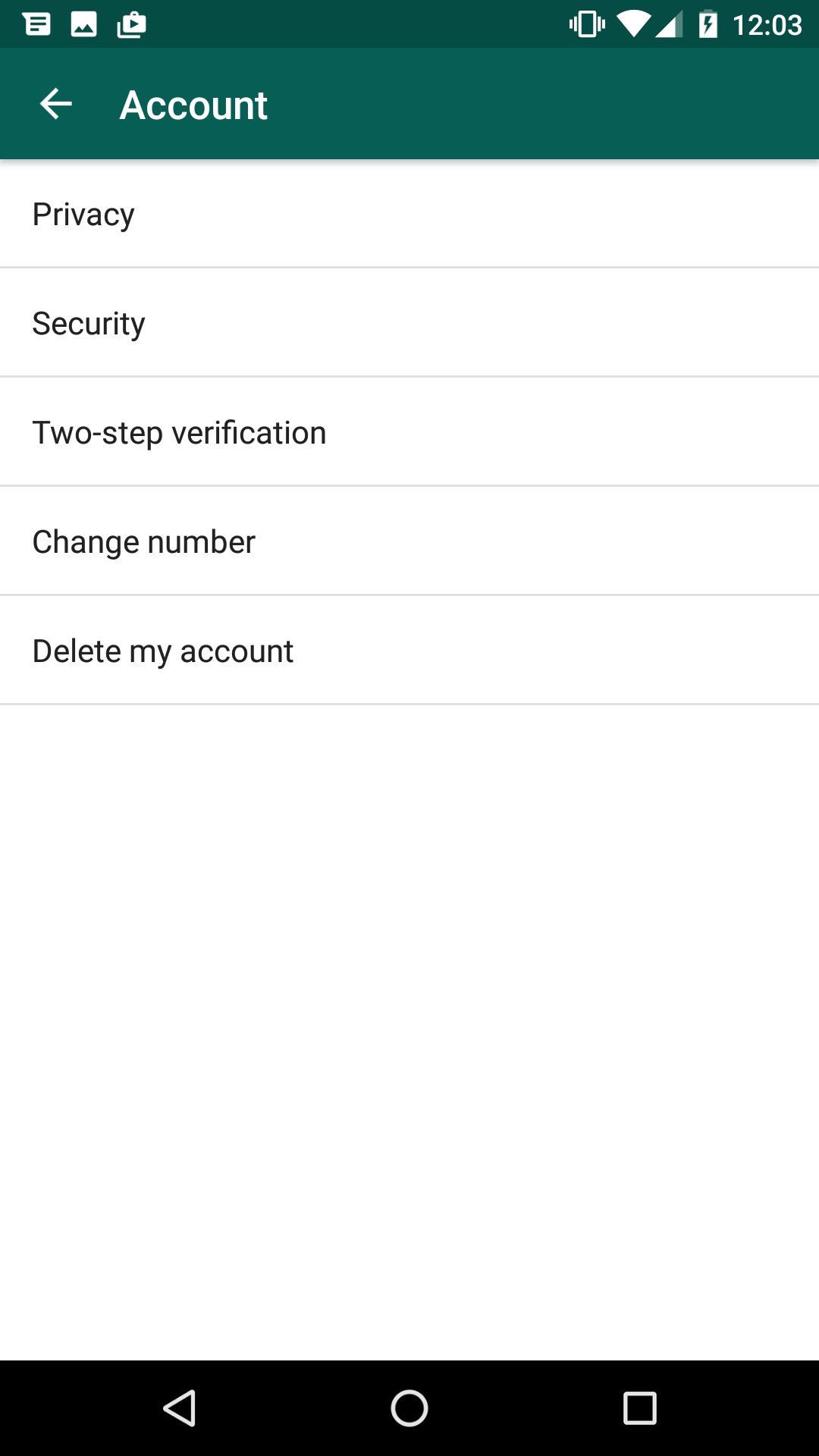}
		\caption{Privacy settings}
		\label{fig:impl-whatsapp-priv-1}
	\end{subfigure}
	\begin{subfigure}{.24\textwidth}
  		\centering
  		\includegraphics[width=.95\linewidth]{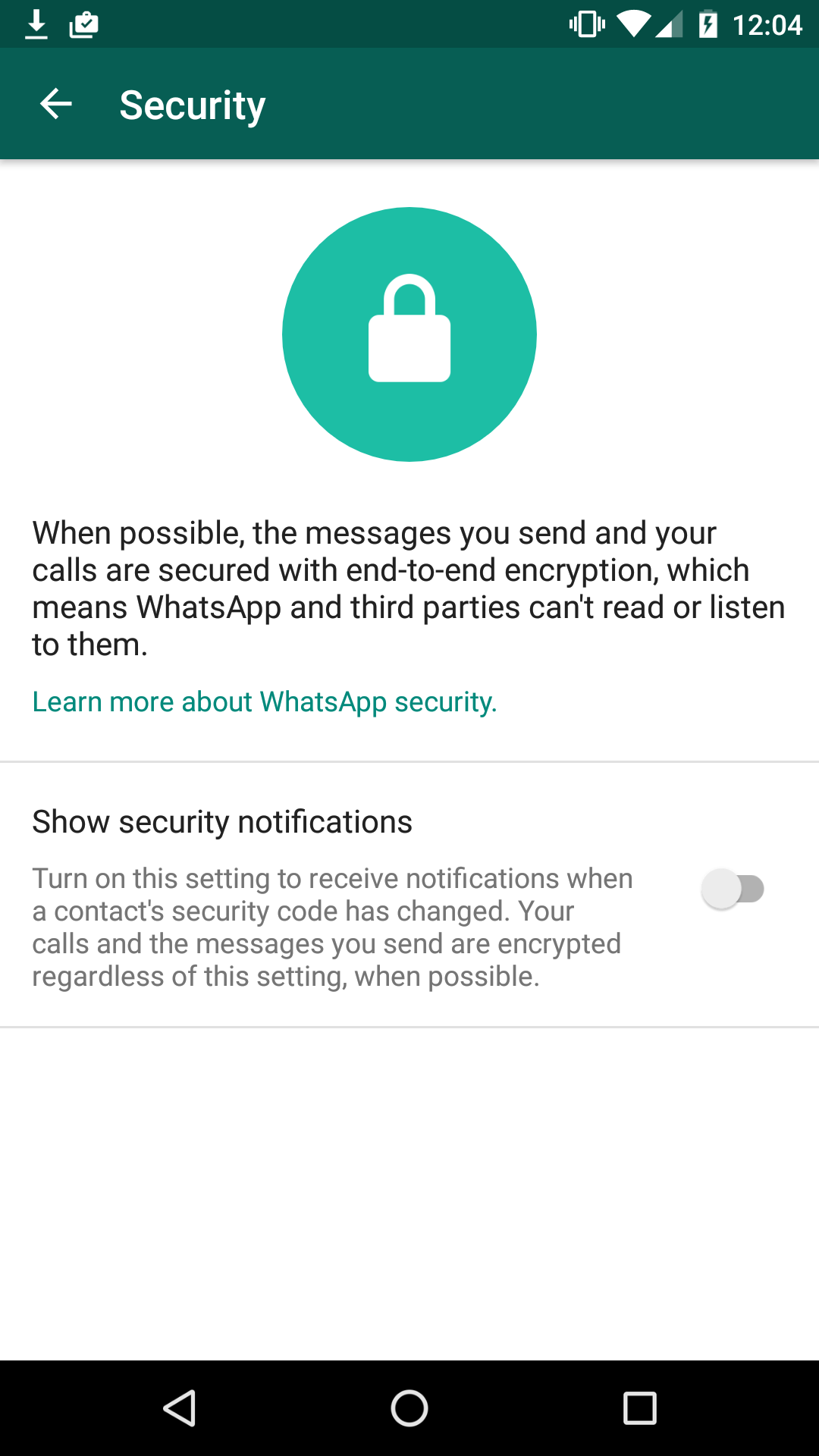}
  		\caption{Security notification}
  		\label{fig:impl-whatsapp-priv-2}
	\end{subfigure}
	\begin{subfigure}{.24\textwidth}
  		\centering
  		\includegraphics[width=.95\linewidth]{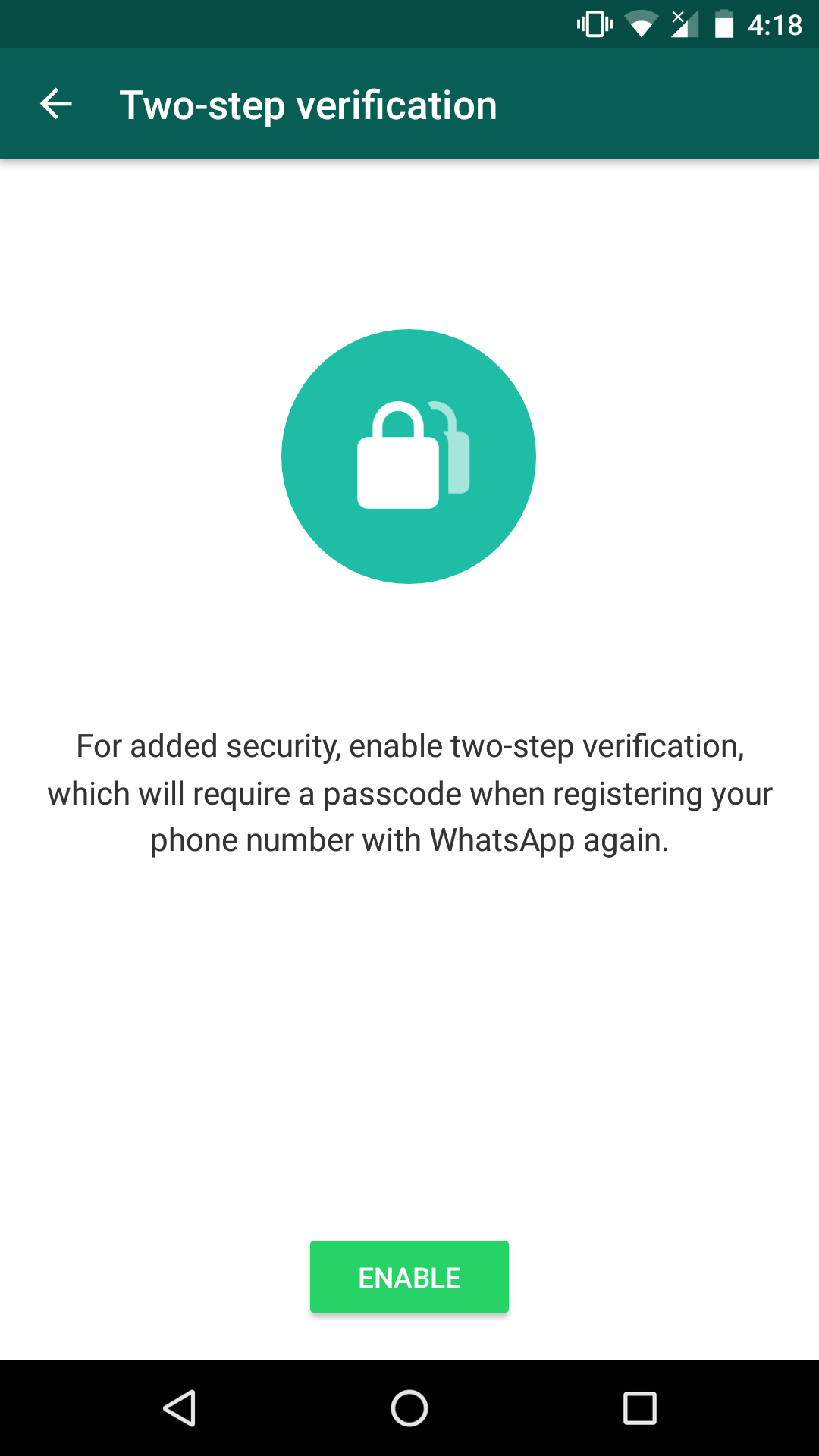}
  		\caption{Two step verification}
  		\label{fig:impl-whatsapp-2step-1}
	\end{subfigure}%
	\caption{WhatsApp: other security implementations}
	\label{fig:impl-whatsapp-privacy-1}
\end{figure}


As shown in Fig.~\ref{fig:impl-whatsapp-privacy-1}\subref{fig:impl-whatsapp-priv-1} (setting page), there are options for changing the number of the account or even delete the account.

If Alice chooses the ``Security'' menu item and then activates the ``show security notification''  option (shown in Fig.~\ref{fig:impl-whatsapp-privacy-1}\subref{fig:impl-whatsapp-priv-2}), then when Bob re-installs the application or receives a new device the application shows a notification to Alice. Otherwise, if the option is turned off, then Alice does not receive any notification.

Fig.~\ref{fig:impl-whatsapp-privacy-1}\subref{fig:impl-whatsapp-2step-1} shows the two-step verification settings that WhatsApp has implemented, where the user needs to enter an additional passphrase when registering the account with the same number on a new device or after a fresh re-install.

\subsubsection{Case 3: Wire} \label{impl:wire-intro}

Wire is an application that implements end-to-end encryption using the Proteus protocol, which is heavily based on the Signal protocol, but re-implemented in-house.\footnote{Proteus Protocol, by Wire Swiss GmbH available at \url{https://github.com/wireapp/proteus}}
Wire was started in 2012 by developers who previously worked at Microsoft
and Skype,
and finally released their own instant messaging application in 2014.\footnote{``Skype Co-Founder Backs Wire, A New Communications App Launching Today On iOS, Android And Mac, by Sarah Perez in Tech Crunch on December 2014 available at \url{https://techcrunch.com/2014/12/02/skype-co-founder-backs-wire-a-new-communications-app-launching-today-on-ios-android-and-mac/}} 
The first version did not offer end-to-end encryption until March 2016, when they launched the encryption on instant messaging and their video calling feature \cite{wire-e2e-article}.
%
Wire offers the same features as the other applications, such as text, video, voice, photo and music messages, and is supported on multiple platforms, from smartphones to personal computers, and is also open sourced.\footnote{\url{https://github.com/wireapp}}

\paragraph{Initial Set Up:}\label{impl:wire-setup}

\begin{figure}[t]
\centering
\begin{subfigure}{.24\textwidth}
  		\centering
  		\includegraphics[width=.95\linewidth]{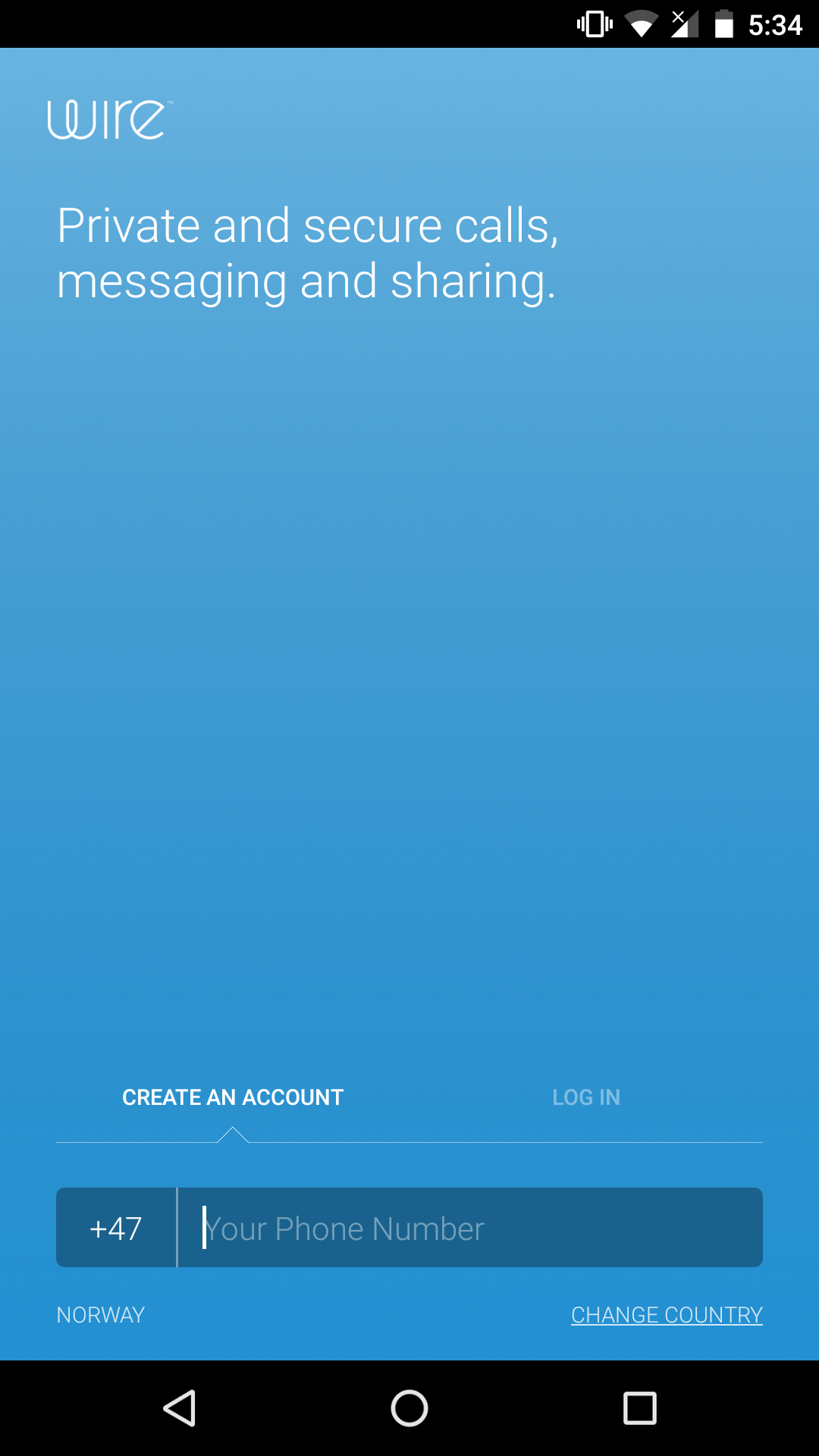}
  		\caption{Phone number registration}
  		\label{fig:impl-wire-init-1}
	\end{subfigure}%
	\begin{subfigure}{.24\textwidth}
  		\centering
  		\includegraphics[width=.95\linewidth]{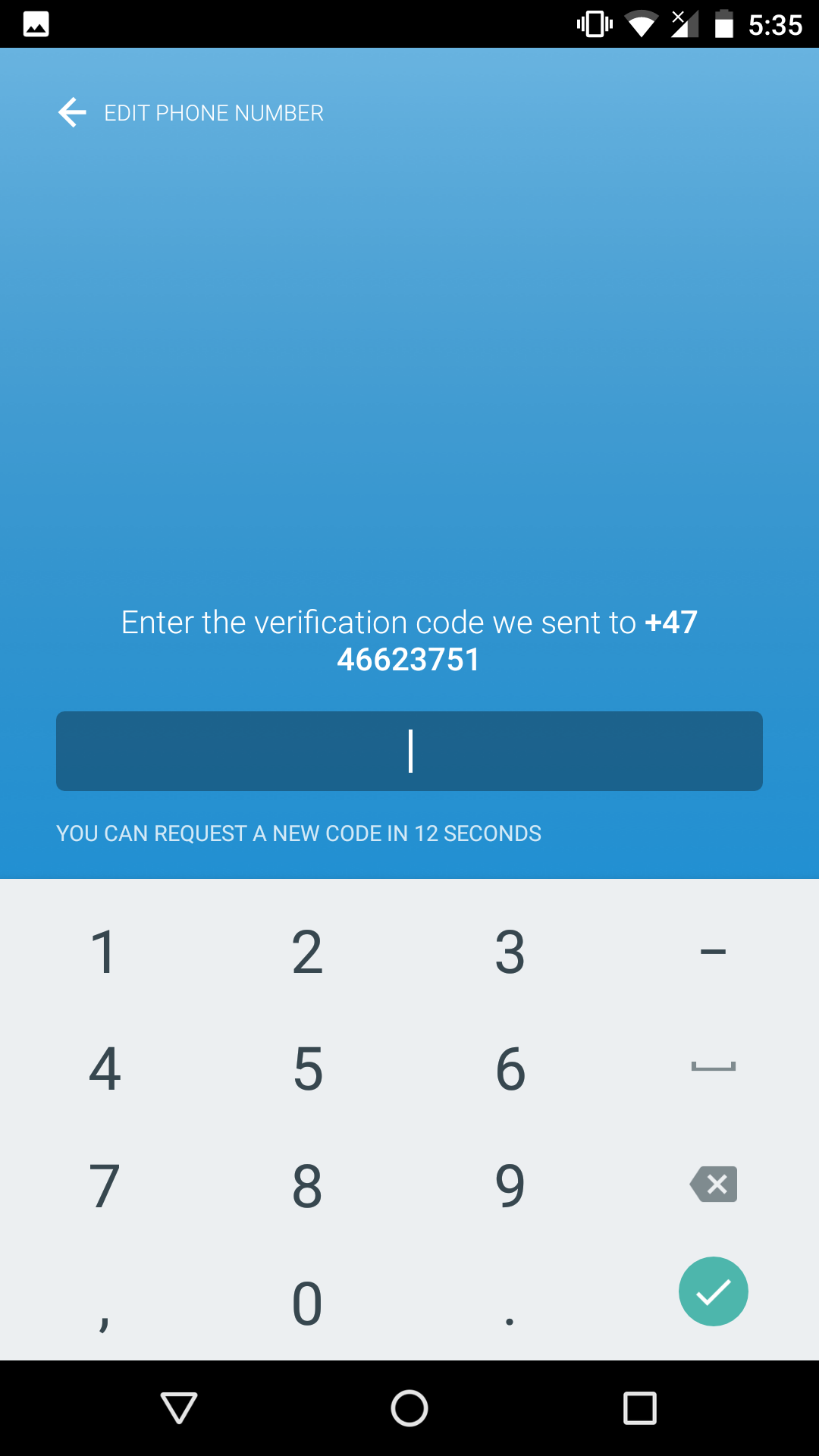}
  		\caption{Phone verification}
  		\label{fig:impl-wire-init-2}
	\end{subfigure}
	\begin{subfigure}{.24\textwidth}
		\centering
		\includegraphics[width=.95\linewidth]{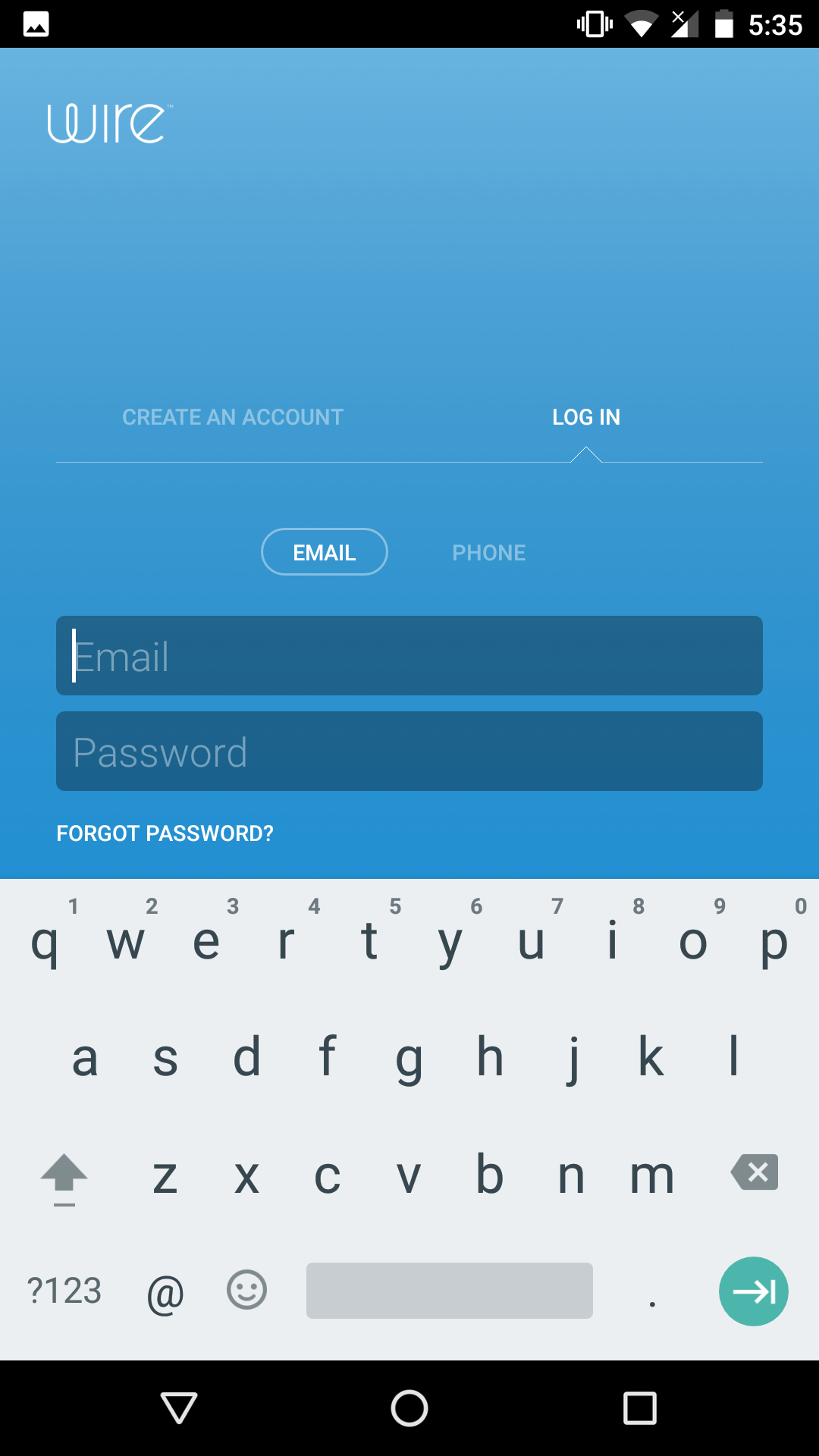}
		\caption{User login with e-mail}
		\label{fig:impl-wire-init-3}
	\end{subfigure}
	\begin{subfigure}{.24\textwidth}
		\centering
		\includegraphics[width=.95\linewidth]{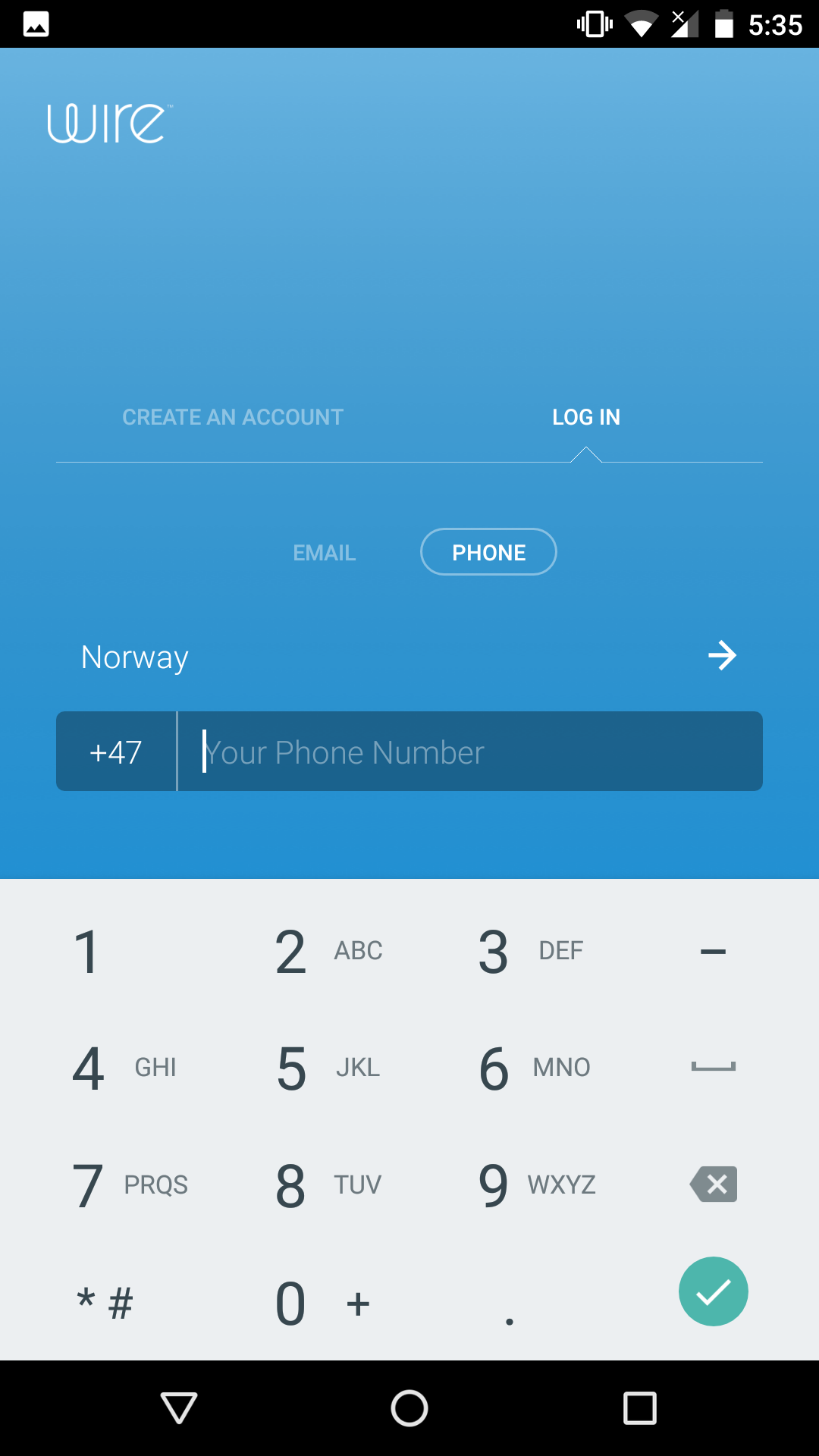}
		\caption{Login with phone number}
		\label{fig:impl-wire-init-4}
	\end{subfigure}
	\caption{Wire: registration process}
	\label{fig:impl-wire-init}
\end{figure}

The Wire app has a different registration process than the other applications. 
The first page, as shown in Fig.~\ref{fig:impl-wire-init}\subref{fig:impl-wire-init-1}, asks to register a phone number. However, one can also create an account through the Wire web application by using just an email address. 
Fig.~\ref{fig:impl-wire-init}\subref{fig:impl-wire-init-2} shows the verification process, where the user needs to enter the verification code (which is received in an SMS) manually. If the user never receives the verification code, she can ask Wire to call the user to receive it. Wire application does not read the code received in the SMS automatically.  Figures~\ref{fig:impl-wire-init}\subref{fig:impl-wire-init-3} and \ref{fig:impl-wire-init}\subref{fig:impl-wire-init-4} demonstrate options to log in with an email and/or a phone number, respectively. When a user reinstalls the application or changes her device, she does not need to go through the registration again.

\vspace{1mm}
\paragraph{Message After a Key Change:}

\begin{figure}[t]
\centering
\begin{subfigure}{.24\textwidth}
  		\centering
  		\includegraphics[width=.95\linewidth]{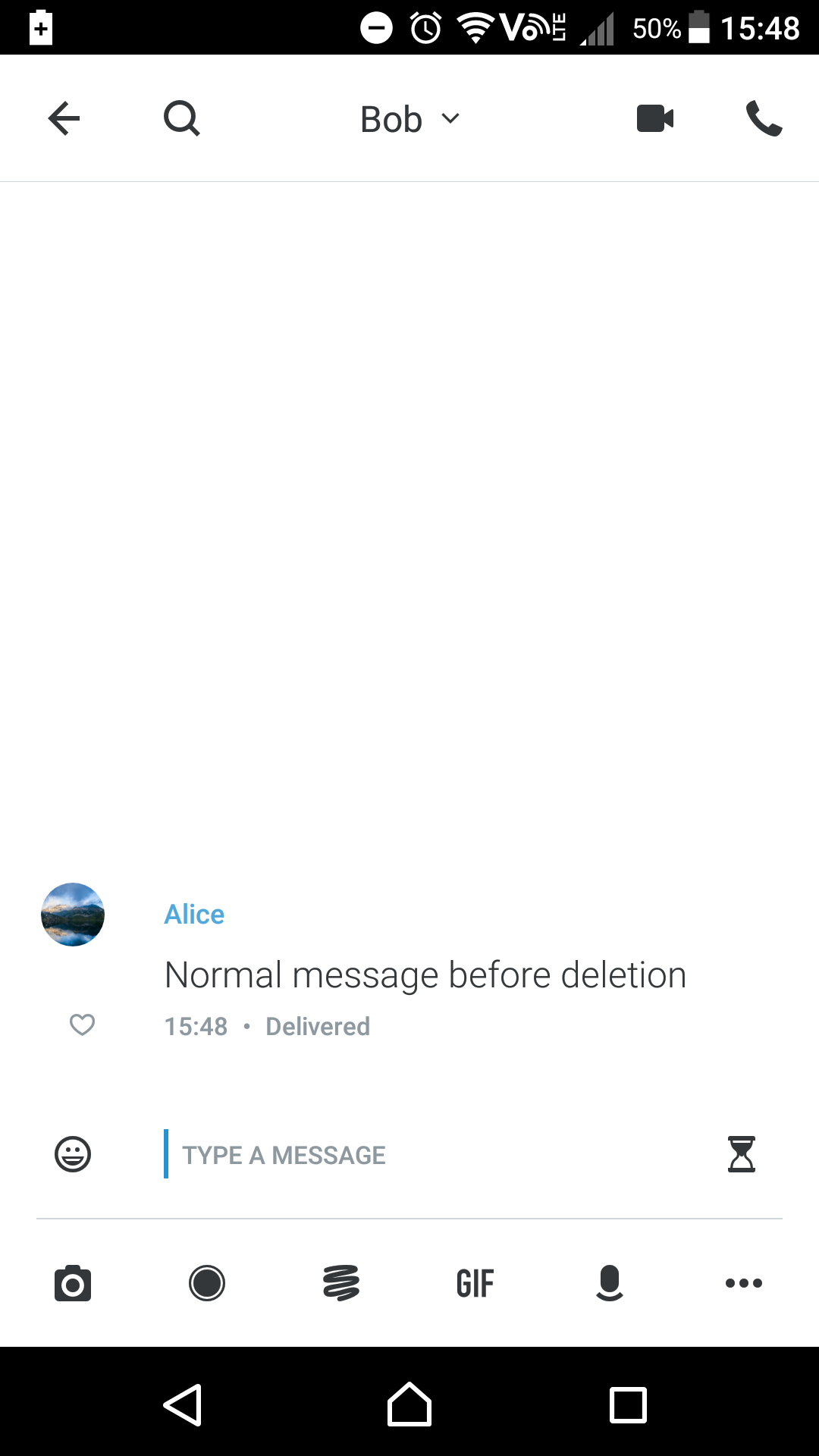}
  		\caption{Alice's first message before deletion}
  		\label{fig:impl-wire-kc-transit-1}
	\end{subfigure}%
	\begin{subfigure}{.24\textwidth}
  		\centering
  		\includegraphics[width=.95\linewidth]{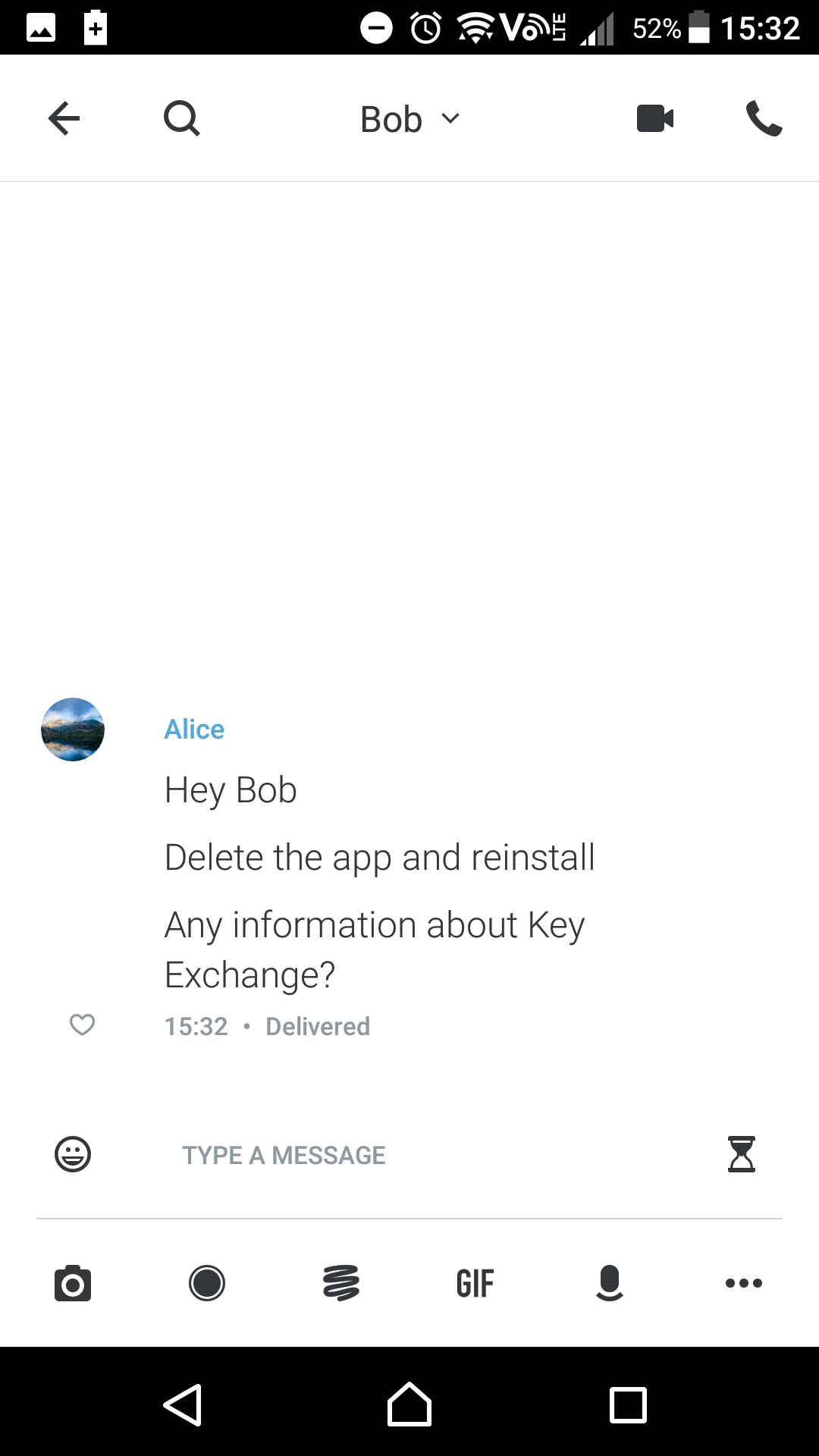}
  		\caption{After Bob has reinstalled}
  		\label{fig:impl-wire-kc-2}
	\end{subfigure}
	\begin{subfigure}{.24\textwidth}
		\centering
		\includegraphics[width=.95\linewidth]{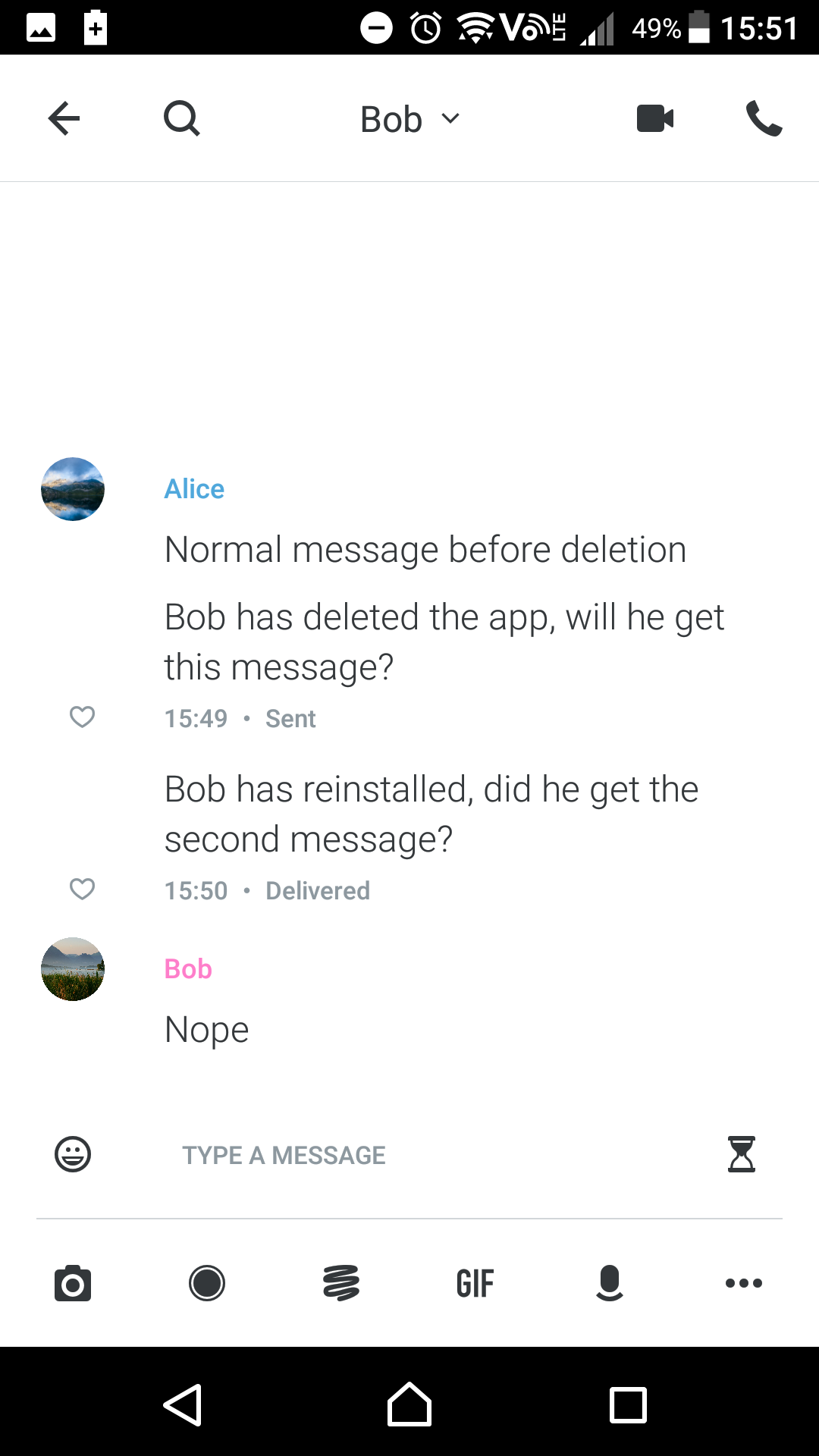}
		\caption{No notification after Bob reinstalls}
		\label{fig:impl-wire-kc-transit-2}
	\end{subfigure}
	\begin{subfigure}{.24\textwidth}
  		\centering
  		\includegraphics[width=.95\linewidth]{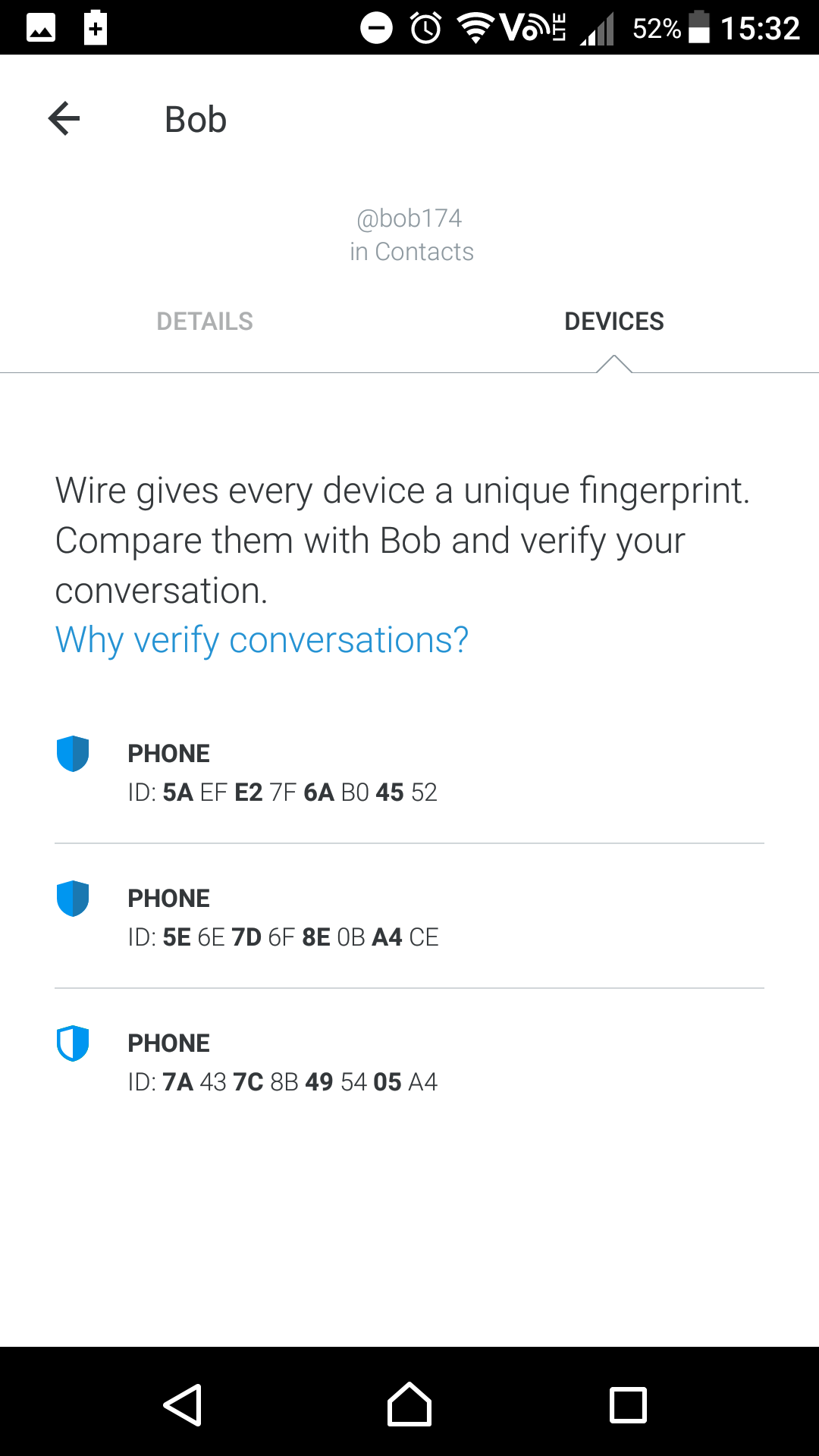}
  		\caption{Bob's device keys}
  		\label{fig:impl-wire-kc-4}
	\end{subfigure}	
	\caption{Wire: Key change while a message is in transit; and Message after a key change}
	\label{fig:impl-wire-kc-transit}
\end{figure}

Fig.~\ref{fig:impl-wire-kc-transit}\subref{fig:impl-wire-kc-transit-1} shows Alice's initial contact with Bob.
Wire uses a text under each message to explain if the message is delivered to the recipient or not.
If the receiver re-installs his application (which results in changing his cryptographic keys), then the sender (i.e., Alice) will not be notified by the Wire application about the new cryptographic keys of the receiver. As shown in 
Fig.~\ref{fig:impl-wire-kc-transit}\subref{fig:impl-wire-kc-2}, 
Alice sends two messages to Bob, where Bob has reinstalled his application, but Alice does not get any notification by Wire that Bob has new cryptographic keys; it may look to Alice that Bob has the same keys as before.
Alice can check Bob's account information to see if Bob has got new cryptographic keys. 
Fig.~\ref{fig:impl-wire-kc-transit}\subref{fig:impl-wire-kc-4} 
shows Bob's device keys under his account, for three different devices, because Wire allows multiple devices to be associated to one account. This means that Alice needs to verify each device to know that the conversation is secure with end-to-end encryption. The two top devices have a full blue shield which means they are verified, while the bottom device only has a half shield because it has not been verified yet.
%

%

\paragraph{Key Change While a Message is in Transit:}

Fig.~\ref{fig:impl-wire-kc-transit}\subref{fig:impl-wire-kc-transit-1} shows the initial message from Alice to Bob.
%
If Alice sends a message when Bob has deleted the app, the message will be labeled by just ``sent'' and not delivered (Fig.~\ref{fig:impl-wire-kc-transit}\subref{fig:impl-wire-kc-transit-2}). However, if Alice sends another message when Bob has reinstalled the app, then the message will be labeled as ``delivered''. This test demonstrates that Wire does not notify Alice about Bob's new keys, and at the same time does not deliver messages encrypted with old cryptographic keys to devices with new keys.

\paragraph{Verification Process Between Participants:}

\begin{figure}[t]
\centering
	\begin{subfigure}{.24\textwidth}
  		\centering
  		\includegraphics[width=.95\linewidth]{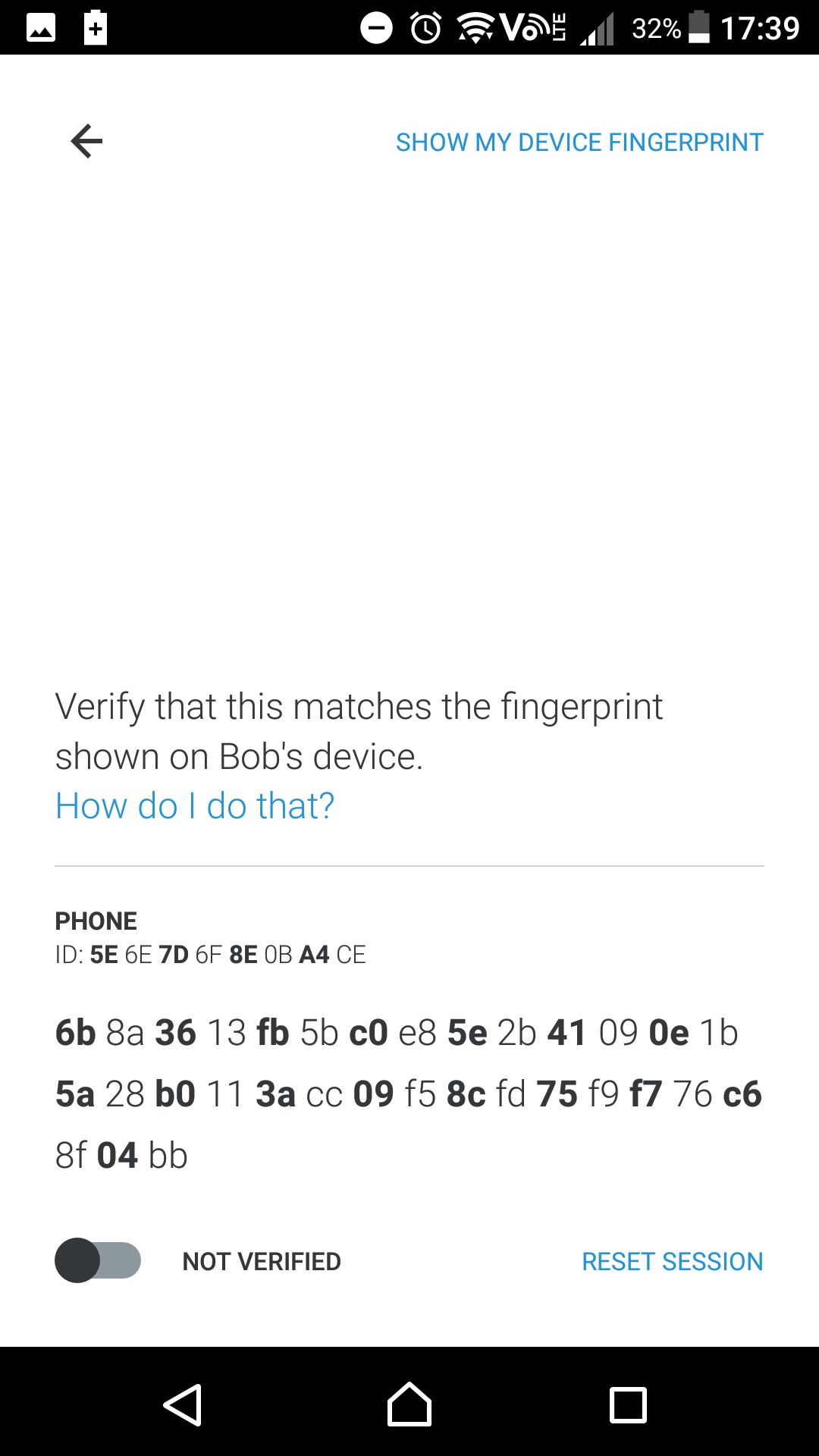}
  		\caption{Bob's public keys for one device}
  		\label{fig:impl-wire-verify-2}
	\end{subfigure}%
\begin{subfigure}{.24\textwidth}
  		\centering
	\includegraphics[width=.95\linewidth]{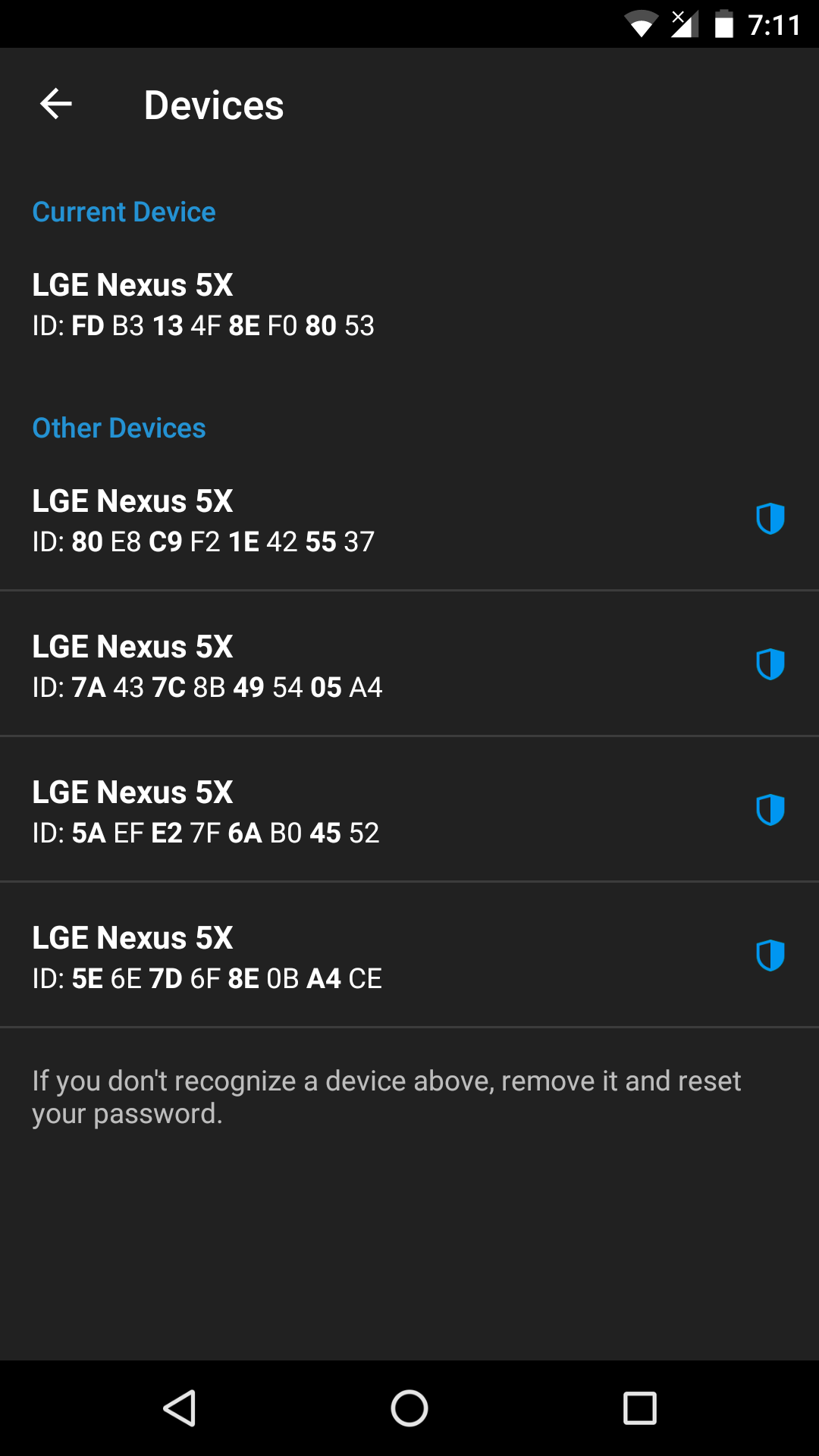}
	\caption{Other security implementations}
	\label{fig:impl-wire-devices}
	\end{subfigure}%
	\caption{Wire: verification process}
	\label{fig:impl-wire-verify}
\end{figure}

Wire's verification process does not offer the same options for verifying each participant as the other applications. However, because Wire allows several devices to be associated with one account each user has access to the whole list of devices of another conversation party.
%
When Alice wants to verify one of Bob's devices 
she can see Bob's profile, and particularly the tab displaying information about all his devices (Fig.~\ref{fig:impl-wire-kc-transit}\subref{fig:impl-wire-kc-4}). 
%
If Alice taps on one of the devices from the list, as show in Fig.~\ref{fig:impl-wire-verify}\subref{fig:impl-wire-verify-2}, she can get some information about the phone's ID number and the public keys of that device. Alice can either call Bob over the phone or meet him in person and then verify the keys. When the verification is done, Alice needs to toggle the ``not verified'' switch to specify that this particular device is verified.

\paragraph{Other Security Implementations:}

Wire does not have the extra security implementations that Signal or WhatsApp have. 
The few options include a way to change how the message conversation looks and the possibility to add an email to the account for easier log in.
The user can look at the devices which have been used with her account (as shown in Fig.~\ref{fig:impl-wire-verify}\subref{fig:impl-wire-devices}), and if there are any devices which the user does not own or recognize, she can delete that specific device. After deleting a device, the user is prompted to change her password.


\subsubsection{Case 4: Viber} \label{impl:viber-intro}

Viber is another instant messaging application that was launched in 2010, and has become quite popular, with 800 million overall users and 266 million monthly active users \cite{viber-stat}. Viber has properties similar to the other applications, where users are capable of forming groups, send messages, call each other and send pictures, videos or voice messages to other users of Viber.\footnote{Viber, by Rakuten Inc. \url{https://www.viber.com/en/about}}
Viber works on smartphones and personal computers, which makes it cross-platform. 
Viber did not have end-to-end encryption in the beginning, but introduced it in April 2016 for both one-to-one and group conversations.\footnote{``Giving Our Users Control Over Their Private Conversations'', by Michael Schmilov from Viber on April 19 2016, at \url{https://www.viber.com/en/blog/2016-04-19/giving-our-users-control-over-their-private-conversations}}
Viber does not use the Signal protocol, but implements its own protocol, which has the same concepts as the double-ratchet protocol used by Signal (as stated by its developers).

\paragraph{Initial Set Up:}\label{impl:viber-setup}
\begin{wrapfigure}[15]{r}{.49\textwidth}
\vspace{-6ex}\centering
\begin{subfigure}{.24\textwidth}
  		\centering
  		\includegraphics[width=.95\linewidth]{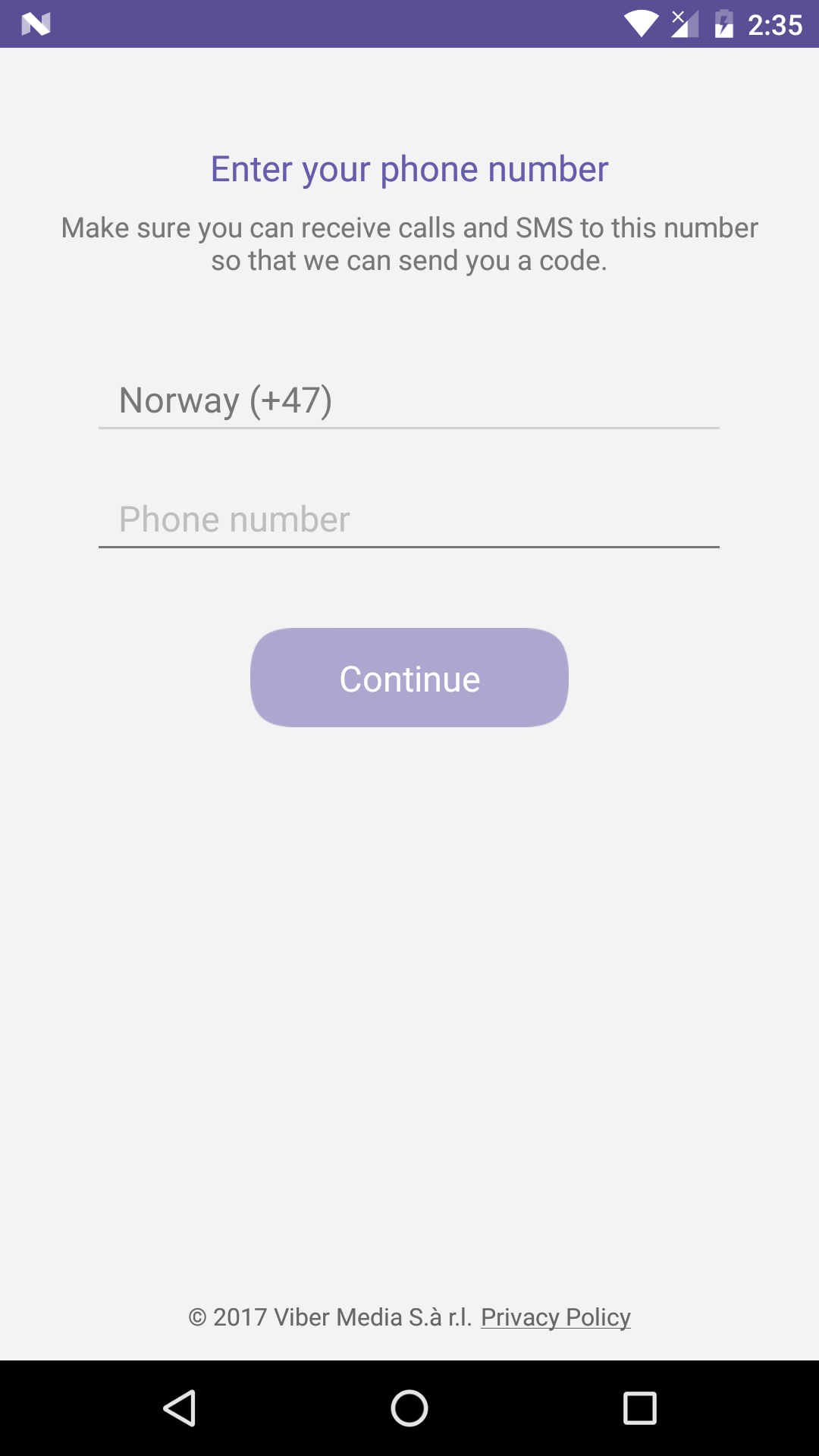}
  		\caption{Registration with phone number}
  		\label{fig:impl-viber-init-1}
	\end{subfigure}%
	\begin{subfigure}{.24\textwidth}
  		\centering
  		\includegraphics[width=.95\linewidth]{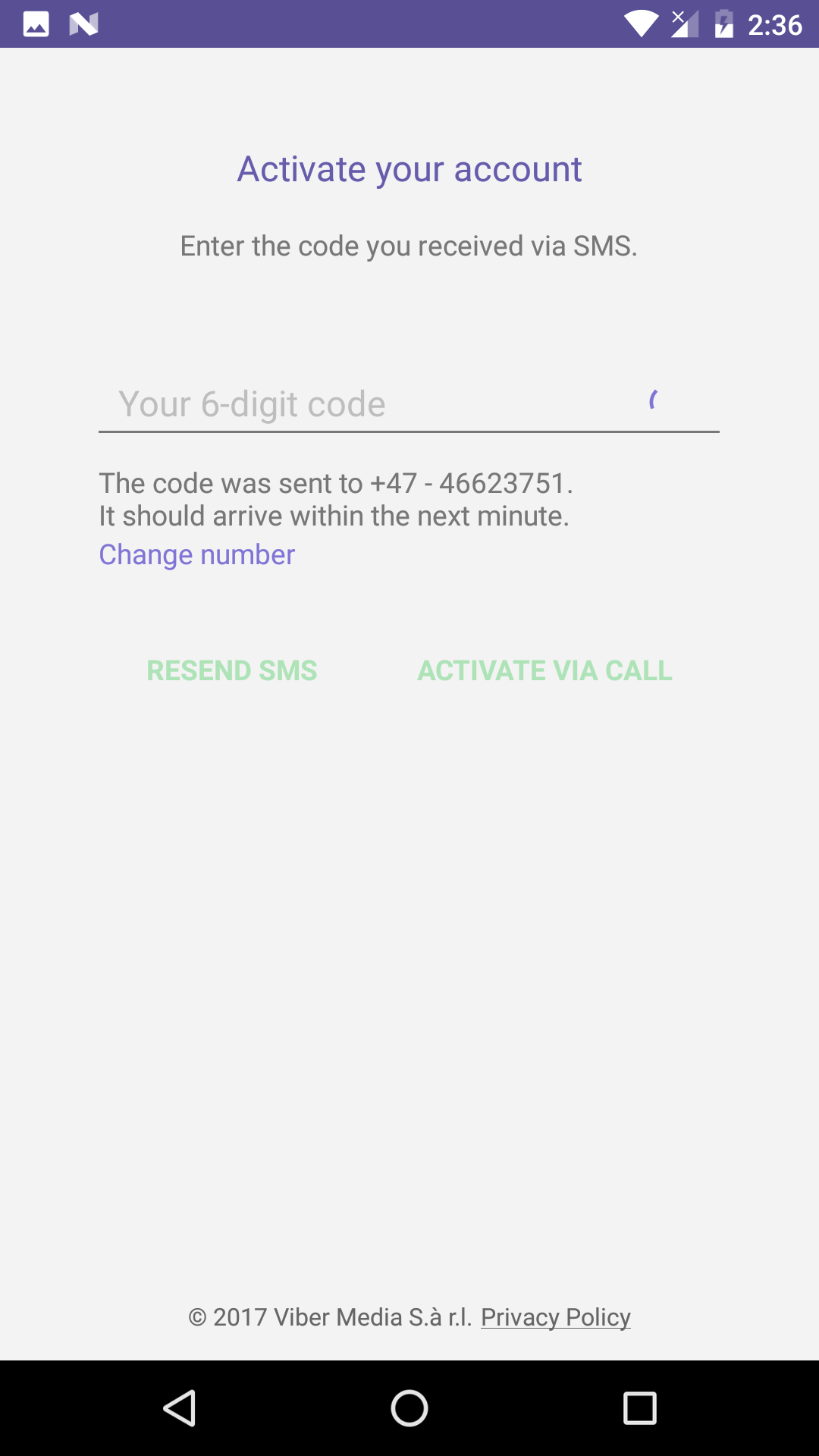}
  		\caption{Verification of phone number}
  		\label{fig:impl-viber-init-2}
	\end{subfigure}
	\caption{Viber: registration process}
	\label{fig:impl-viber-init}
\end{wrapfigure}

The user registration process of Viber is the same as that in the previous applications. Fig.~\ref{fig:impl-viber-init}\subref{fig:impl-viber-init-1} shows the user input for the user's phone number in the registration screen.
Fig.~\ref{fig:impl-viber-init}\subref{fig:impl-viber-init-2} shows the activation process of the user account. The user can either give Viber access to the SMS inbox to enter the verification code automatically or do it manually otherwise. If the SMS with the verification code does not arrive within one minute, the user can ask the application to either resend a new verification code or get the code through a phone call.

\paragraph{Message After a Key Change:}

\begin{figure}[t]
\centering
	\begin{subfigure}{.24\textwidth}
  		\centering
  		\includegraphics[width=.95\linewidth]{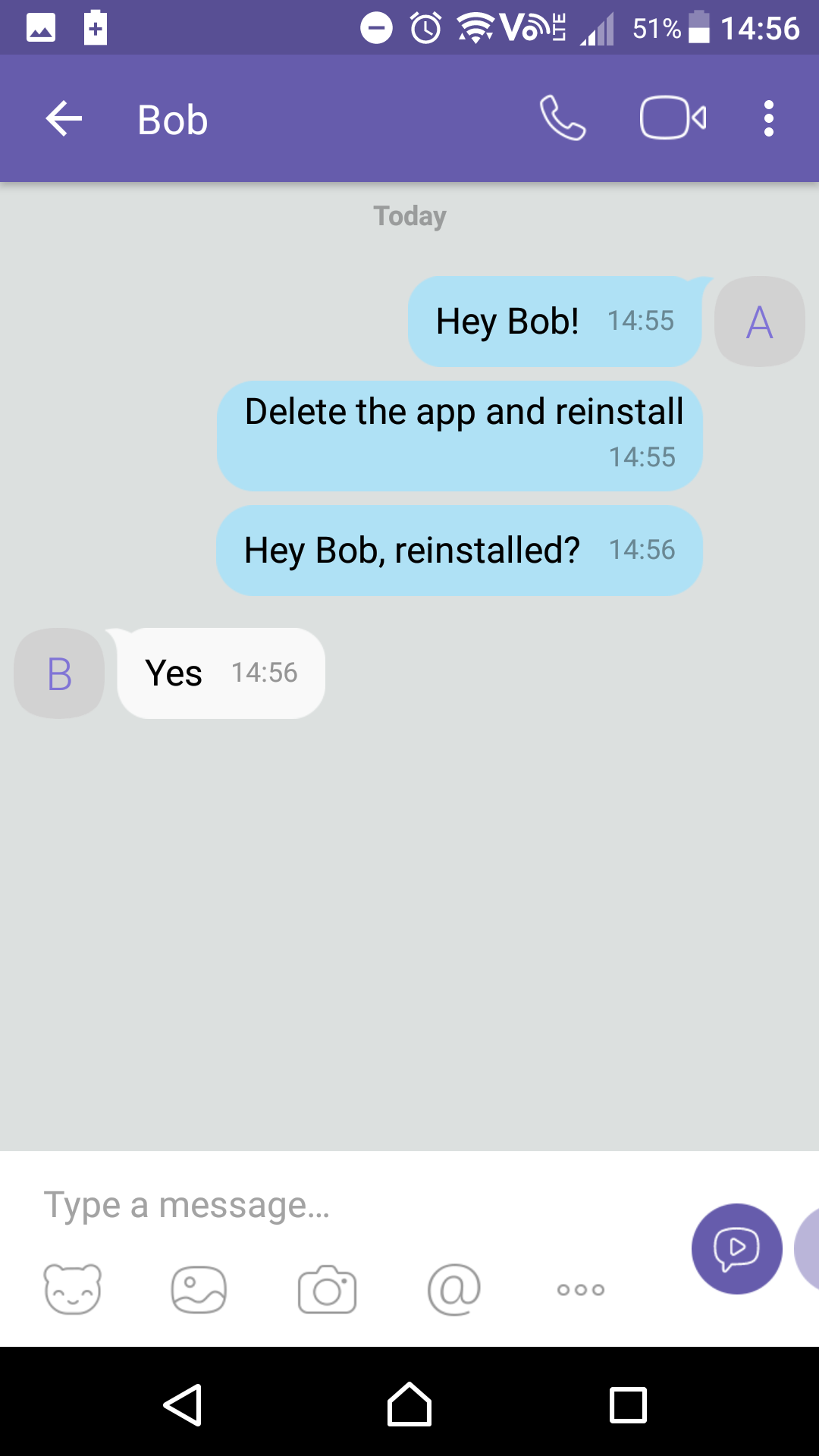}
  		\caption{No notification about key changes}
  		\label{fig:impl-viber-kc-2}
	\end{subfigure}
	\begin{subfigure}{.24\textwidth}
		\centering
		\includegraphics[width=.95\linewidth]{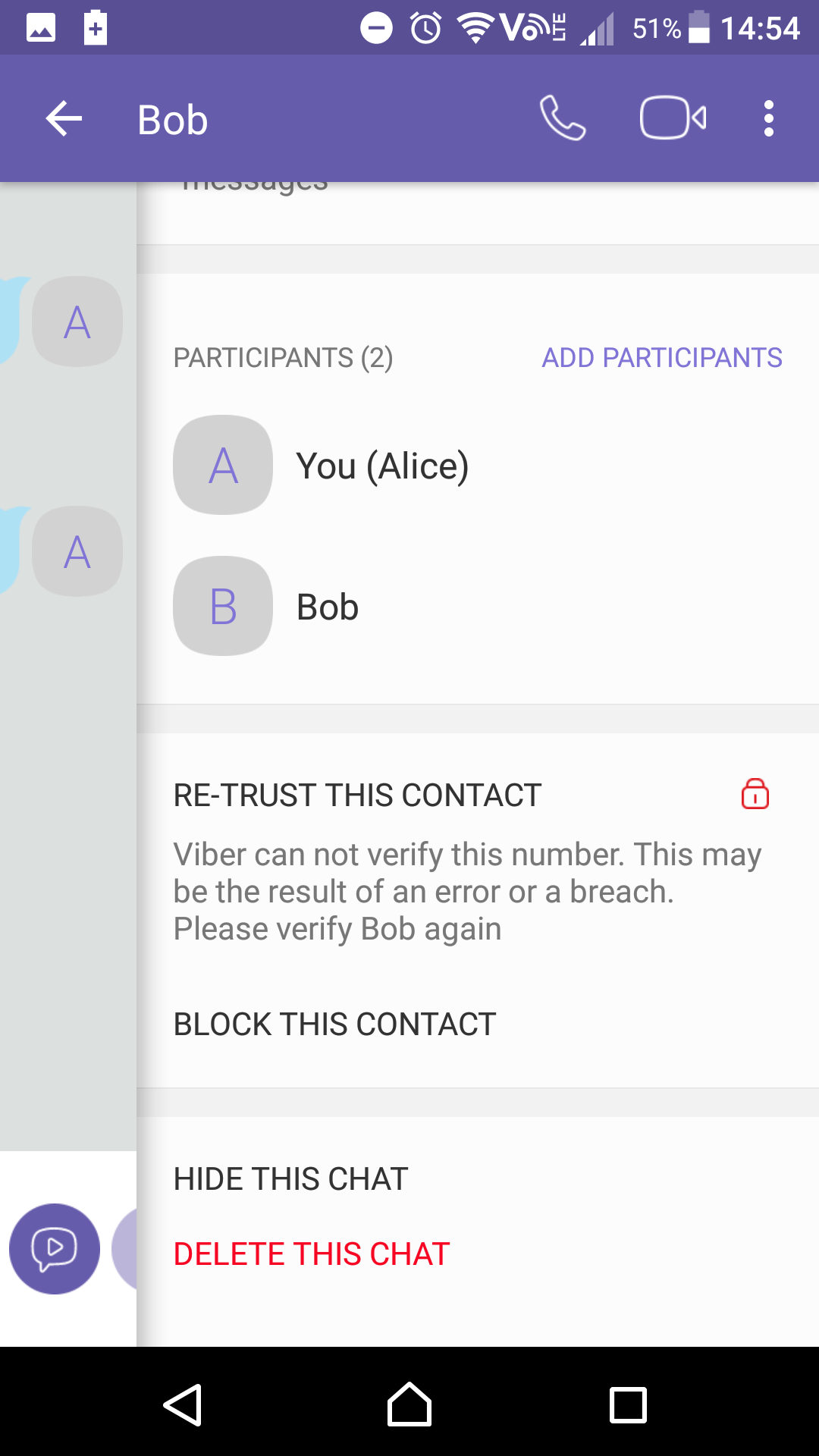}
		\caption{Bob needs to be re-trusted}
		\label{fig:impl-viber-kc-3}
	\end{subfigure}
	\begin{subfigure}{.24\textwidth}
  		\centering
  		\includegraphics[width=.95\linewidth]{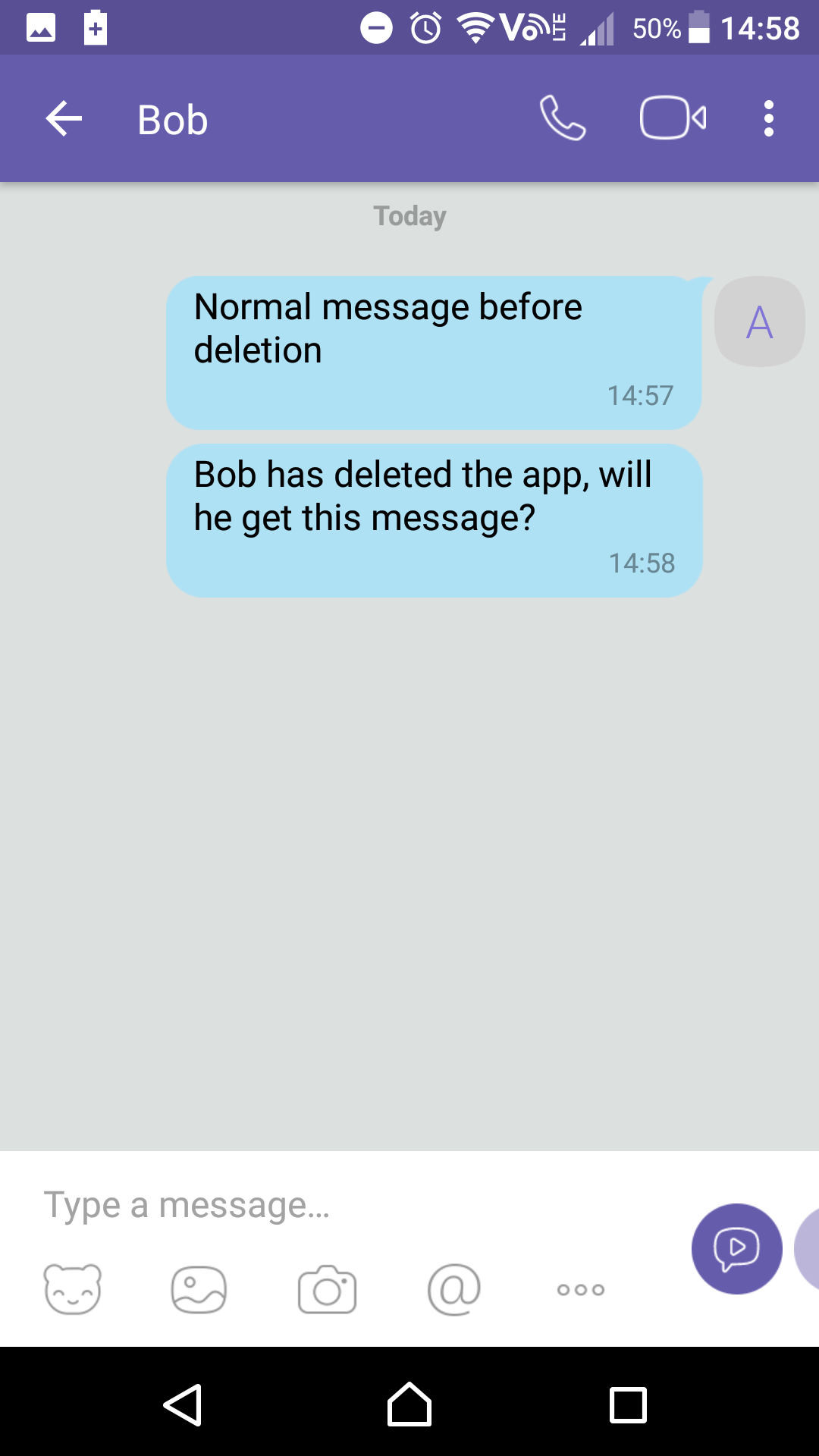}
  		\caption{Message after Bob has deleted}
  		\label{fig:impl-viber-kc-transit-2}
	\end{subfigure}
	\begin{subfigure}{.24\textwidth}
		\centering
		\includegraphics[width=.95\linewidth]{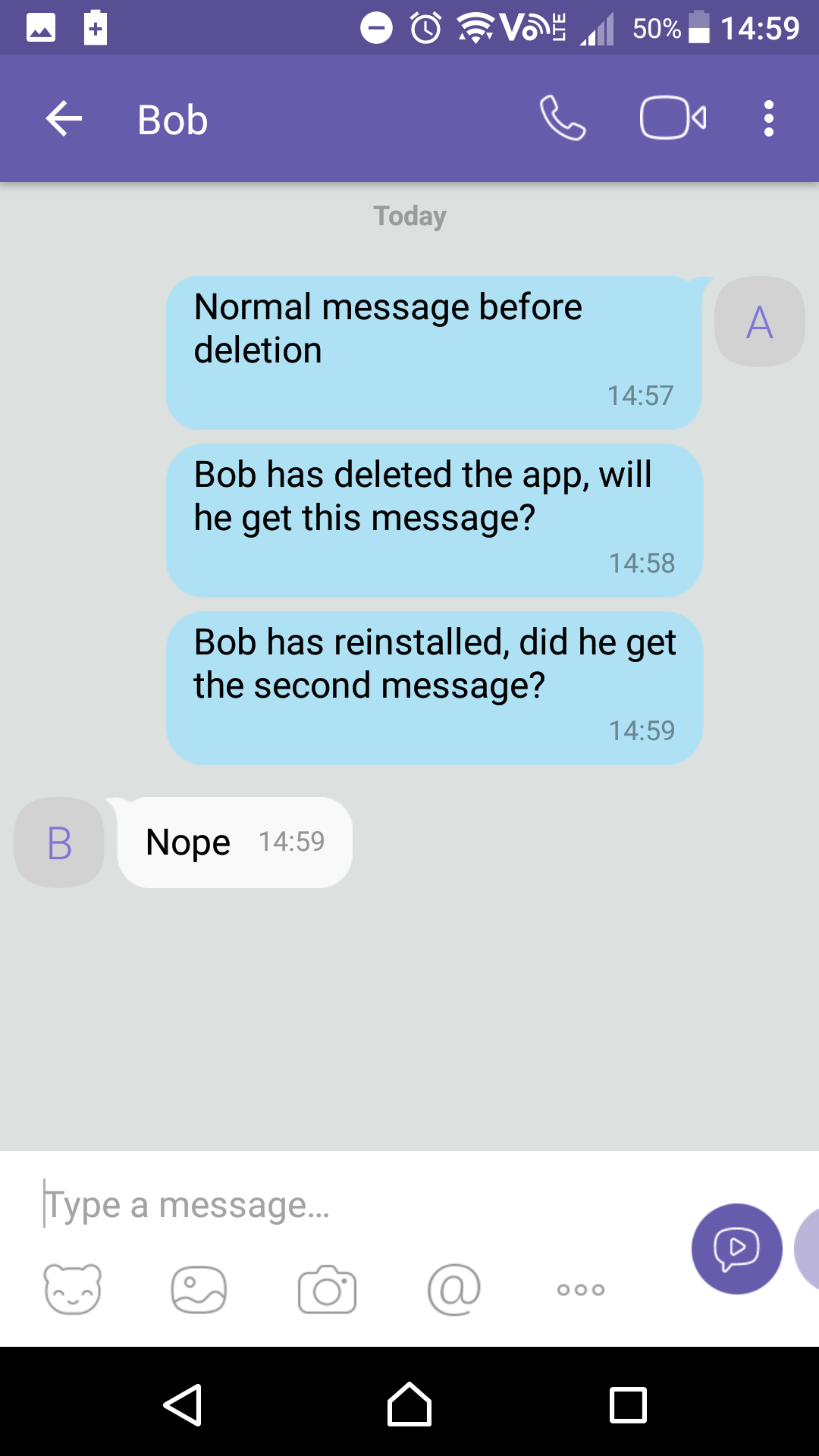}
		\caption{Bob did not get the second message}
		\label{fig:impl-viber-kc-transit-3}
	\end{subfigure}
	\caption{Viber: Message after a key change; and Key change while a message is in transit.}
	\label{fig:impl-viber-kc}
\end{figure}

Viber does not notify the participants when the cryptographic keys change during a conversation. 
The first two messages in 
Fig.~\ref{fig:impl-viber-kc}\subref{fig:impl-viber-kc-2} show Alice initiating the conversation with Bob.
%
After Bob has reinstalled the application, Alice sends him another message. However, Fig.~\ref{fig:impl-viber-kc}\subref{fig:impl-viber-kc-2} (third message) shows that Viber does not give any notification to Alice that Bob has generated new cryptographic keys. 
The only way for Alice to find out this is by 
checking the details of the conversation, by swiping from right to left. As shown in Fig.~\ref{fig:impl-viber-kc}\subref{fig:impl-viber-kc-3}, if the ``Trust this contact'' tab has changed to ``Re-trust this contact'', then Alice can infer that Bob has new cryptographic keys that should be verified.

%
\paragraph{Key Change While a Message is in Transit:}


Key changes in transit are handled the same as key changes after the reinstall of the application explained before. 
%
Fig.~\ref{fig:impl-viber-kc}\subref{fig:impl-viber-kc-transit-2} shows that Alice sends a second message to Bob before he has reinstalled his application, and there is no information given to Alice if the message is sent or read by Bob.
Fig.~\ref{fig:impl-viber-kc}\subref{fig:impl-viber-kc-transit-3} shows the third message from Alice to Bob after he has reinstalled his application. Alice never receives any notification from Viber that Bob has new cryptographic keys nor that he has not received the second message, 
and Viber does not re-encrypt and re-send messages later on.


\paragraph{Verification Process Between Participants:}

\begin{figure}[t]
\centering
\begin{subfigure}{.24\textwidth}
  		\centering
  		\includegraphics[width=.95\linewidth]{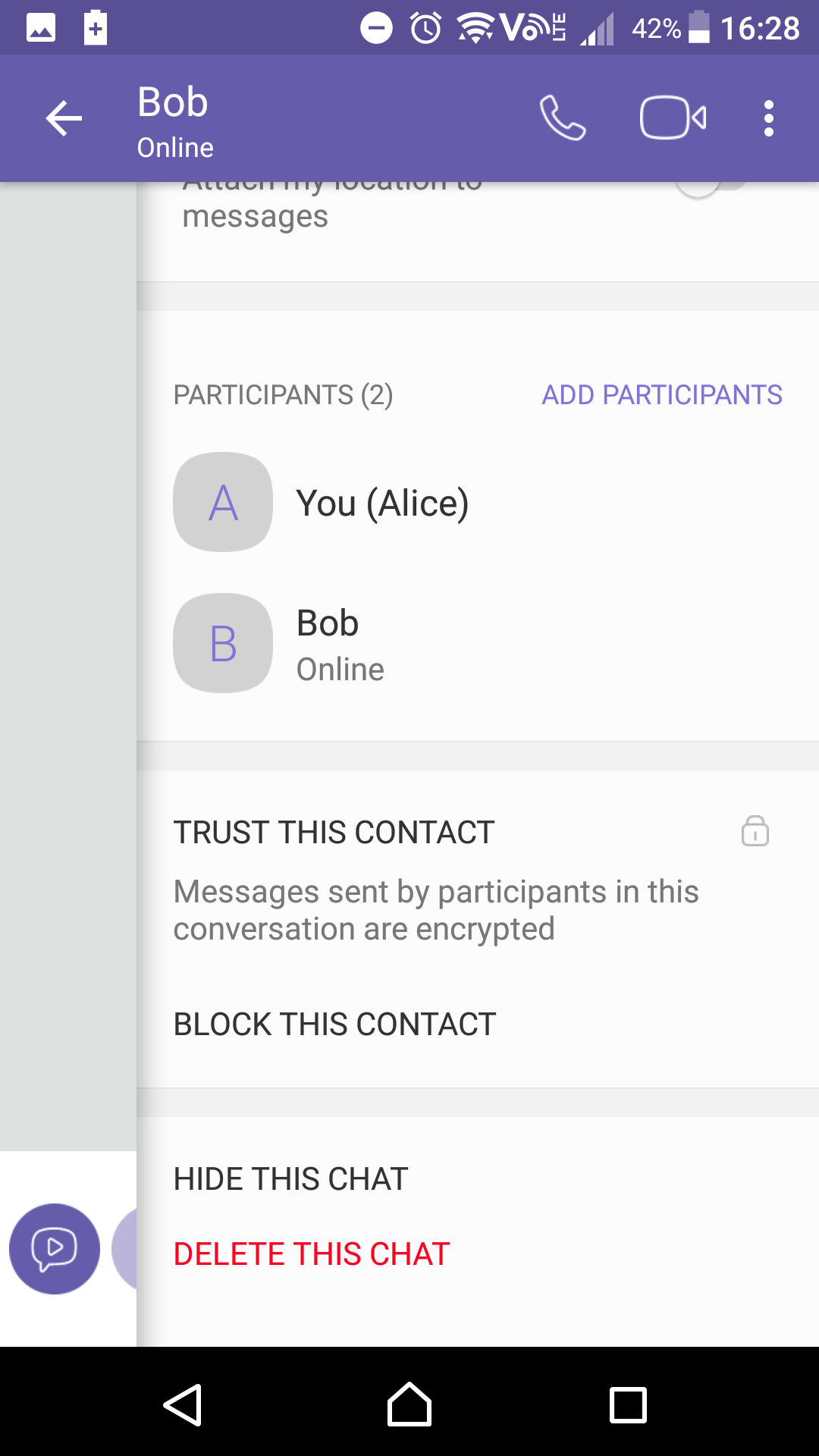}
  		\caption{Conversation info}
  		\label{fig:impl-viber-verify-1}
	\end{subfigure}%
	\begin{subfigure}{.24\textwidth}
  		\centering
  		\includegraphics[width=.95\linewidth]{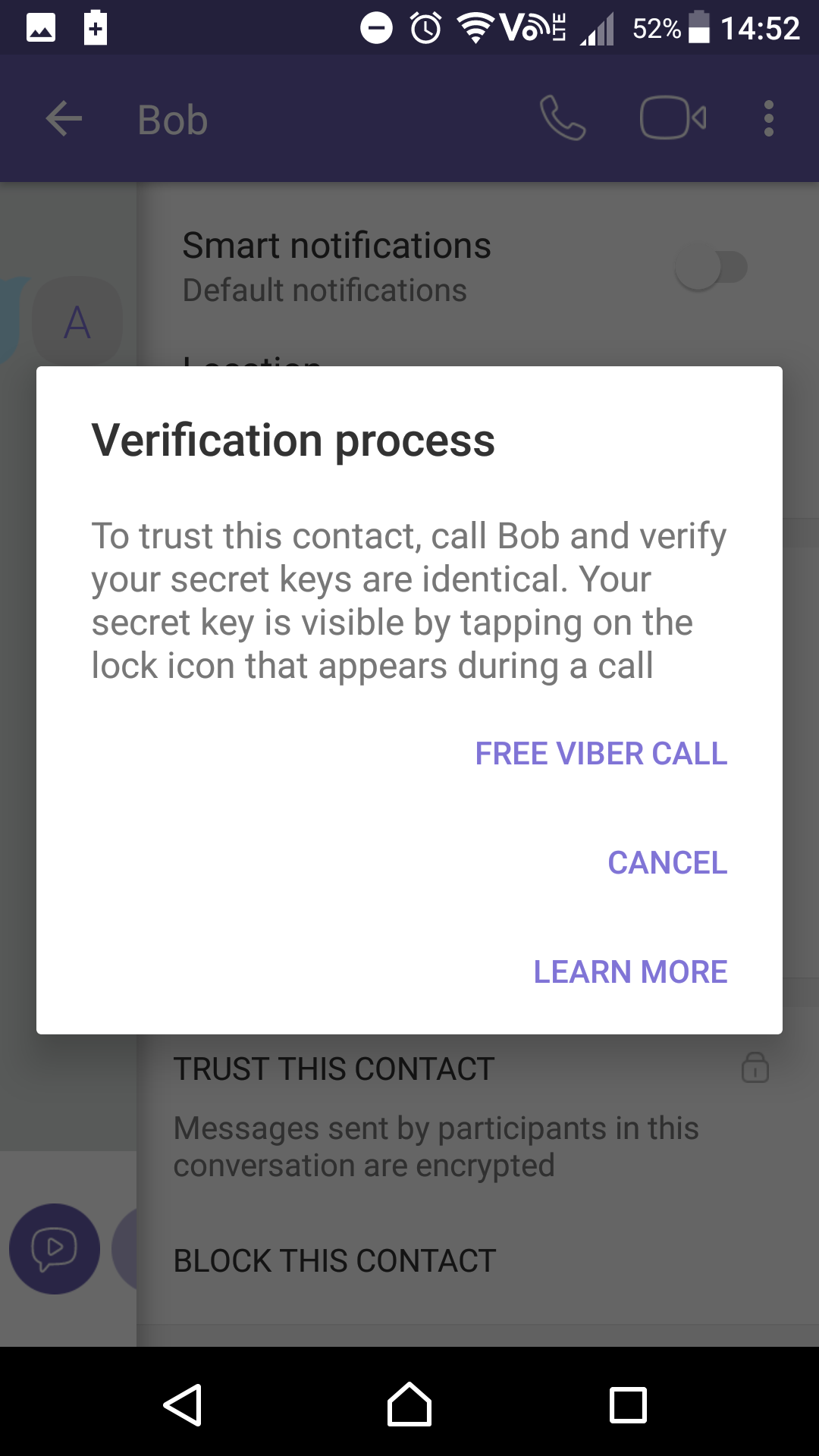}
  		\caption{Verify a contact}
  		\label{fig:impl-viber-verify-2}
	\end{subfigure}
	\begin{subfigure}{.24\textwidth}
		\centering
		\includegraphics[width=.95\linewidth]{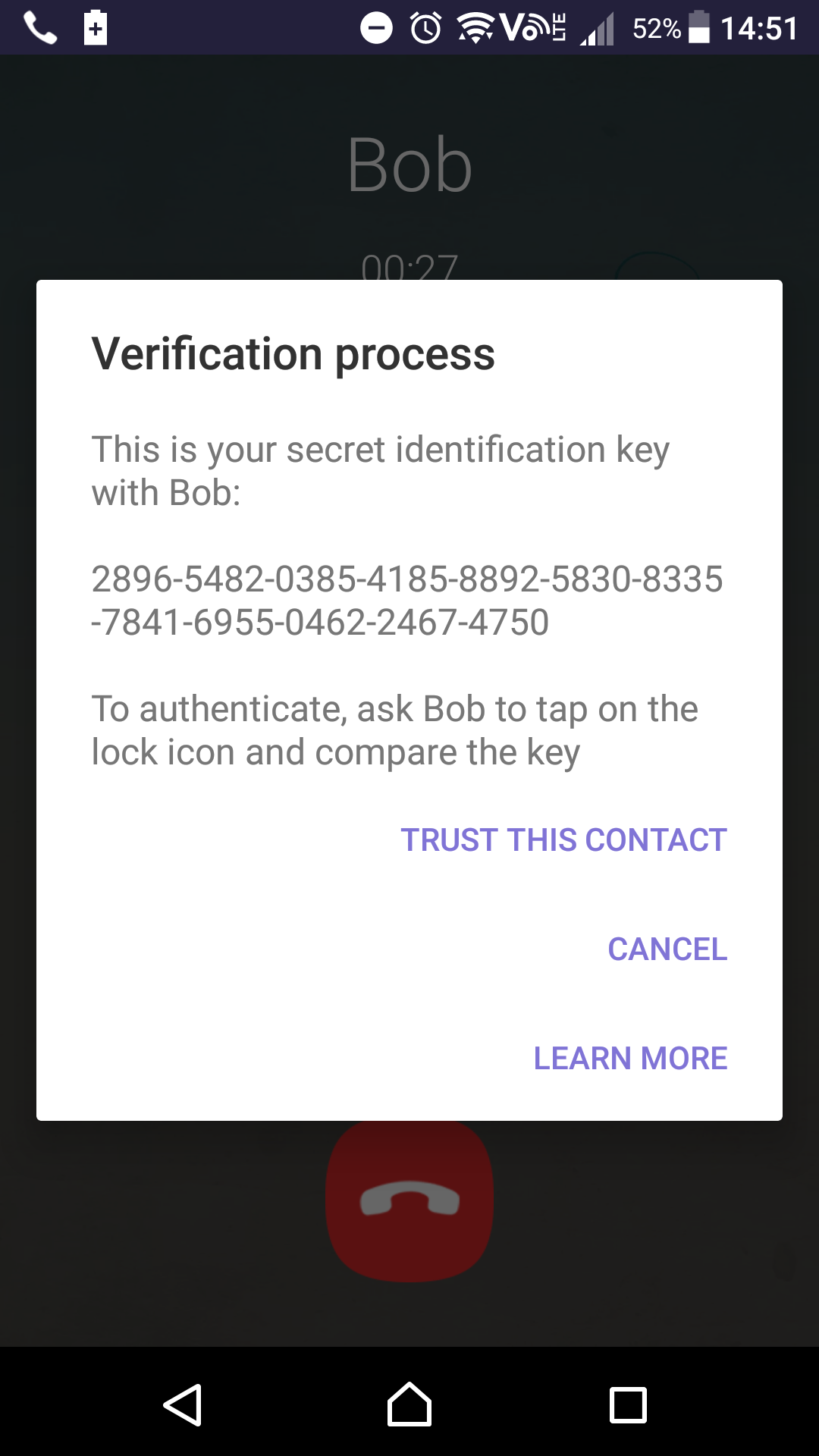}
		\caption{Alice verifying Bob}
		\label{fig:impl-viber-verify-3}
	\end{subfigure}
	\begin{subfigure}{.24\textwidth}
		\centering
		\includegraphics[width=.95\linewidth]{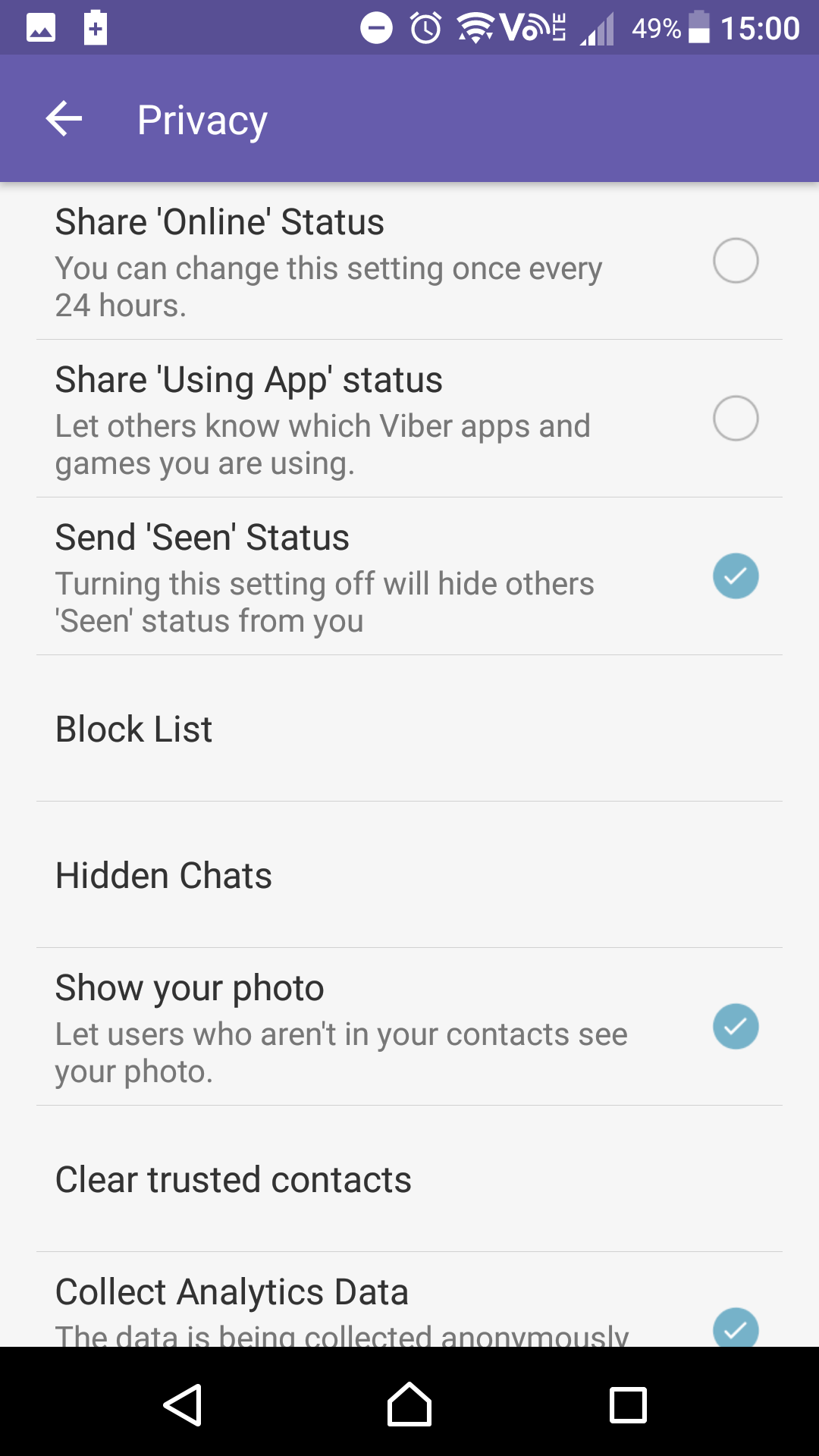}
		\caption{Viber: Privacy settings}
		\label{fig:impl-viber-priv-1}
	\end{subfigure}
	\caption{Viber: Verification process; and Privacy settings.}
	\label{fig:impl-viber-verify}
\end{figure}

The process of verifying a contact in Viber is quite straight forward. If Alice wants to verify Bob, she goes to one of their conversations, swipes (Fig.~\ref{fig:impl-viber-verify}\subref{fig:impl-viber-verify-1}) to get the information tab, and then goes to the ``Trust this contact'' option.
Fig.~\ref{fig:impl-viber-verify}\subref{fig:impl-viber-verify-2} shows the popup notification box after Alice clicks the ``Trust this contact'' option. The only verification option Alice can use to verify Bob is by calling Bob and then read the cryptographic keys over the phone.
Fig.~\ref{fig:impl-viber-verify}\subref{fig:impl-viber-verify-3} shows when Alice calls Bob and wants to verify, the popup message displays the cryptographic keys that both Alice and Bob share. When they have verified each other, they press the ``Trust this contact'' button.

\paragraph{Other Security Implementations:}

Viber does not have extra security implementations. Fig.~\ref{fig:impl-viber-priv-1} shows the privacy settings where the only security implementation is the ``Clear trusted contacts'' which clears all the contacts that Alice has verified throughout the time she had the account.


\subsubsection{Case 5: Riot} \label{impl:riot-intro}

Riot is a new chat client that is built on top of the Matrix\footnote{\url{https://matrix.org/}} protocol for its end-to-end encrypted capabilities. Matrix is an open standard for decentralized communications which uses bridged networks and cross-platform possibilities plus full end-to-end encryption that is based on the Double Ratchet protocol from Signal. 

Riot uses servers (the same as Signal), but one does not need to rely on servers under the control of the Matrix team (unlike Signal). Riot and Matrix are open source, which means anyone can set up their own servers with the Matrix implementation and use its end-to-end encryption. This is good for companies that want to have secure chat between employees but do not want to rely on anything outside their own network. Riot also provides group chat, voice (VoIP) and video calling, file transfer and integration with other applications such as Slack\footnote{\url{https://slack.com/}} or IRC\footnote{\url{https://www.wikiwand.com/en/Internet_Relay_Chat}}. 

\paragraph{Initial Set Up:} \label{impl:riot-setup}
\begin{wrapfigure}[14]{r}{.49\textwidth}
\vspace{-5ex}\centering
\begin{subfigure}{.24\textwidth}
  		\centering
  		\includegraphics[width=.95\linewidth]{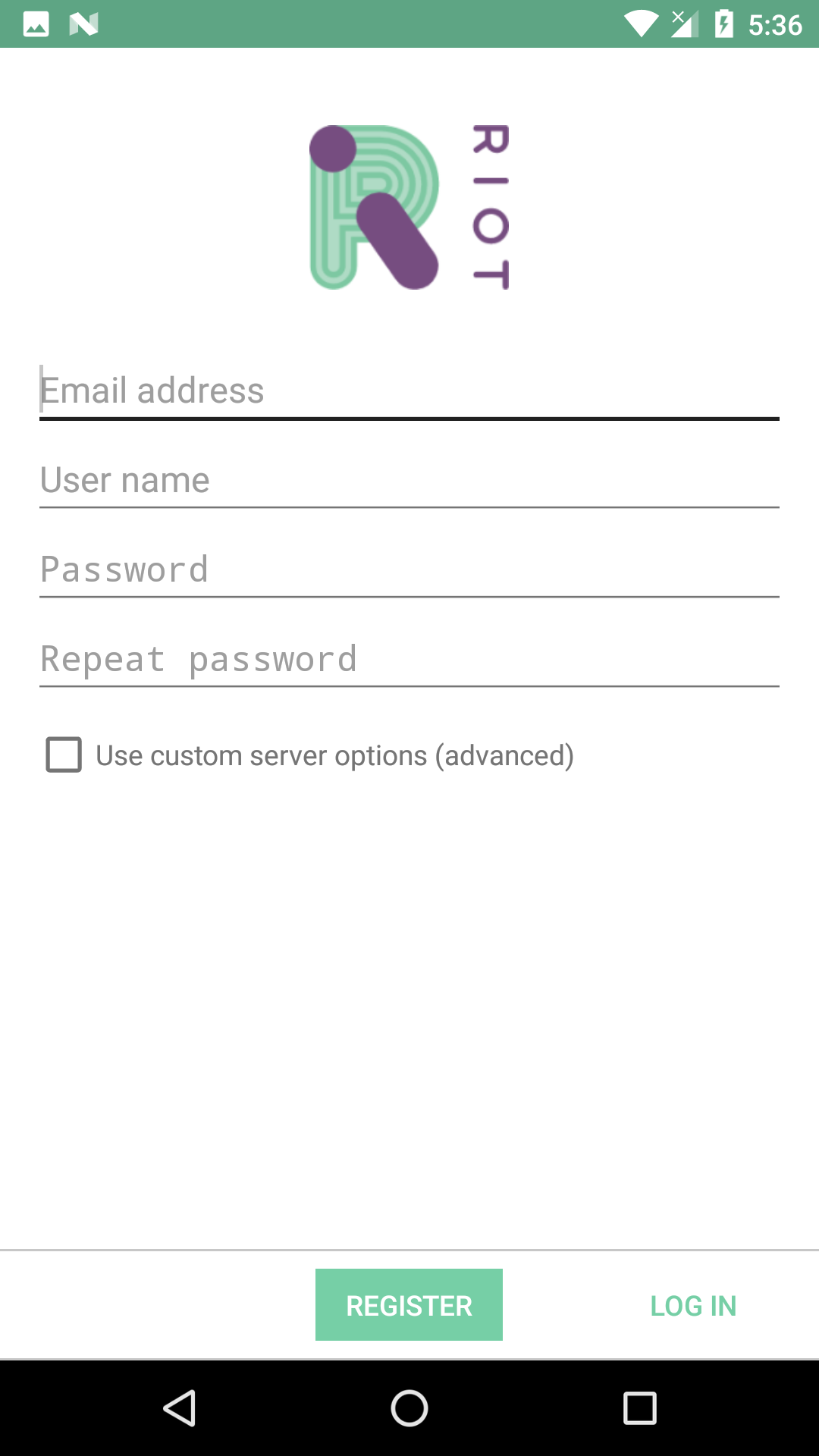}
  		\caption{Registration by email}
  		\label{fig:impl-riot-init-1}
	\end{subfigure}%
	\begin{subfigure}{.24\textwidth}
		\centering
		\includegraphics[width=.95\linewidth]{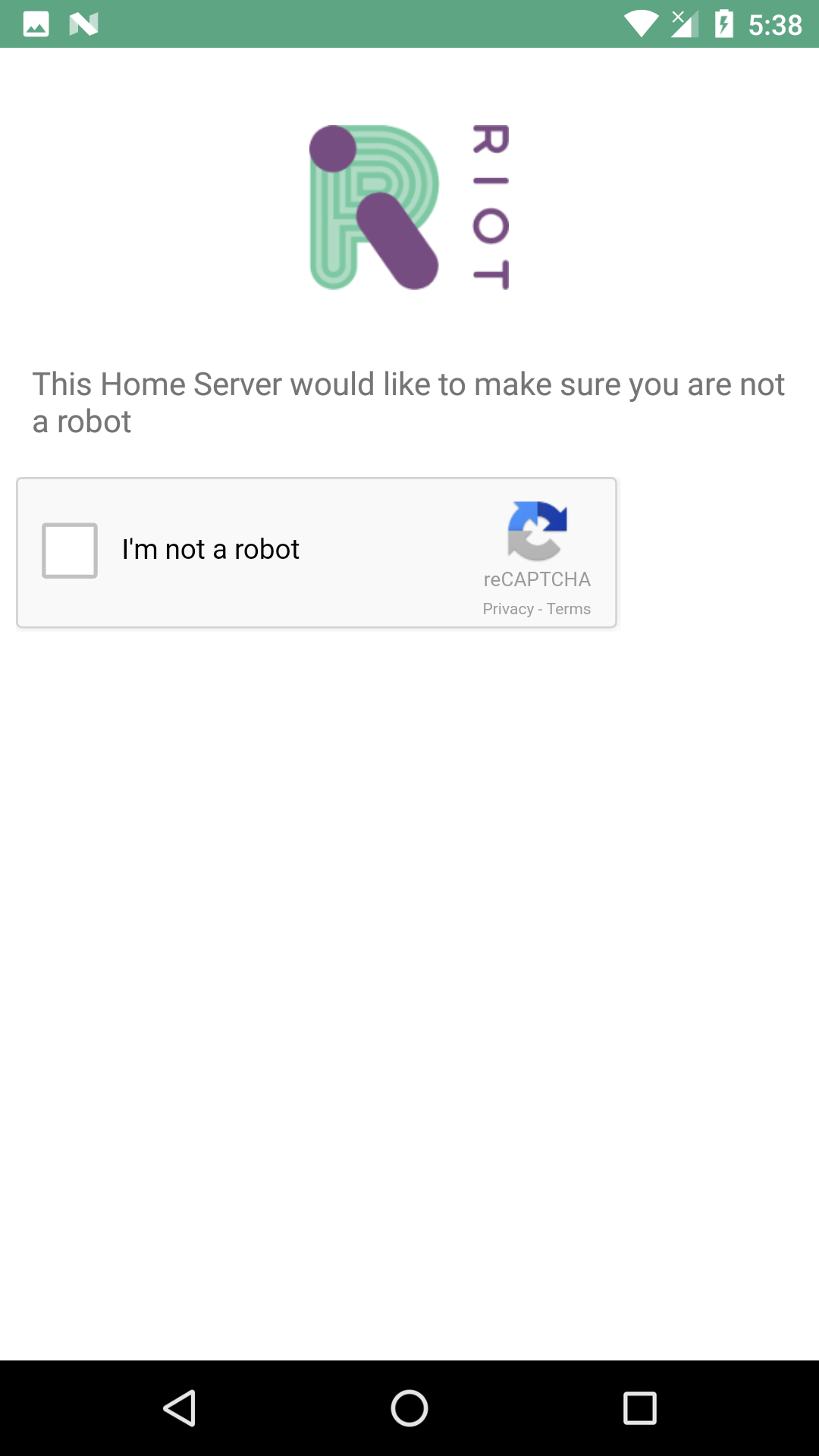}
		\caption{Captcha code}
		\label{fig:impl-riot-init-2}
	\end{subfigure}
	\caption{Riot: registration process}
	\label{fig:impl-riot-init}
\end{wrapfigure}

Riot is the only messaging client that does not rely on a phone number, but a user registers an account with an email address (Fig.~\ref{fig:impl-riot-init}\subref{fig:impl-riot-init-1}).
When a user registers through the app, Riot sends a confirmation link to the entered email address and the user has to click on that link, after which
a Captcha verification will be requested for an extra layer of security as shown in Fig.~\ref{fig:impl-riot-init}\subref{fig:impl-riot-init-2}.

\paragraph{Message After a Key Change:}

\begin{figure}[t]
\centering
 \begin{subfigure}{.24\textwidth}
   		\centering
   		\includegraphics[width=.95\linewidth]{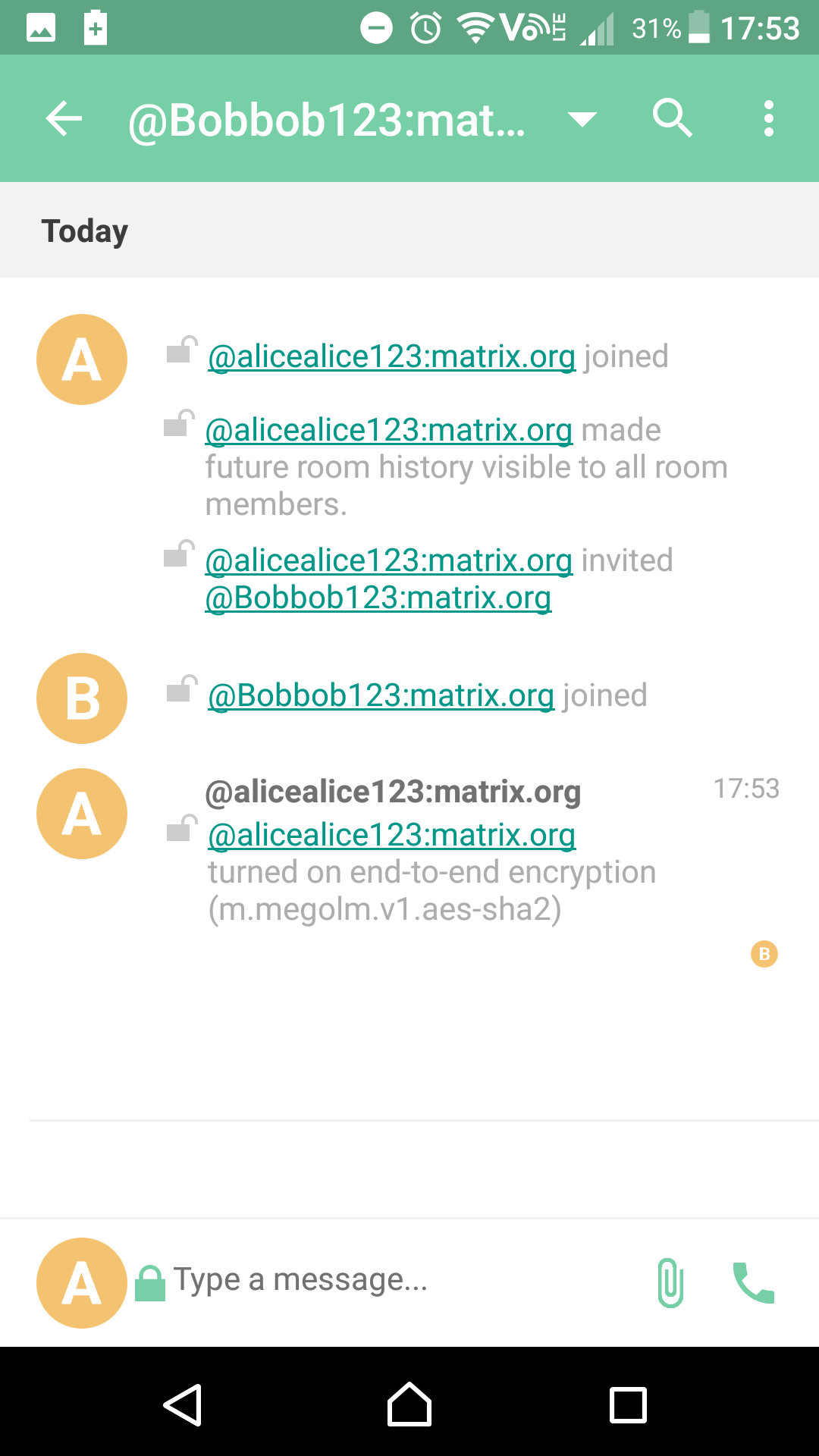}
   		\caption{Initial conversation, Alice's view}
   		\label{fig:impl-riot-kc-1}
 	\end{subfigure}%
	\begin{subfigure}{.24\textwidth}
		\centering
		\includegraphics[width=.95\linewidth]{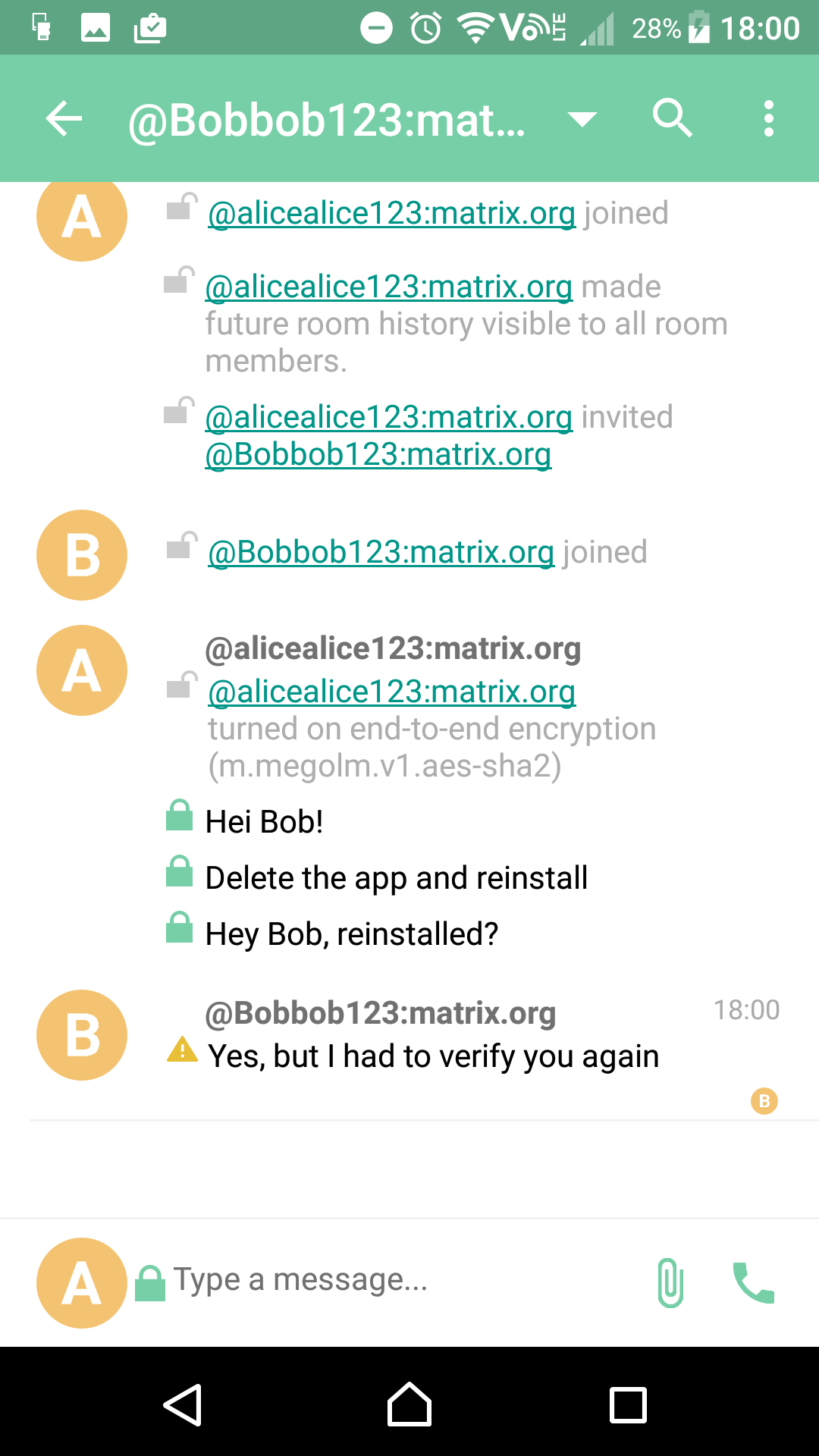}
		\caption{Message to Bob after he reinstalled}
		\label{fig:impl-riot-kc-3}
	\end{subfigure}
%
	\begin{subfigure}{.24\textwidth}
  		\centering
  		\includegraphics[width=.95\linewidth]{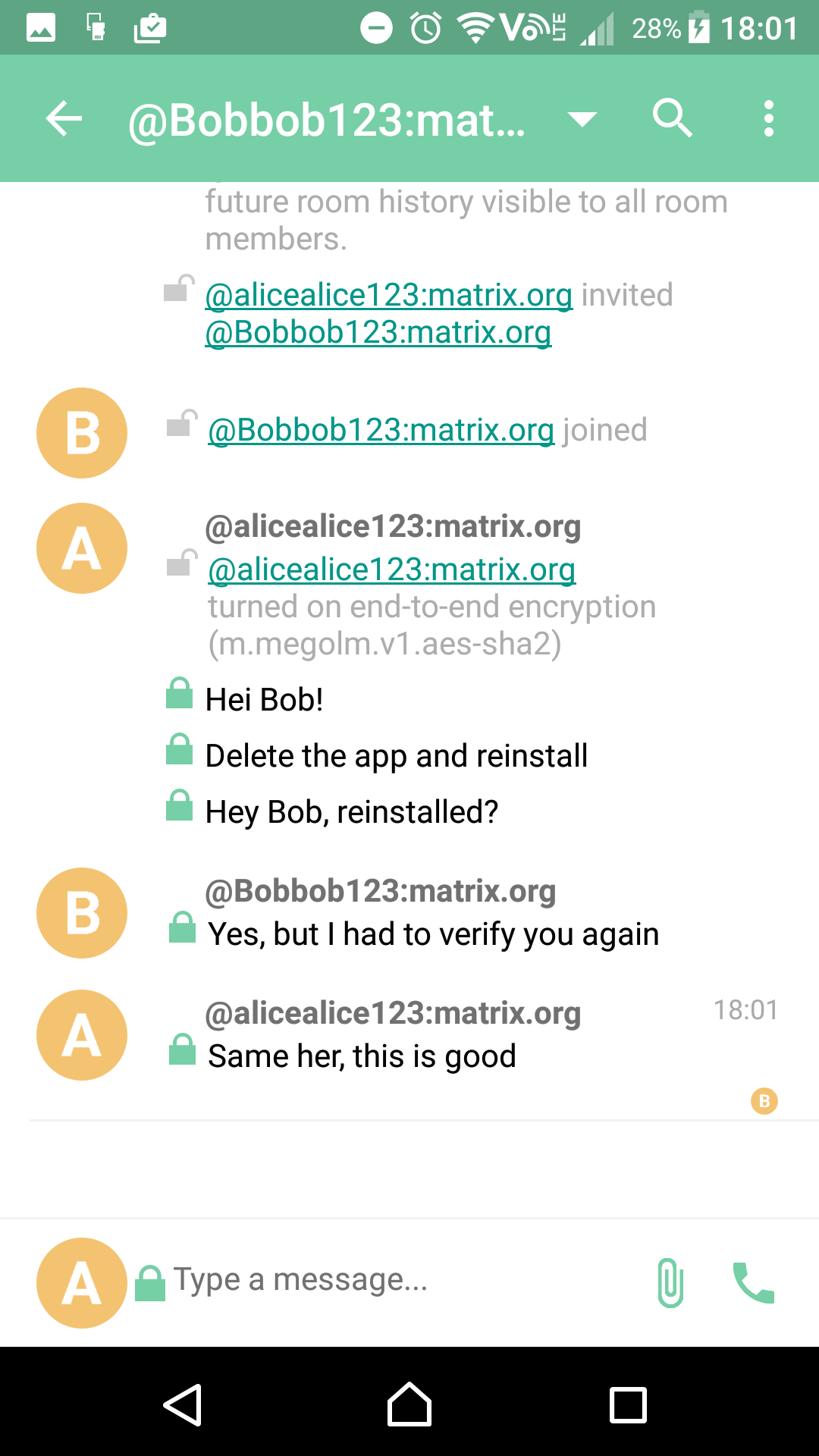}
  		\caption{New messages are verified}
  		\label{fig:impl-riot-kc-5}
	\end{subfigure}
	\begin{subfigure}{.24\textwidth}
		\centering
		\includegraphics[width=.95\linewidth]{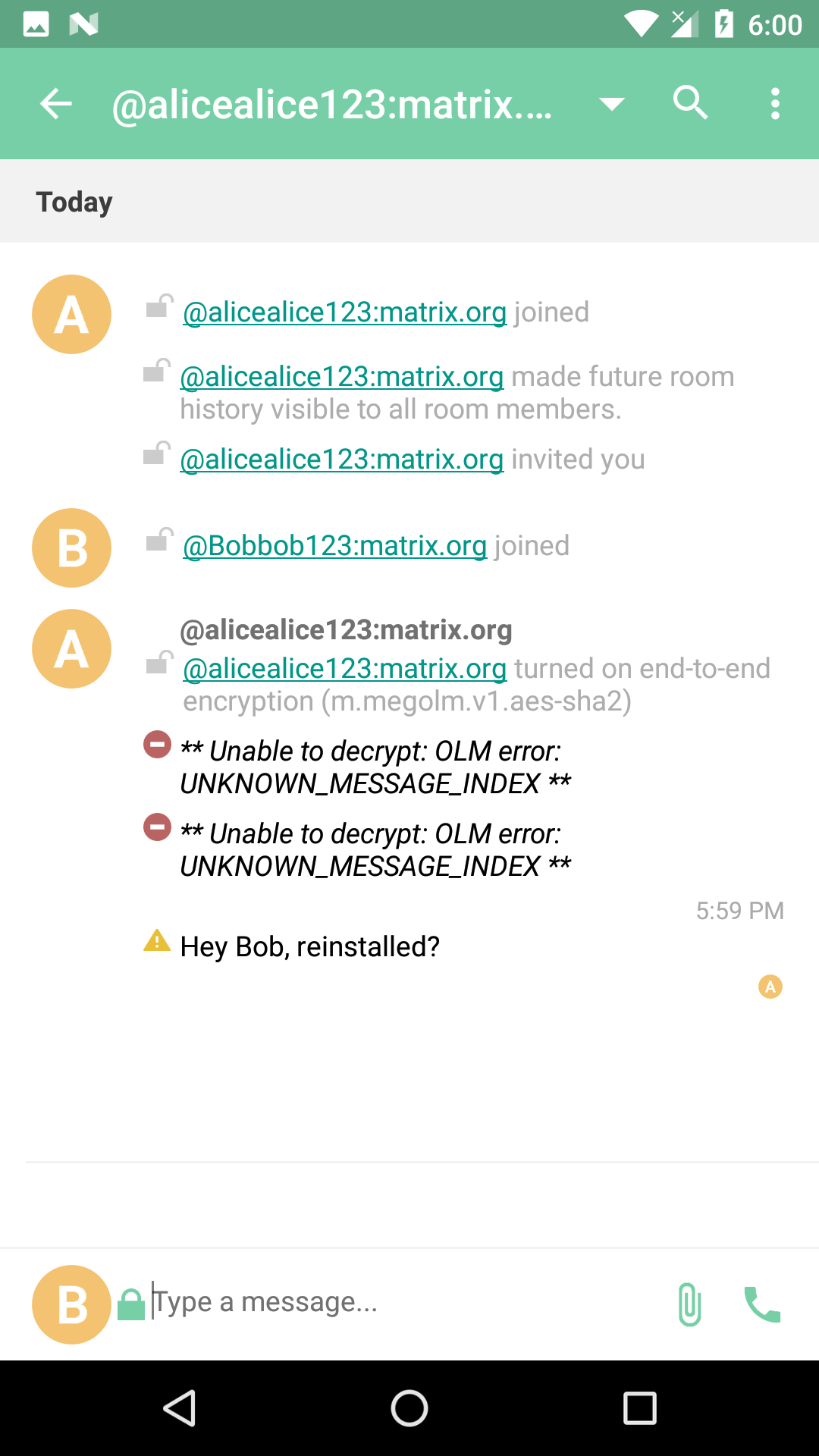}
		\caption{Bob's view of previous messages}
		\label{fig:impl-riot-kc-6}
	\end{subfigure}
	\caption{Riot: message after a key change}
	\label{fig:impl-riot-kc}
\end{figure}

\begin{wrapfigure}[16]{r}{.49\textwidth}
\vspace{-4ex}\centering
	\begin{subfigure}{.24\textwidth}
  		\centering
  		\includegraphics[width=.95\linewidth]{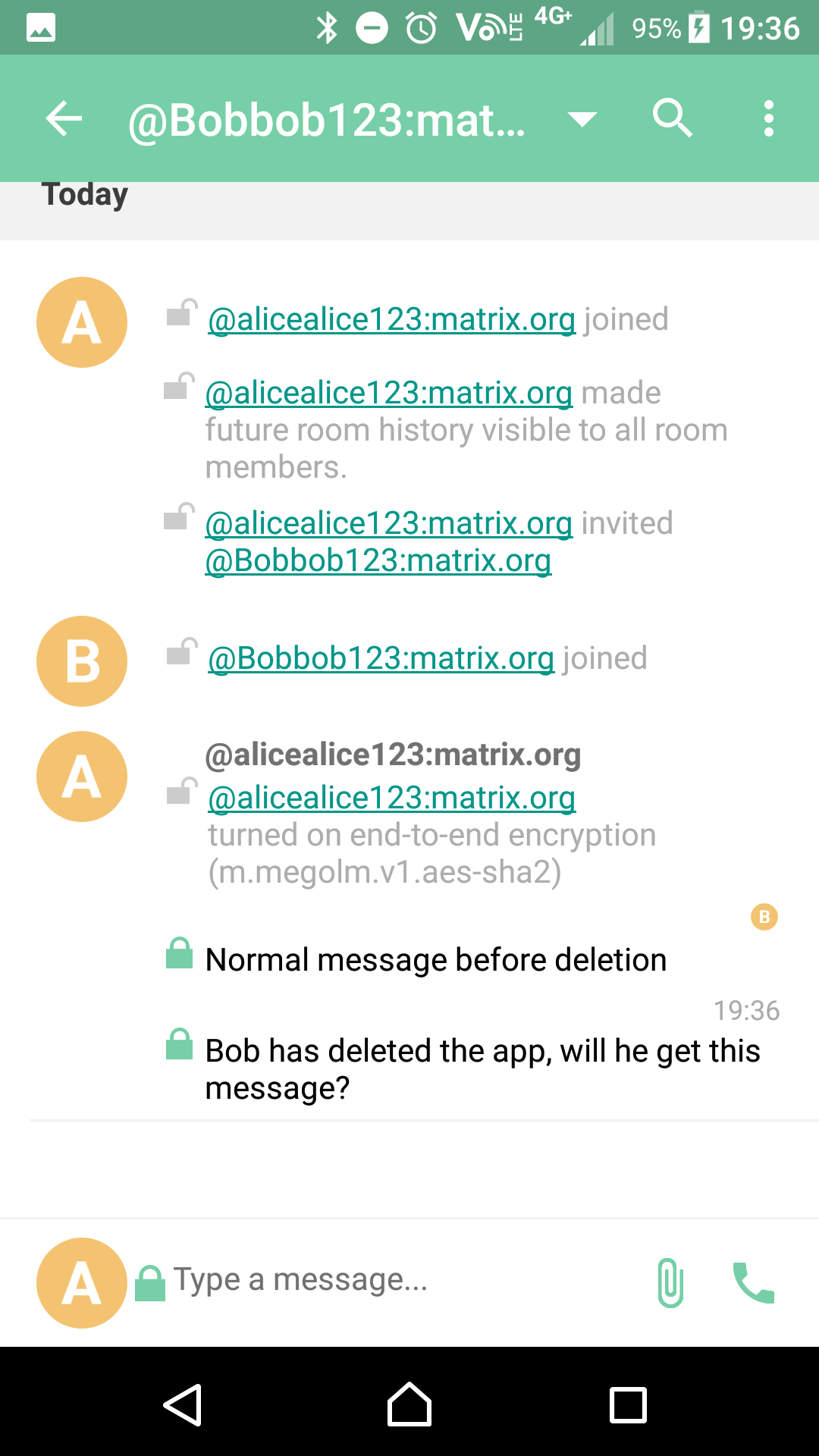}
  		\caption{Message before Bob reinstalled}
  		\label{fig:impl-riot-kc-transit-2}
	\end{subfigure}
	\begin{subfigure}{.24\textwidth}
		\centering
		\includegraphics[width=.95\linewidth]{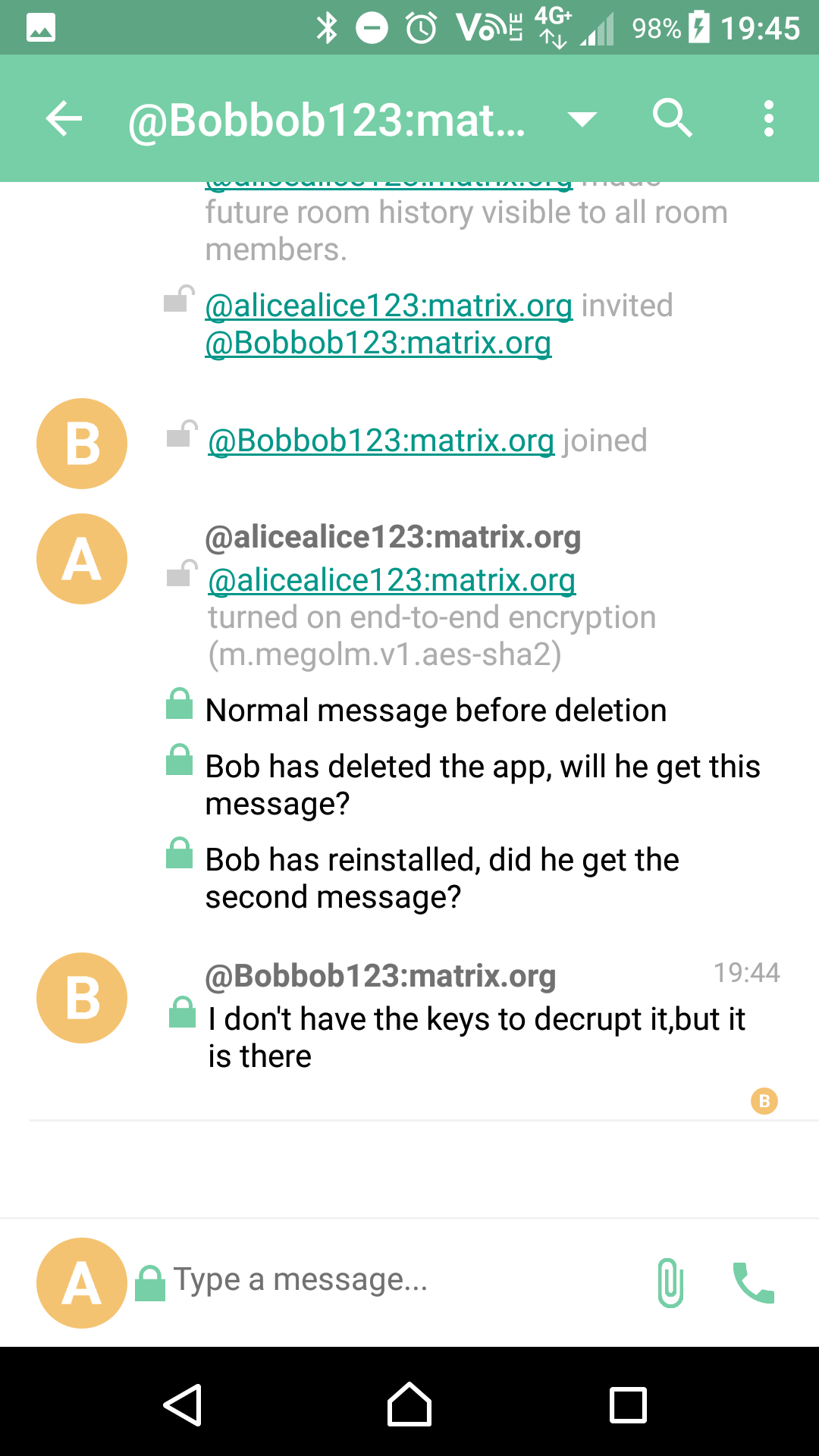}
		\caption{Bob's message after reinstall}
		\label{fig:impl-riot-kc-transit-3}
	\end{subfigure}
	\caption{Riot: key change while a message is in transit}
	\label{fig:impl-riot-kc-transit}
\end{wrapfigure}

\begin{figure}[t]
\centering
\begin{subfigure}{.24\textwidth}
  		\centering
  		\includegraphics[width=.95\linewidth]{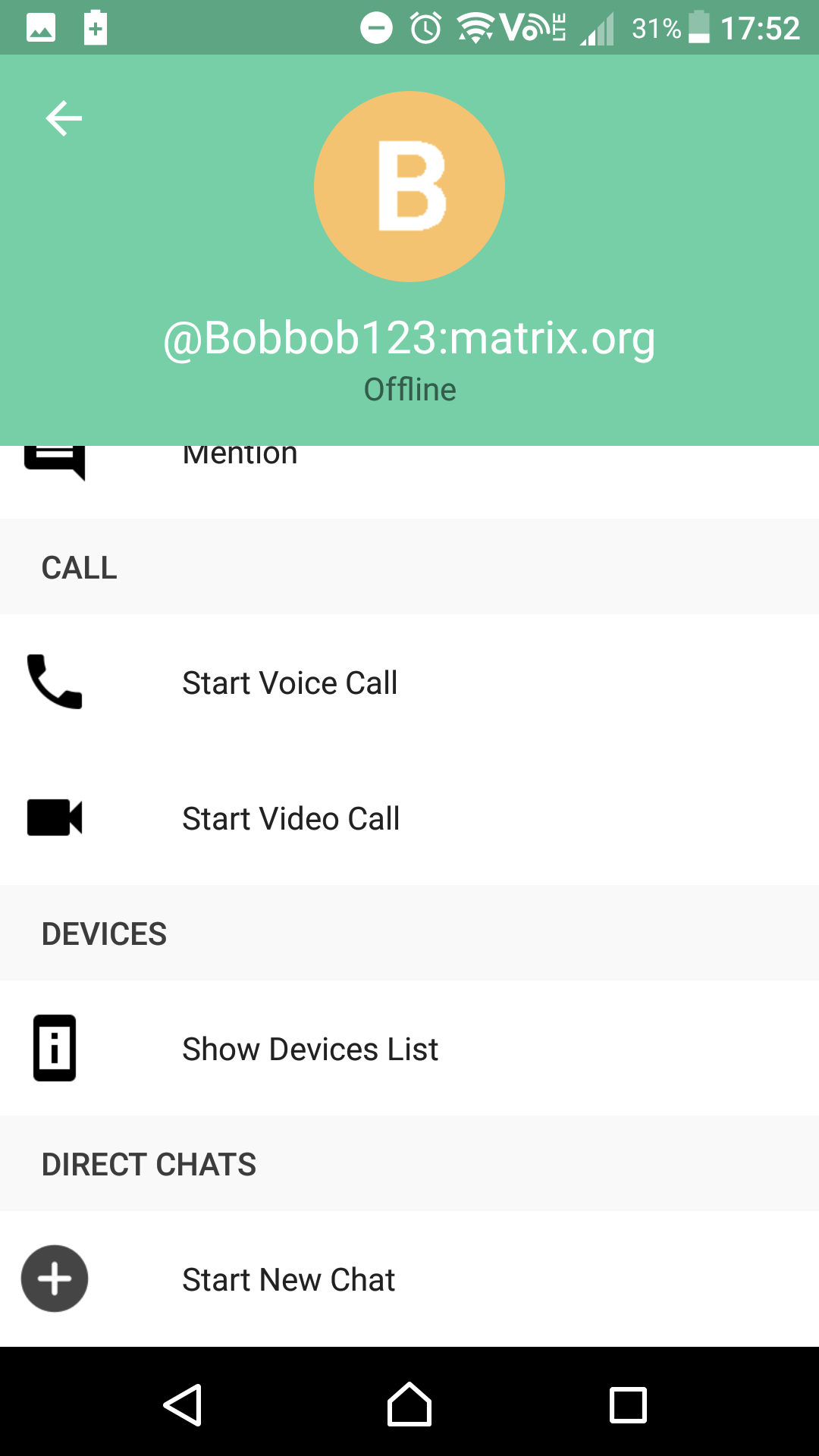}
  		\caption{Profile details}
  		\label{fig:impl-riot-verify-1}
	\end{subfigure}%
	\begin{subfigure}{.24\textwidth}
  		\centering
  		\includegraphics[width=.95\linewidth]{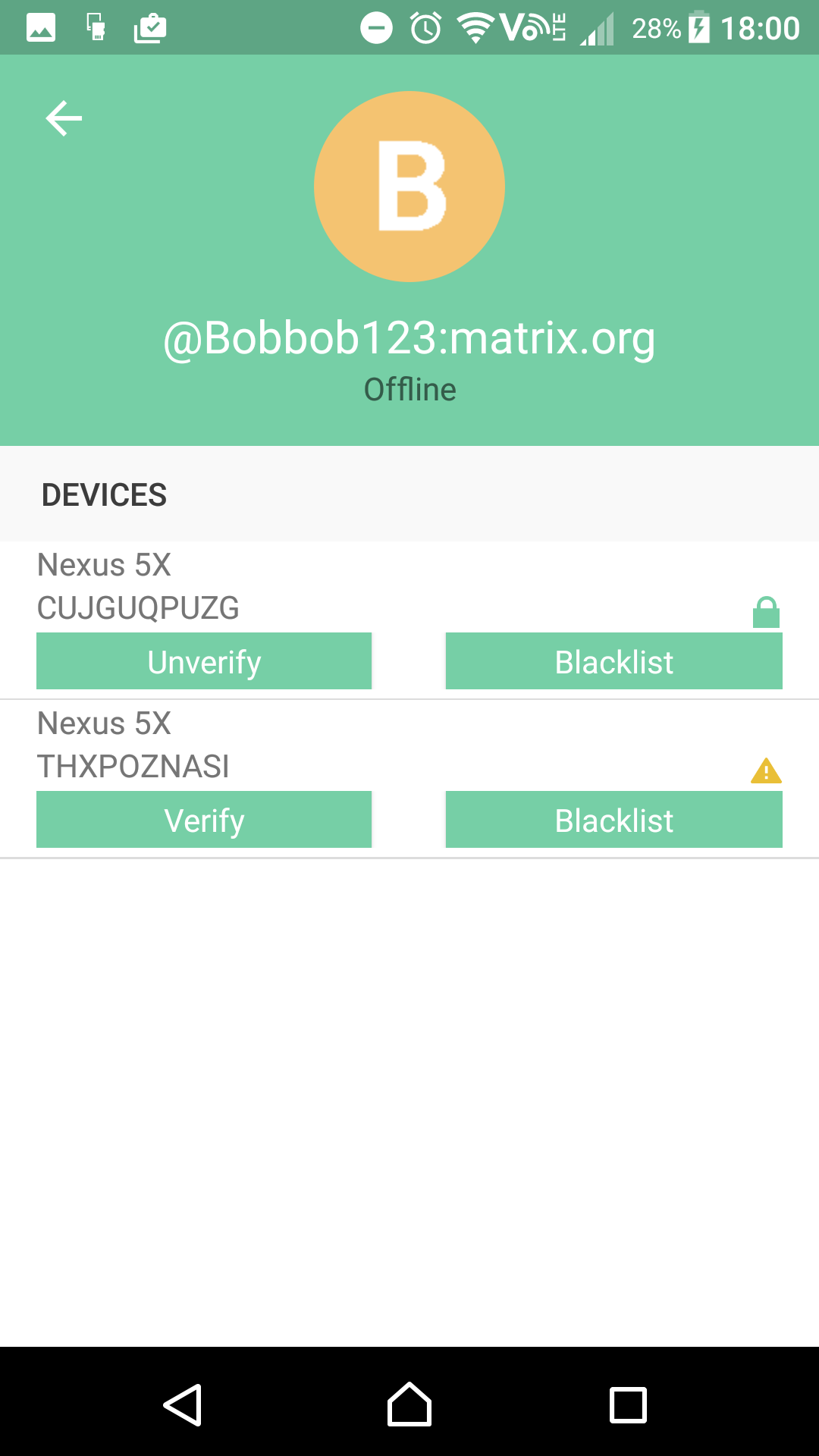}
  		\caption{Verifying Bob the 2nd time}
  		\label{fig:impl-riot-kc-4}\label{fig:impl-riot-verify-2}
	\end{subfigure}%
	\begin{subfigure}{.24\textwidth}
		\centering
		\includegraphics[width=.95\linewidth]{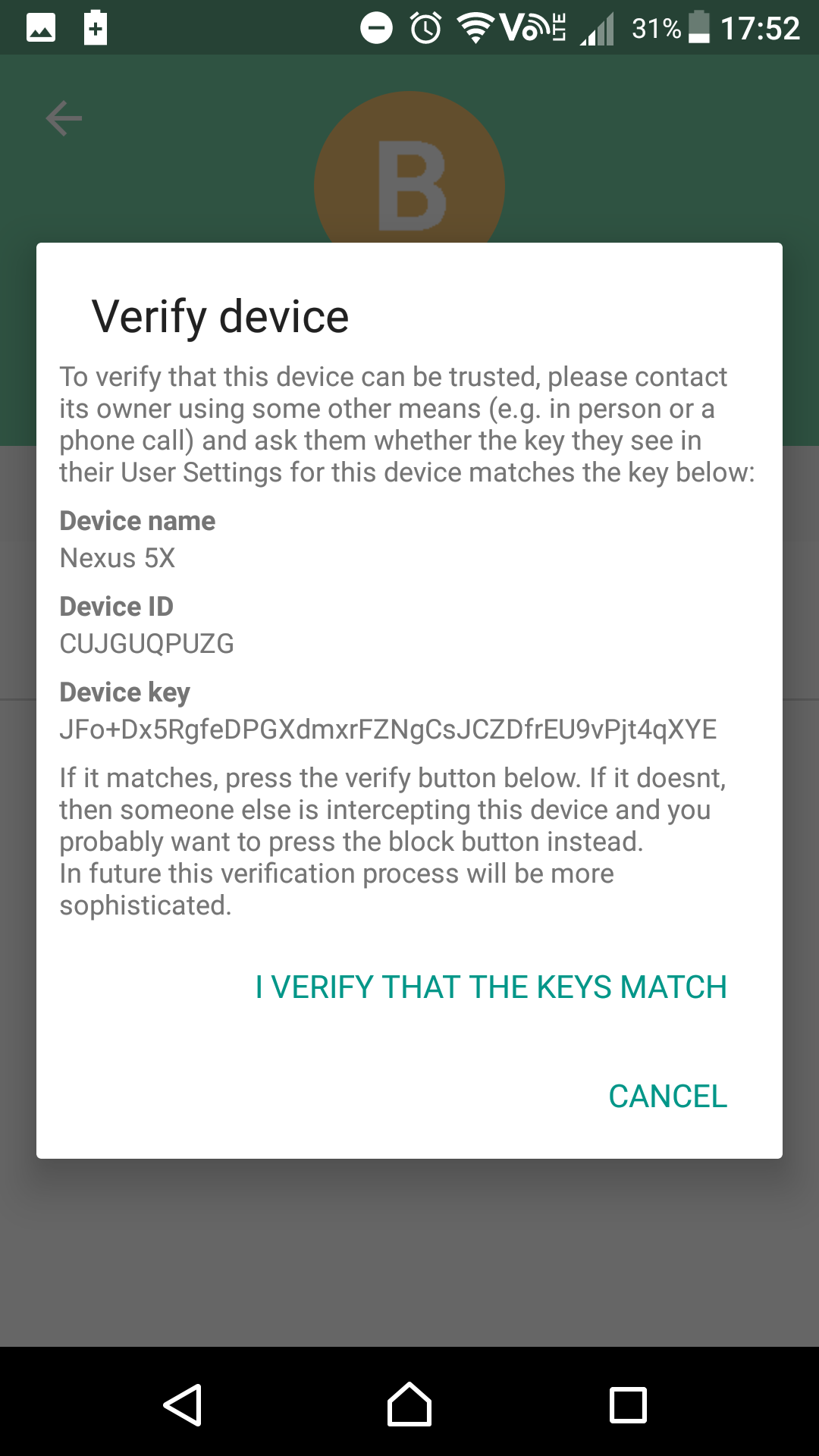}
		\caption{Verifying Bob}
		\label{fig:impl-riot-verify-3}
	\end{subfigure}
	\begin{subfigure}{.24\textwidth}
		\centering
	\includegraphics[width=.95\linewidth]{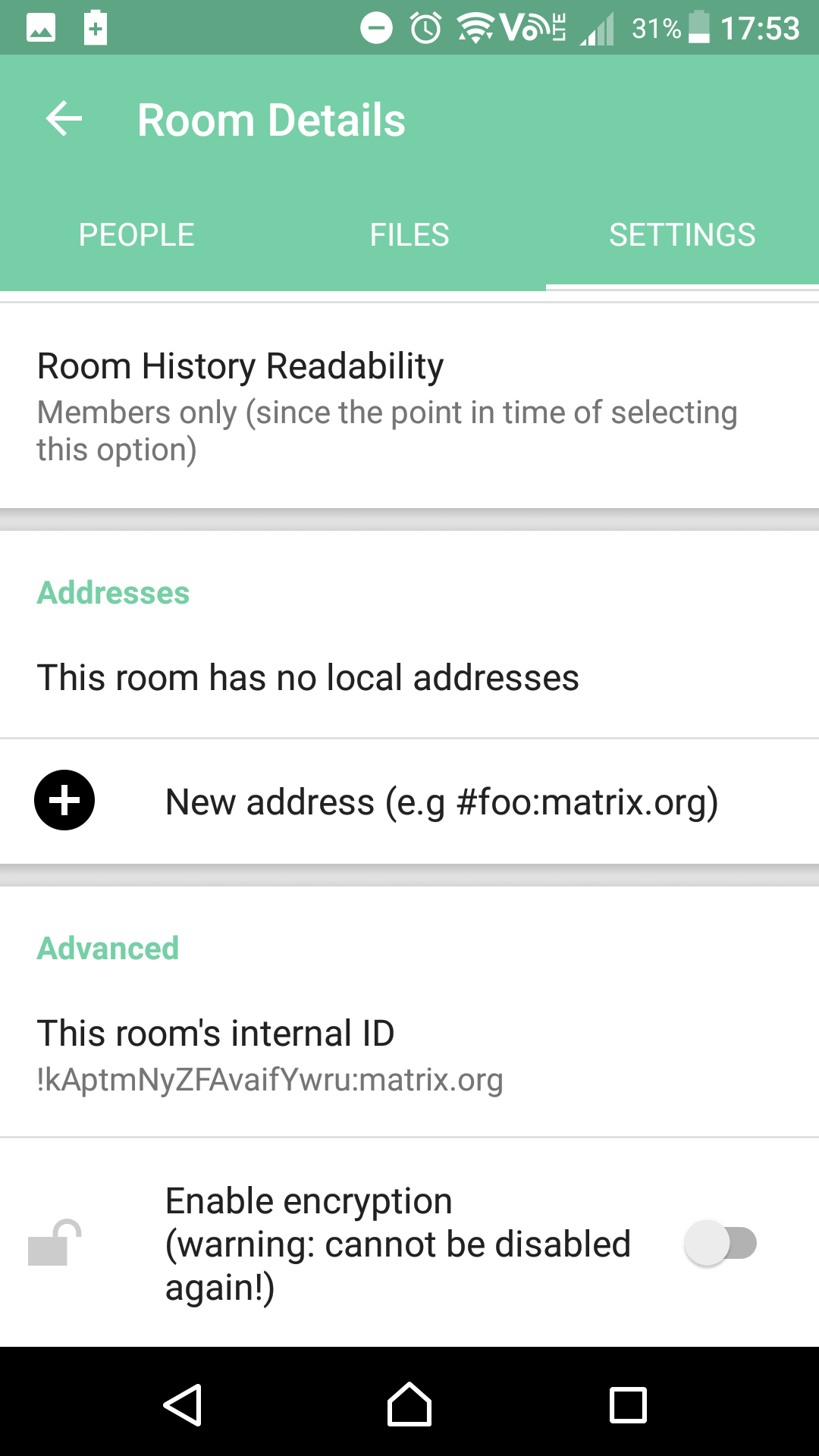}
	\caption{Riot: Other security implementations}
	\label{fig:impl-riot-settings}
	\end{subfigure}
	\caption{Riot: verification process}
	\label{fig:impl-riot-verify}
\end{figure}

Riot is not the typical instant messaging application such as Signal or WhatsApp. Their vision is to make an application which works in the same way as Slack or IRC, where there are chat rooms to join and talk to others. Therefore, Alice starts a chat room, invites Bob and then activates end-to-end encryption. It should be noted that end-to-end encryption is still in beta form, and thus is not turned on by default. 
Fig.~\ref{fig:impl-riot-kc}\subref{fig:impl-riot-kc-1} 
shows the chat room, which in the beginning has open locks on each of the messages from Alice and Bob that have been sent before the encryption was toggled on. How the end-to-end encryption is toggled on is shown in the ``other security implementations'' part (Fig.\ref{fig:impl-riot-verify}\subref{fig:impl-riot-settings}). When Alice sends her initial message to Bob,
the lock is changed to closed (Fig.\ref{fig:impl-riot-kc-transit}\subref{fig:impl-riot-kc-transit-2}) since E2E encryption is on.
Fig.~\ref{fig:impl-riot-kc}\subref{fig:impl-riot-kc-3} shows that 
if Bob reinstalls his application, he has to re-verify Alice because his device keys are changed. 
Alice can also see that she has to re-verify Bob as the Bob's message has a yellow notification triangle.
%
When Alice verifies Bob, the messages are then listed with a correct closed lock, which means that the messages are encrypted correctly (Fig.~\ref{fig:impl-riot-kc}\subref{fig:impl-riot-kc-5}). In Riot, when a new user (or existing user with new keys) enters the chat room, she cannot read/access the previous messages (see Fig.~\ref{fig:impl-riot-kc}\subref{fig:impl-riot-kc-6}).

\paragraph{Key Change While a Message is in Transit:}

Riot handles key changes in the same way as the previous subsection regardless of whether the message is in transit or not. 
When Bob deletes the application, Alice sends the second message to Bob (Fig.~\ref{fig:impl-riot-kc-transit}\subref{fig:impl-riot-kc-transit-2}). Alice also sends a third message to Bob when Bob reinstalls the application. As shown in Fig.~\ref{fig:impl-riot-kc-transit}\subref{fig:impl-riot-kc-transit-3}, Bob can read just the third message but not the second. This shows that Riot handles the key changes in the same way as before, Bob can see there were some messages sent, but cannot decrypt since he lost the old keys.
%

\paragraph{Verification Process Between Participants:}

The verification process is rather easy in the Riot application. Moreover, Riot gives considerable amounts of information to the user about the users they interact with. When Alice wants to verify Bob's devices, she needs to look at his profile account to find Bob's list of devices by clicking on the ``Device'' tab, as shown in Fig.~\ref{fig:impl-riot-verify}\subref{fig:impl-riot-verify-1}. 
One of Bob's devices in Fig.~\ref{fig:impl-riot-verify}\subref{fig:impl-riot-verify-2} has the yellow notification triangle, then that specific device has not been verified by Alice yet. She can either verify the device or put it into the blacklist, as in Fig.~\ref{fig:impl-riot-verify}\subref{fig:impl-riot-verify-3}, which means that specific device is no longer able to send any messages or invites to Alice.

If Alice decides to verify Bob's device, she clicks on the verify button and sees the verification popup from Fig.~\ref{fig:impl-riot-verify}\subref{fig:impl-riot-verify-3}. The popup is informative for users, but for some end-users it may have too much information and could look cluttered. The popup states that the verification information and process will become more sophisticated in future versions when the application starts to reach the end of the beta period. For the verification, Alice can either call Bob or meet him in person and then exchanges the device keys. Finally, Alice needs to press the ``I verify that the keys match'' button.

\paragraph{Other Security Implementations:}\label{Riot_Sec}

Riot does not have that many extra security implementations, but since the application is only in beta, they may be implemented in future versions. Fig.~\ref{fig:impl-riot-settings} shows the settings page providing details about a chat room. The administrator of the room (the one who initialized the room) is the only user who can change the settings of the room. The last setting shown in the figure is the option to enable encryption in that specific room. Once encryption is enabled, it cannot be disabled throughout the conversation.


\subsubsection{Case 6: Telegram} \label{impl:telegram-intro}

Telegram is an instant messaging platform which was started in 2013 after the NSA scandal. It has been developed for smartphones, tablets and even computers.\footnote{Telegram FAQ \url{https://telegram.org/faq}}
Telegram allows one-to-one and group communications, and the possibility to send files to people in the contact list. The difference between Telegram and the other secure IM applications is that it only offers opt-in secure messaging, while normal conversations are cloud chats that are not end-to-end encrypted. Their motivation is to offer seamless cloud chat synchronization between all connected devices.\footnote{Seamless chat cloud synd, tweet by Pavel Durov in 2015 at \url{https://twitter.com/durov/status/678305311921410048}}

For secure chatting, Telegram implements its own cryptographic protocol called the MTProto Protocol\footnote{MTProto Protocol, by Nikolai Durov in Telegram Documentation, available at \url{https://core.telegram.org/mtproto}}.
The same protocol is also used for normal cloud chats to encrypt the communication between the server and the client.

\begin{wrapfigure}[18]{r}{.49\textwidth}
\vspace{-1ex}\centering
\begin{subfigure}{.24\textwidth}
  		\centering
  		\includegraphics[width=.95\linewidth]{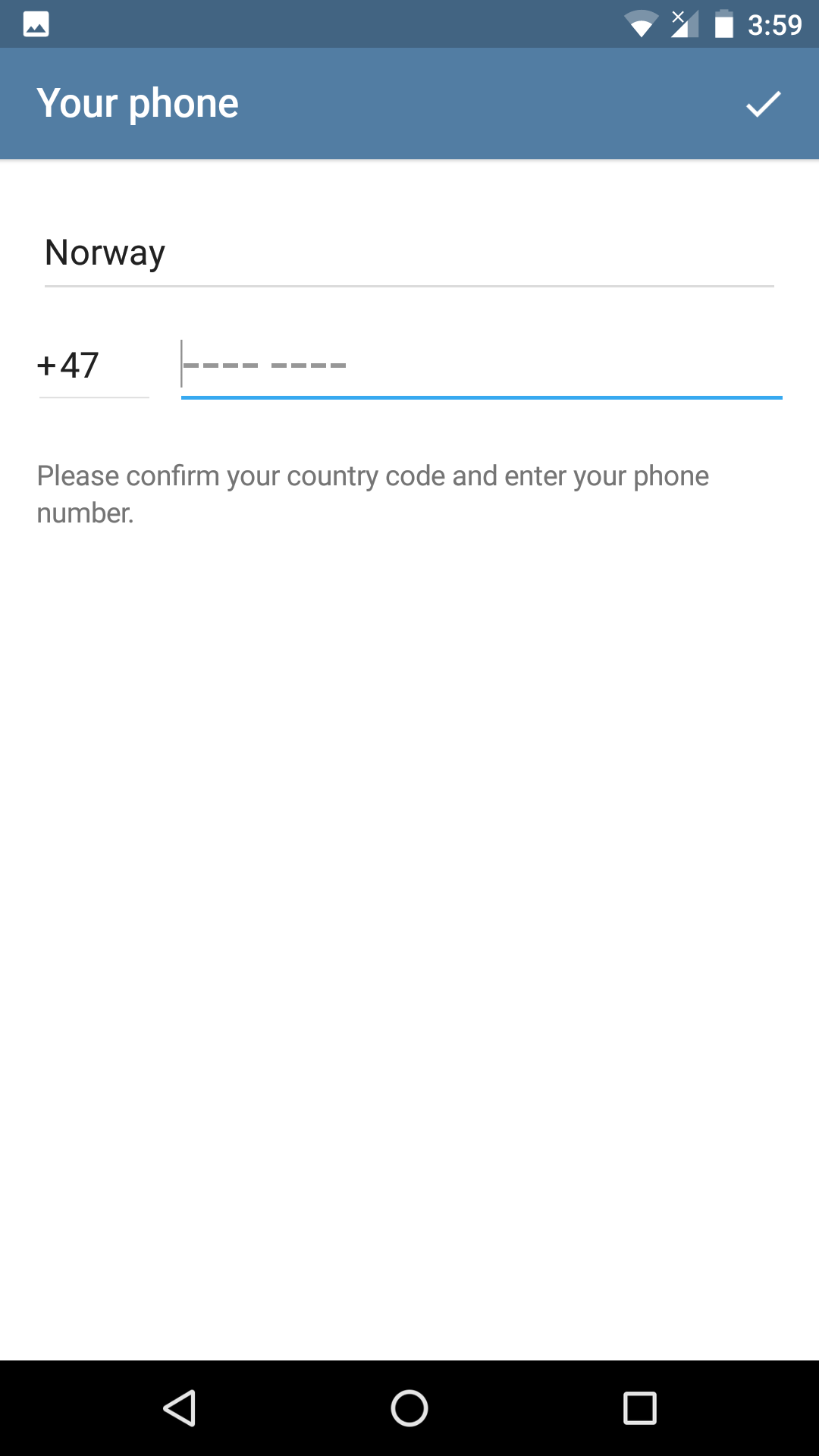}
  		\caption{Phone number registration}
  		\label{fig:impl-tg-reg-1}
	\end{subfigure}%
	\begin{subfigure}{.24\textwidth}
  		\centering
  		\includegraphics[width=.95\linewidth]{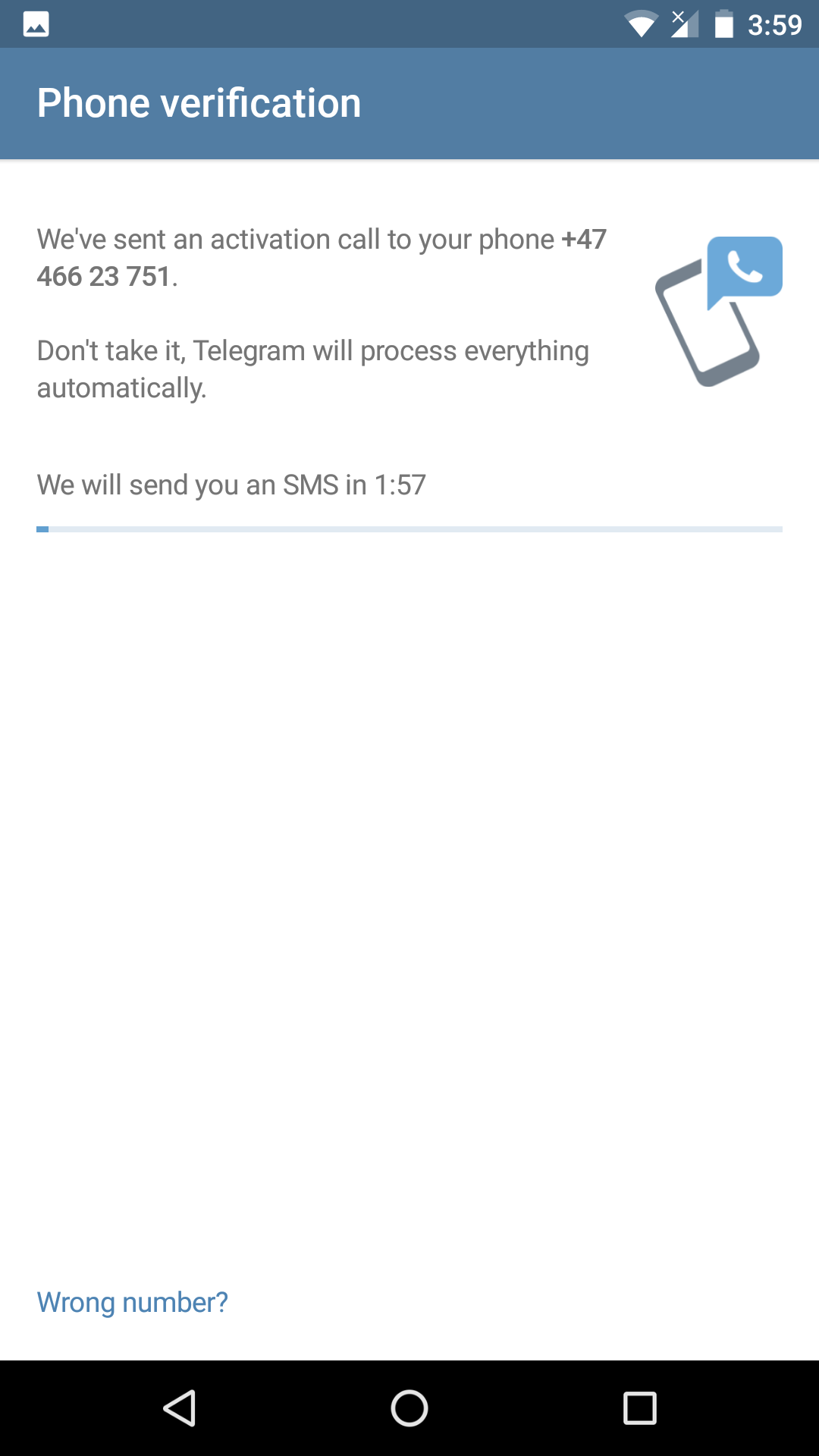}
  		\caption{Verification of phone number}
  		\label{fig:impl-tg-reg-2}
	\end{subfigure}
	\caption{Telegram: registration process}
	\label{fig:impl-tg-reg}
\end{wrapfigure}
For end-to-end encrypted chats, it is not allowed to screenshot inside the secret chat conversation. Hence, in order to provide images for different test scenarios, we used an external camera.

\paragraph{Initial Set Up:} \label{impl:telegram-setup}

The initial setup of the Telegram application and user registration is the same for the other applications. Fig.~\ref{fig:impl-tg-reg}\subref{fig:impl-tg-reg-1} shows that the user needs to enter her phone number for the registration process. As shown in Fig.~\ref{fig:impl-tg-reg}\subref{fig:impl-tg-reg-2}, Telegram sends an activation code through SMS, which can either be input manually or give Telegram access to take it automatically. If the verification message is not received in two minutes, a new SMS will be sent. The user also can ask Telegram to call her and activate it through phone call.

\paragraph{Message After a Key Change:}

\begin{figure}[h]
\centering
	\begin{subfigure}{.24\textwidth}
  		\centering
  		\includegraphics[width=.95\linewidth]{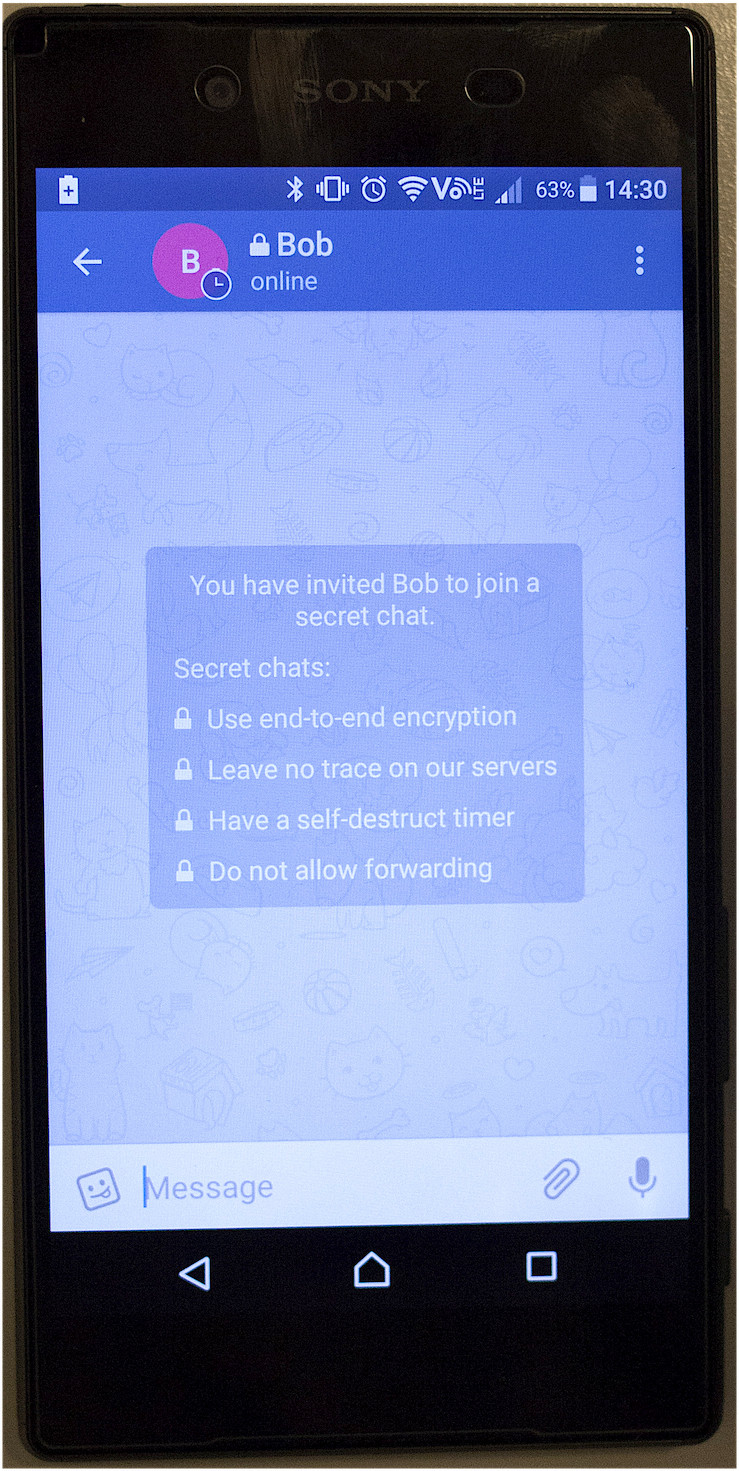}
  		\caption{Initial contact}
  		\label{fig:impl-tg-init}
	\end{subfigure}%
	\begin{subfigure}{.24\textwidth}
  		\centering
  		\includegraphics[width=.95\linewidth]{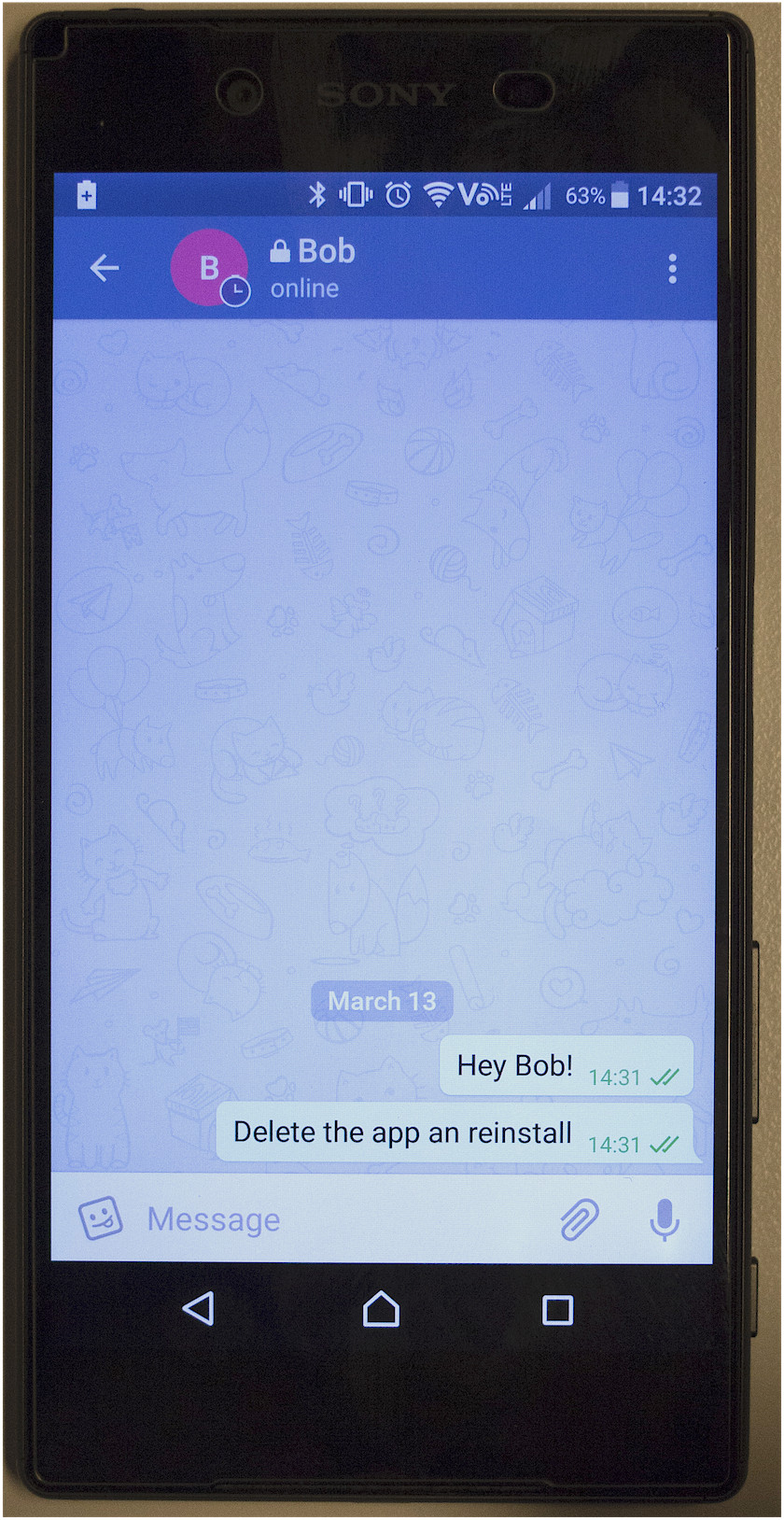}
  		\caption{Alice: Initial message}
  		\label{fig:impl-tg-kc-1}
	\end{subfigure}%
	\begin{subfigure}{.24\textwidth}
  		\centering
  		\includegraphics[width=.95\linewidth]{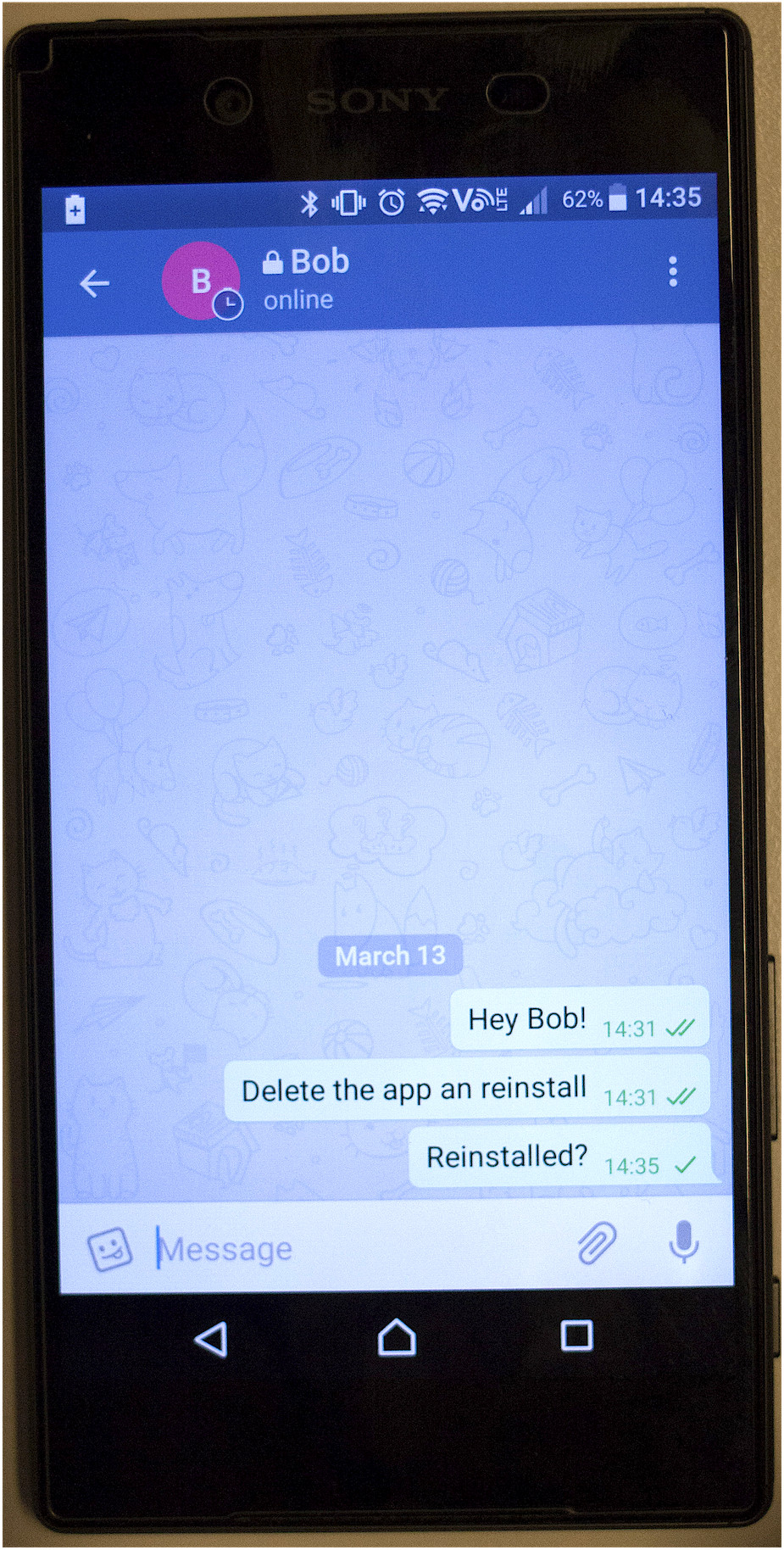}
  		\caption{Message after reinstall}
  		\label{fig:impl-tg-kc-2}
	\end{subfigure}
	\begin{subfigure}{.24\textwidth}
  		\centering
  	\includegraphics[width=.95\textwidth]{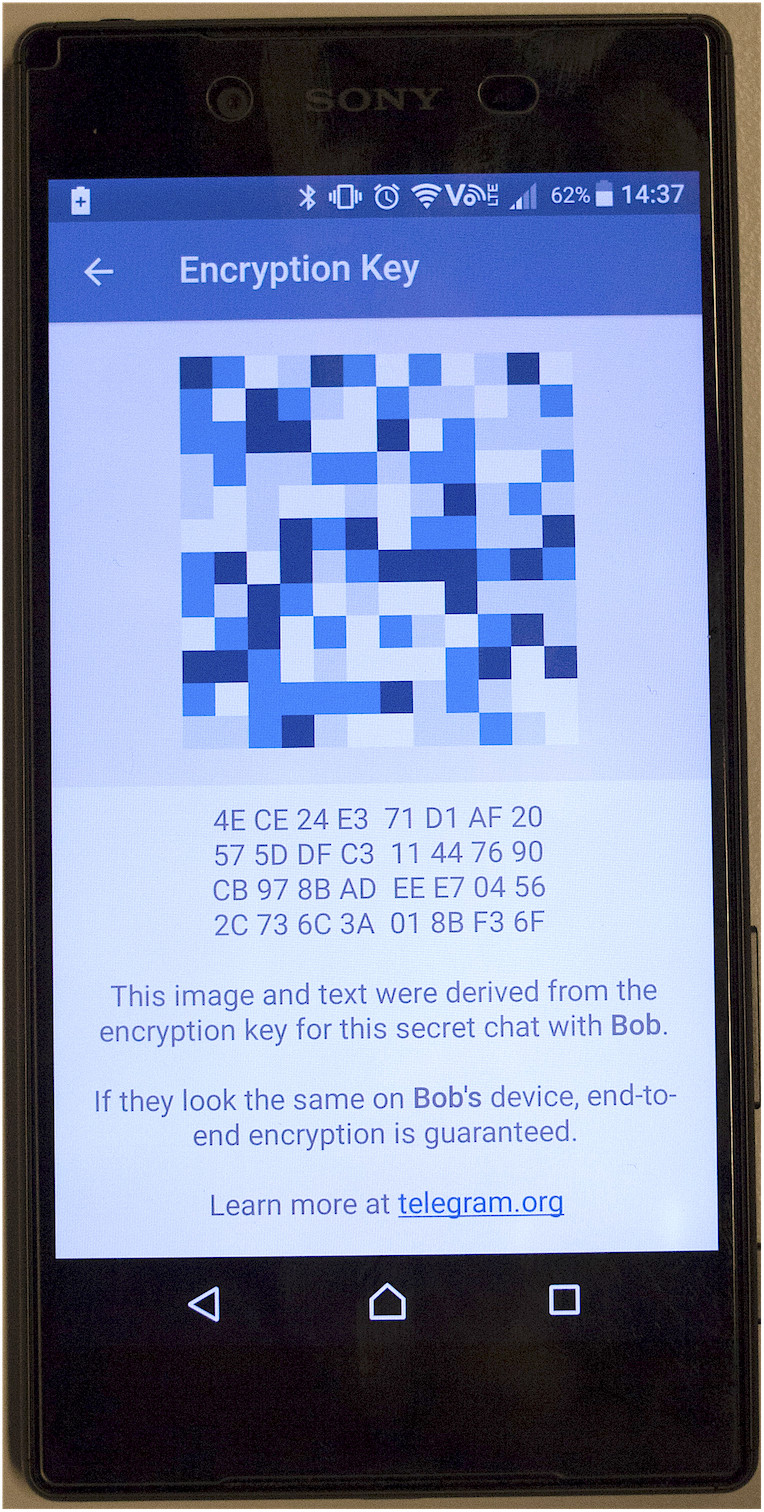}
	\caption{Telegram: Verification process}
	\label{fig:impl-tg-verify}
	\end{subfigure}
	\caption{Telegram: message after a key change}
	\label{fig:impl-tg-kc}
\end{figure}

The end-to-end encryption is not enabled in Telegram by default. Normal messages, which are called \textit{cloud chats} on Telegram, are not encrypted. Fig.~\ref{fig:impl-tg-kc}\subref{fig:impl-tg-init} shows the first view Alice sees when initiating a secret conversation with Bob, which says that the chat is end-to-end encrypted and the messages cannot be forwarded for security reasons.
Fig.~\ref{fig:impl-tg-kc}\subref{fig:impl-tg-kc-1} shows the first messages that have been sent from Alice to Bob, and the double checkmarks illustrate that Bob has received and read the message (a single checkmark means that the message has been sent). 
%
If Bob reinstalls his application and meanwhile Alice sends a message to Bob, then Bob will never receive that message (even after finishing the re-installation of the application) as shown in Fig.~\ref{fig:impl-tg-kc}\subref{fig:impl-tg-kc-2}. Thus, Telegram does not use the previous device keys after re-installing the application by one of the participants. Hence, Alice needs to start a new secret chat with Bob and send the previous undelivered messages again. Telegram does not store keys, or any other information that could reveal that two users have ever had a secret chat. Therefore, Telegram cannot check whether one of the users has reinstalled the application or not. 

\paragraph{Key Change While a Message is in Transit:}

As described in the last scenario, Telegram does not store any information about Alice or Bob having a secret chat; all is done by the client and nothing is sent to the cloud of Telegram. Therefore, there makes no sense to test this scenario, as it will have the same outcome as the one before.

\paragraph{Verification Process Between Participants:}

The verification process between Alice and Bob is rather difficult when using secret chats in Telegram. If Alice wants to verify Bob's encryption keys, she needs to open the specific secure chats settings page and then click on the ``Encryption Key'' button. Telegram just supports messaging and does not support calling for the verification. Figure \ref{fig:impl-tg-verify} shows the verification page, with an image, which is derived from the encryption key, and the encryption key below. There is no way for Alice to 
arrive 
to the conclusion that it is the right image for the conversation.



\paragraph{Other Security Implementations:}

\begin{wrapfigure}[19]{r}{.49\textwidth}
\vspace{-1ex}\centering
\begin{subfigure}{.24\textwidth}
  		\centering
  		\includegraphics[width=.95\linewidth]{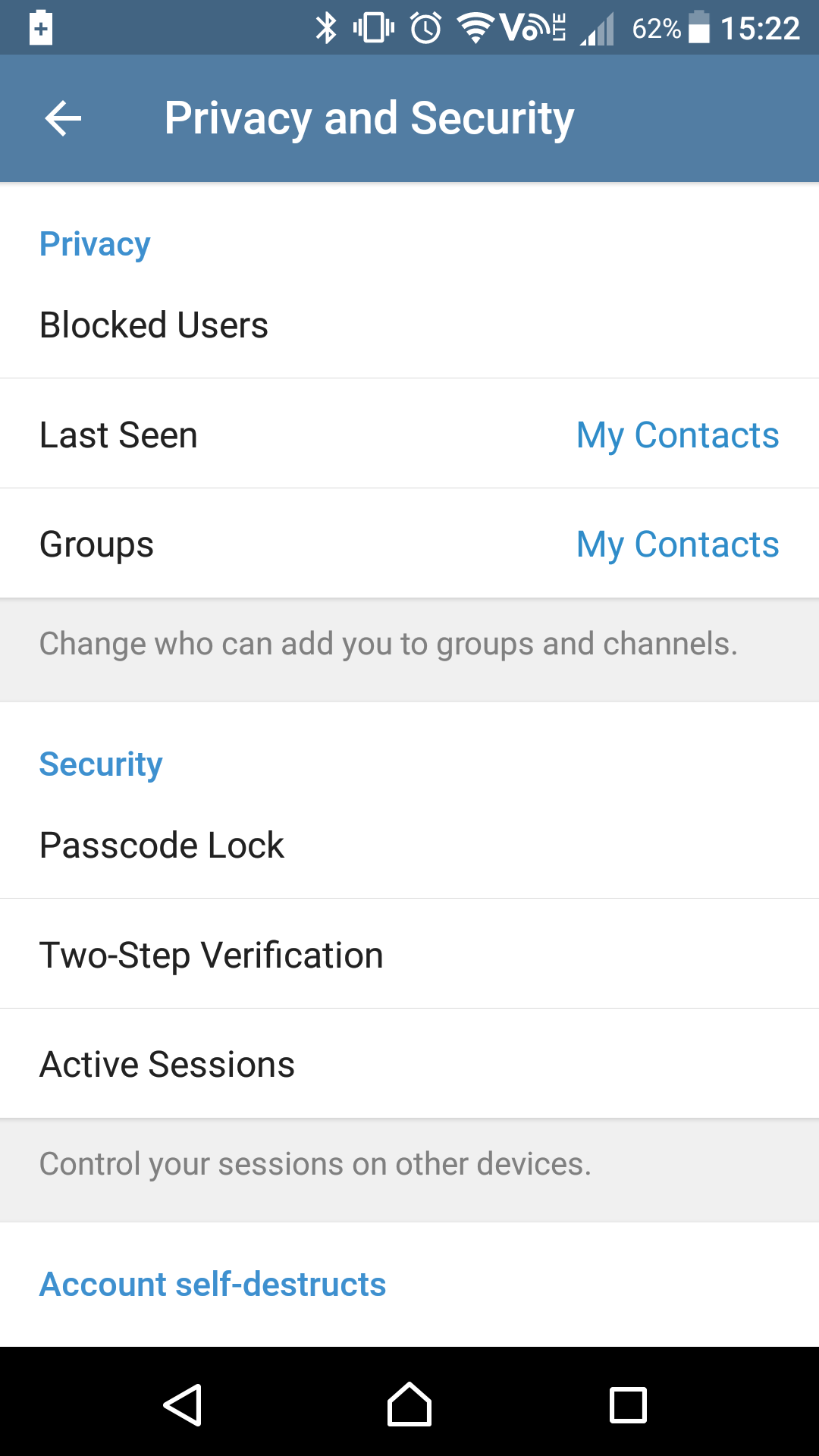}
  		\caption{Privacy and Security settings}
  		\label{fig:impl-tg-priv-1}
	\end{subfigure}%
	\begin{subfigure}{.24\textwidth}
		\centering
		\includegraphics[width=.95\linewidth]{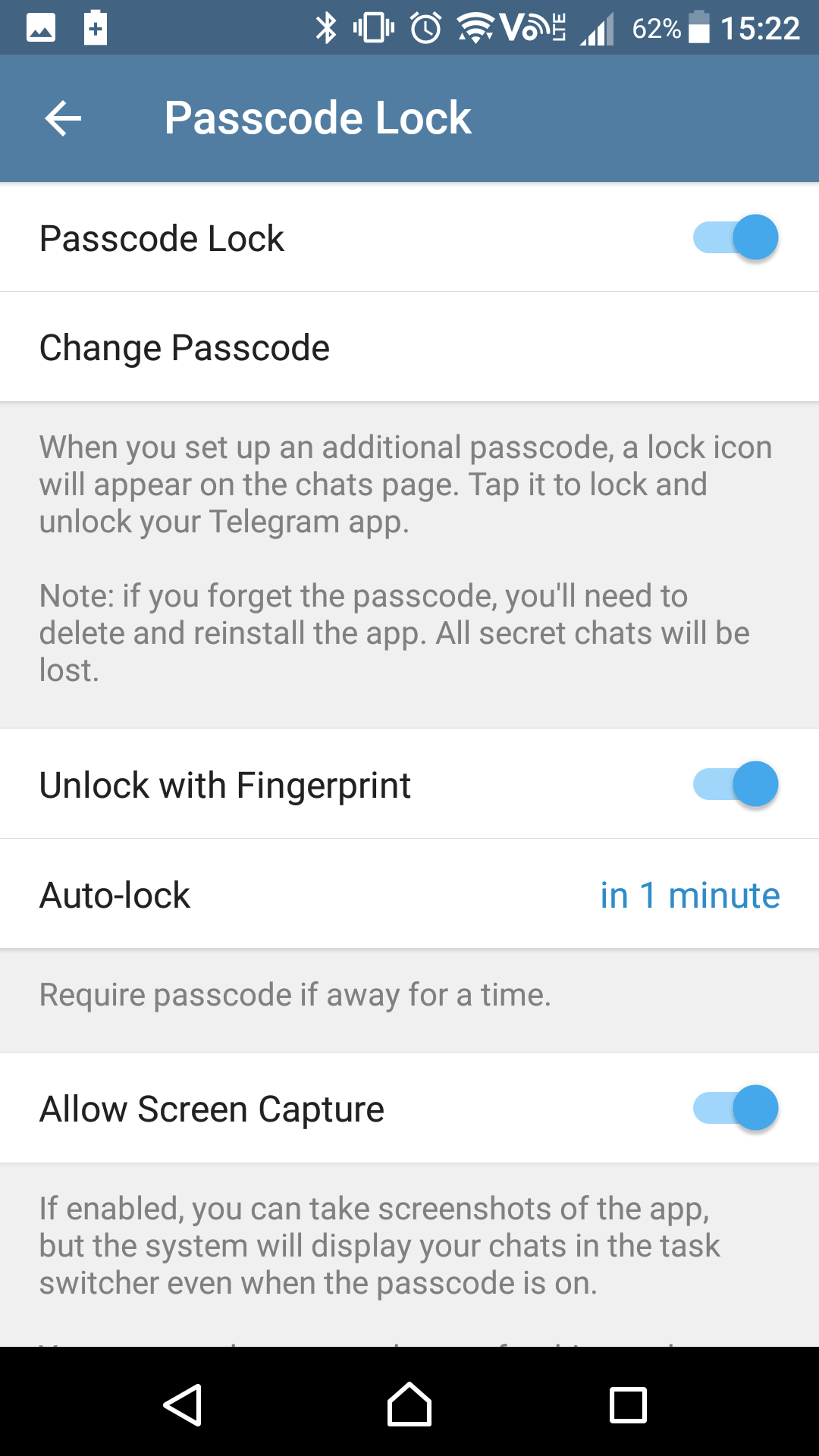}
		\caption{Passcode settings}
		\label{fig:impl-tg-priv-2}
	\end{subfigure}
	\caption{Telegram: other security implementations}
	\label{fig:impl-tg-priv}
\end{wrapfigure}

Telegram supports a few other security features. Inside the settings page, there is one option to look at the ``Privacy and Security'' settings for the application (Fig.~\ref{fig:impl-tg-priv}\subref{fig:impl-tg-priv-1}) 
Telegram supports two-step verification, i.e., when a user wants to log in on another device or after a reinstall, then they need to write a second, personally chosen, password after the activation code received by SMS. ``Active sessions'' is a list of devices the user has logged into. The last option, ``Account self-destructs'', is a security measurement where if the user has not used their account in the last six months, the account gets deleted by Telegram. The length of the counter for self-destructing can be changed to one month, three months, six months or one year.

Fig.~\ref{fig:impl-tg-priv}\subref{fig:impl-tg-priv-2} shows options under the ``Passcode Lock''. 
This function locks the whole application with a passcode that the user chooses. Telegram has implemented the possibility to unlock the application by fingerprint if the user has added a fingerprint in the operating system. A user has the chance to change when the application should auto-lock, from one minute to five hours. The last option shown is the ``Allow screen capture'', which (if enabled) allows users to screenshot anything inside the application. However, for secure chats it is never allowed to screenshot.


\section{Summary of Results}\label{results}

The results of the tests are summarised in Table~\ref{tbl-results}, listing what properties each application provides and which are missing.

\begin{table*}[!tp]
\centering
\caption{Overview of the results from the analysis of secure messaging applications test scenarios.}\label{tbl-results}
\begin{tabular}{@{\hspace{0ex}}l@{\hspace{0ex}}c@{\hspace{1ex}}c@{\hspace{1ex}}c@{\hspace{1ex}}c@{\hspace{1ex}}c@{\hspace{1ex}}c@{\hspace{0ex}}}
\hline
\multirow{2}{*}{Test Scenario and Properties}                                   & \multicolumn{6}{l}{Application}                        \\ \cline{2-7} 
                                                                                & Signal       & WhatsApp & Wire   & Viber    & Riot   & Telegram \\ \hline
					  \textbf{Setup and Registration}       &              &          &        &          &        &   	\\
					  Phone Registration                    & \pr{100}       & \pr{100}   & \pr{100} & \pr{100} & \pr{0} & \pr{100}   \\
                                          E-mail Registration                   & \pr{0}       & \pr{0}   & \pr{100} & \pr{0} & \pr{100} & \pr{0}   \\
                                          Access SMS Inbox                      & \pr{100}       & \pr{100}   & \pr{0} & \pr{100} & \pr{0} & \pr{100}   \\
                                          Contact list Upload                   & \pr{100}       & \pr{100}   & \pr{100} & \pr{100} & \pr{100} & \pr{100}   \\
                                          Verification by SMS                   & \pr{100}       & \pr{100}   & \pr{100} & \pr{100} & \pr{0} & \pr{100}   \\ \vspace{2mm}
                                          Verification by Phone Call            & \pr{100}       & \pr{100}   & \pr{100} & \pr{100} & \pr{0} & \pr{100}   \\
                                          \textbf{Initial Contact}              &      	       &    	  &        &        &        &          \\
					  Trust-On-First-Use                    & \pr{100}       & \pr{100}   & \pr{0} & \pr{0} & \pr{0} & \pr{0}   \\\vspace{2mm}
                                          Notification About E2E Encryption     & \pr{0}       & \pr{100}   & \pr{0} & \pr{0} & \pr{100} & \pr{100}   \\
                                          \textbf{Message After a Key Change}   &              &          &        &        &        &    \\
					  Notification about key changes        & \pr{100}       & \pr{100}   & \pr{0} & \pr{0} & \pr{100} & \pr{0}   \\\vspace{2mm}
                                          Blocking message                      & \pr{100}       & \pr{0}   & \pr{0} & \pr{0} & \pr{0} & \pr{0}   \\
                                          \textbf{Key Change While a Message Is}&   &          &        &        &        &           \\
					  \textbf{In Transit}&   &          &        &        &        &           \\
					  Re-encrypt and Send Message           & \pr{0}       & \pr{100}   & \pr{0} & \pr{0} & \pr{0} & \pr{0}   \\\vspace{2mm}
                                          Details About Transmission of Message & \pr{100}       & \pr{100}   & \pr{100} & \pr{0} & \pr{100} & \pr{100}   \\
                                          \textbf{Verification Process}         &              &          &        &        &        &    \\\vspace{1mm}                                         
					  QR-Code                               & \pr{100}       & \pr{100}   & \pr{0} & \pr{0} & \pr{0} & \pr{100}   \\
                                          Verify By Phone Call                  & \pr{100}       & \pr{100}   & \pr{100} & \pr{100} & \pr{100} & \pr{0}   \\
                                          Share Keys Through 3rd Party          & \pr{100}       & \pr{100}   & \pr{0} & \pr{0} & \pr{0} & \pr{0}   \\\vspace{2mm}
                                          Verified Check                        & \pr{0}       & \pr{0}   & \pr{100} & \pr{100} & \pr{100} & \pr{0}   \\
                                          \textbf{Other Security Implementations}&    	       &   	  &        &        &        &    \\		
					  Two-Step Verification                 & \pr{0}       & \pr{100}   & \pr{0} & \pr{0} & \pr{0} & \pr{100}   \\
                                          Passphrase/Code                       & \pr{100}       & \pr{0}   & \pr{0} & \pr{0} & \pr{0} & \pr{100}   \\
                                          Screen Security                       & \pr{100}       & \pr{0}   & \pr{0} & \pr{0} & \pr{0} & \pr{100}   \\
                                          Clear Trusted Contacts                & \pr{0}       & \pr{0}   & \pr{0} & \pr{100} & \pr{0} & \pr{0}   \\
                                          Delete Devices From Account          & \pr{0}       & \pr{0}   & \pr{100} & \pr{100} & \pr{100} & \pr{100}   \\ \hline
\multicolumn{7}{l}{\pr{100}: Has the property; \pr{0}: Does not have the property.}
\end{tabular}
\end{table*}

From this overview one can conclude that all applications provide mostly the same properties related to the setup and registration phase. 
All applications except Riot support registration with the user's phone numbers, whereas Riot needs an e-mail address. Wire supports both phone number and e-mail address, which can also be used later to log in.
Access to the SMS inbox is not mandatory in any of the applications, but it is set up as default to make it easier for the user, for otherwise one would need to enter the verification code manually. 
As Riot does not use a phone number for verification, it does not need access to SMS. Wire also does not have access to the SMS inbox because they believe that it is easy for the user to enter the verification code by hand.

Uploading the contacts list to a server is required by all applications because it enables to find if any of the contacts is already using the application. However, if a user, in order to remain anonymous, does not want to upload her contacts list, she needs instead to only give out her phone number to particular persons in order to communicate.

All the applications (except Riot that does not use a phone number) have the same properties when it comes to verification by SMS and phone call. They all first give the user the option to verify by reading the SMS and if the user never receives the SMS, then they can ask the application to call them.

Signal and WhatsApp have a Trust-On-First-Use method, where users trust the other participants in a conversation without verifying first. The other applications ask users to verify each other first in order to be assured that the conversation is secure.

A notification at the start of the conversation would be useful to a new user who does not know what end-to-end encryption is, and having this only in the beginning would not bother the users. Only half of the applications have this notification implemented.

The differences in the way applications are handling key changes are quite big. 
Only the Signal application had both blocking messages and showed a notification that the other user in the conversation did not have the same cryptographic keys after a reinstall. The blocking message functionality would not allow the sender to send a message before they verify the new cryptographic keys of the receiver if the receiver has generated new cryptographic keys during the conversation. Applications that do not give any notification or block sending of messages could be target for man-in-the-middle attacks, since one of the participants would never get the notification of key changes and thus could not detect any inconsistencies in the hijacked conversation. 
Wire and Viber are particularly vulnerable to this. 
The secret chats of Telegram do not work if cryptographic keys change because the application does not store any information about secret chats. The participants would need to re-start the conversation.

The only application that re-encrypts the messages and sends them again after the receiver gets new cryptographic keys was WhatsApp. This is a useful usability property, which we hope it would be implemented by the rest of the applications as well. There is one problem with the way WhatsApp re-encrypts and sends the message again: it never asks the user if it is the correct receiver because the keys have changed.

The ``details about the transmission of a message'' property questions whether the applications show to the sender that the message is either sent, delivered, or seen by the receiver. If the message is never delivered because of changes to the receivers cryptographic keys, then the message is only tagged as ``sent'' for the sender, but if the message is re-encrypted and sent correctly, then the message details should also indicate ``seen'' by the receiver. The only application that does not show any information about whether past messages are sent or delivered is Viber.

Signal and WhatsApp have the easiest verification process, using a QR-code in conjunction with a built-in scanner. However, both have shortcomings since they do not have a check for revealing whether the particular user is already verified. Wire, Viber, and Riot confirmed when a user is already verified, but they did not have the useful QR-code nor any way of sharing the keys outside the application. Telegram was the only application which only offered a QR-code but no way of actually scanning the code. Users had to read the secret keys that are shared between them, whereas the image encoding has no technical way of comparing it, besides by only looking at it.
WhatsApp and Telegram have two-step verification capabilities, which means that whenever a user reinstalls the application or changes devices, they need to enter a second password after the normal verification code from the provider, in order to gain access to their account on the new device.

Signal and Telegram both had a passphrase or code that the user had to enter in order to gain access to the application after some specific timeout expires. Both applications have also implemented screen security to not give potential intruders the ability to screenshot conversations. There is a setting to toggle the security off, but it is on by default in both Signal and Telegram.

The only application which had a list of verified contacts, and the option to delete them, was Viber. Clients such as Wire and Riot, which have a verified check on each contact within a conversation, do not offer this option even if it is not difficult to implement since they already know which contacts and their devices are verified.

The ``delete devices from account'' is only interesting for those applications that support multiple devices. All the applications which supported multiple devices also had a list of devices such that the user could delete a device which is not in use anymore.

\section{Discussions and Recommendations}\label{discussion}

Instead of focusing on one test scenario at a time, like in the previous section, this section discusses and evaluates each application as a whole. Moreover, based on the knowledge gained from the test scenarios, some possible improvements for each application are provided, which have to be verified critically using modelling and verification techniques (besides standard software testing) in order to ensure that an improvement does not break other security properties.

\subsection{Signal}\label{subsec_Recommendations_Signal}

%
%

The experiments conducted in Sec.~\ref{impl:signal} demonstrate that the Signal app does not have major weaknesses. However, there are several potential improvements that we discuss in the following. Signal showed good understanding and care for the user experience, with an easy verification process. The users can employ QR-Codes for the verification purpose and/or can call each other and do the verification process through end-to-end encrypted phone calls.

When a party sends a message after changing the keys, the application blocks the message until the sender and receiver verify each other again. This is a useful property, but the application does not reveal the notification immediately when one of the participants in the conversation changes her keys (for example, due to the re-installation of the application).

Overall, the application has both good security when it comes to end-to-end encryption and useful user experience properties which would not cause problems for new users.
%
The following provides recommendations for improvement that are applicable to the Signal app.
We first state the feature in general terms, and then explicit it for the specific app if needed.

\begin{enumerate}
\renewcommand{\labelenumi}{(\Alph{enumi})}
\renewcommand{\theenumi}{(\Alph{enumi})}
\item\label{Re-encrypt_Signal} \textbf{Re-encrypt and send lost messages:} Give the user an option to re-encrypt a lost message and resend it after finishing the re-verification process, so that messages would not get lost during a conversation.
In the Signal application, the sender of a message can know the status of her messages (i.e., sent or delivered/read) due to the existence of checkmarks on each message, making the implementation of this feature easy. 

\item\label{Notif_Key_Ch} \textbf{Notification about key changes:} It is recommended for an application to show a notification message immediately after each change of cryptographic keys.
If the keys of a party in a conversation change, then the Signal application does not show any notification message to the participants immediately. It only gives the notification (that the keys are changed) when one party wants to send a message to another one (after changing the keys). 

\item \textbf{Notification on end-to-end encryption:} Giving a notification at the beginning of each conversation stating that it is now end-to-end encrypted and the possibility to read more about it, could help educate the end-users about what E2E encryption is and why they should care about it. 

\item\label{Verif_Check} \textbf{Verified check:} Offer visual cues so the user can easily know that a party should be verified again to regain the trust properties.
As demonstrated by our experiments, in the Signal application, there is no way to know if a user is already verified or not. However, if the application already keeps the list of verified contacts, then when one of the participants changes her cryptographic keys, it should be easy to implement this feature.

\item\label{Two-step_Signal} \textbf{Two-step verification:} The security of an application would be improved by adding a ``two-step verification'' option, which, e.g., when a user wants to change her device or reinstall the application, she has to enter a \textit{previously chosen password}, besides the code received in the SMS.


\end{enumerate}

\subsection{WhatsApp}\label{subsec_Recommendations_WhatsApp}

The results of the experiments do not show major weaknesses in the WhatsApp application. However, the WhatsApp application can be improved in several ways as discussed below. WhatsApp takes great care to strengthen the security around the user's account and messages; however, it may suffer from impersonation attacks.
Recommendations applicable to WhatsApp.

\begin{enumerate}
\renewcommand{\labelenumi}{(\Alph{enumi})}
\renewcommand{\theenumi}{(\Alph{enumi})}
\setcounter{enumi}{5}

\item\label{BlockIM_Verif} \textbf{Block messaging until verification:} In order to prevent sending private messages to an impersonator, the participants should not be able to send any message before verifying each other. 
The WhatsApp application immediately re-encrypts a lost message when it finds out that the cryptographic keys have changed and the receiver never received the message. Hence, an adversary may impersonate a legitimate contact and consequently WhatsApp re-encrypts and sends lost messages to the adversary who would thus receive private messages in an unauthorized way. Since this process is automatically controlled by the app, even the sender cannot stop re-encrypting and resending of lost messages after a key change. 
In order to overcome this weakness, the application must suspend the process of re-encrypting and resending of lost messages (after a key change) until the new cryptographic keys are verified.
%
%

\item\label{Lock_App_Pass} \textbf{Locking the application using a passphrase/code:} Adding an option that requires a passphrase or code before opening the application, would enhance the security of the user's account from unauthorized access. If an adversary somehow gets access to a user's phone, she would be unable to access the application messages as she does not know the passphrase/code.

\end{enumerate}

\subsection{Wire}\label{subsec_Recommendations_Wire}


The Wire application features several useful security and usability properties. However, there are some properties which are not provided and might cause serious security problems. In this application, if a user uses a new device, Wire does not notify the other participants in a conversation. Thus, an impersonator may join the conversation and receive all the exchanged messages. 
Recommendations applicable to Wire are \ref{Re-encrypt_Signal},\footnote{Note that the Wire application uses a text under each message to show the status of the message (e.g., delivered or sent), which makes easy to implement this feature.} \ref{Two-step_Signal}, \ref{BlockIM_Verif}, \ref{Lock_App_Pass} from above, as well as the new ones below.


\begin{enumerate}
\renewcommand{\labelenumi}{(\Alph{enumi})}
\renewcommand{\theenumi}{(\Alph{enumi})}
\setcounter{enumi}{7}

\item\label{Notification-verification-Wire} \textbf{Notify users regarding verification:} When the application does not verify participants, it should notify users that they have to verify each other manually when initiating a new conversation, to prevent impersonation attacks.


\item \textbf{Notification about the verification of new devices:} Notify the other participants that a user added a new device and they have to verify the new device before sending any more messages (as the new device will also receive these messages).
The Wire application allows a user to use the same account on multiple devices. However, if a user (in a conversation) adds a new device, the other participants (in that conversation) will not be notified about this change. 


\item\label{More_Verif_Opt} \textbf{More verification options:} Provide several different ways of performing the verification. For example, a QR-code or sharing keys with a third-party application are some possible options. Managing keys and authentication credential could greatly benefit from a secure device attached to the smart phone like the OffPAD \cite{OffPAD16demo}, in conjunction with a proxy that knows how to use such a device as in the OTDP architecture of \cite{migdal2017offline}.
Wire provides only calling a person every time, which may become cumbersome for users. 



\item\label{Screen_Sec} \textbf{Screen Security:} The privacy can be improved by providing a ``screen security'' option, which does not allow screenshots to be taken within the conversations. It is worth to mention that the screen security is useful if all participants in a conversation enable it. Because, if a user does not enable this option, then she can take a snapshot and thus breaches the privacy of the other participants in the conversation.
\end{enumerate}

\subsection{Viber}\label{subsec_Recommendations_Viber}

The Viber application provides some good usability and security properties. However, there are several questionable aspects which make us reluctant in recommending this application.
If the cryptographic keys of a user in a conversation change (e.g., because of re-installation), Viber does not notify the other participants in the conversation about such changes. In addition, if a user sends several messages while the receiver is re-installing Viber, then the sender cannot know whether the previous messages have been received or not.
%
%
Many of the observations for Viber are easy to fix or implement, in our opinion, and would drastically improve the application.
These include \ref{Re-encrypt_Signal}, \ref{Notif_Key_Ch}, \ref{Two-step_Signal}, \ref{BlockIM_Verif}, \ref{Lock_App_Pass}, \ref{Notification-verification-Wire}, \ref{More_Verif_Opt}, \ref{Screen_Sec} from above, as well as the new one below.

\begin{enumerate}
\renewcommand{\labelenumi}{(\Alph{enumi})}
\renewcommand{\theenumi}{(\Alph{enumi})}
\setcounter{enumi}{11}
\item \textbf{Labeling the status of messages:} It is important for all messages in a conversation to have a label stating their status (e.g., sent, delivered, read, etc.). 
In Viber, only the last message in a conversation is labeled with status information. 

\end{enumerate}
\subsection{Riot}

The Riot application is still in beta stage, and despite its usability and security properties, it can still be improved. When cryptographic keys change during a conversation, the previous messages are locked for the user (who is reinstalling the application). However, there is an option for the sender to send the locked messages (encrypted with new keys) again. In Riot the users do not receive any notification message regarding the key changes (and the need to re-verify each other). Moreover, Riot does not use end-to-end encryption by default. In addition, Riot does not provide an easy way for the verification of the users.
%
Recommendations applicable to Riot are \ref{Two-step_Signal}, \ref{Lock_App_Pass}, \ref{More_Verif_Opt}, \ref{Screen_Sec} from above.

\subsection{Telegram}

The telegram application has some useful security properties, but the usability features are rather lacking and confusing for people who are not tech savvy.
The biggest flaw in a secure messaging application is that the end-to-end encryption is not on by default. Hence, an explicit secret chat needs to be initiated every time users want to communicate. We think that this should become a norm these days for an application that advertises encrypted conversations.

In Telegram, if a user sends a message while the cryptographic keys have changed, the intended receiver does not get that message because secret chats are locked to one set of keys. Hence, if a user generates new keys, the participants need to start a new secret chat and cannot continue the previous secret chat. This is good for security, but the problem arises since Telegram does not notify users about key changes.

Telegram provides just one way for the verification of users, showing QR-codes in person, which is not good enough for a secure messaging application. We believe that Telegram can be improved by providing more options, such as calling or sharing keys through third-party applications.
Recommendations applicable to Telegram are \ref{Notif_Key_Ch}, \ref{Verif_Check}, \ref{More_Verif_Opt}, \ref{Screen_Sec} from above.
\section{Conclusion and Further work}
\label{conclusion}
The work reported here conducted two analyses about secure messaging protocols and the applications which implement these types of protocols. The first analysis (following the recent article by Unger et al. \cite{sok}) described old and new secure messaging protocols that offer end-to-end encryption and identified types of security and privacy properties they provide, in Section~\ref{review-protocols}. This review of protocols is an important prerequisite for understanding our analysis of the applications implementing them, which is our main contribution presented in Section~\ref{review-applications}. 
The Signal and Matrix protocols are both secure messaging protocols that manage end-to-end encryption well, but none of them could offer every security property. The analysis concluded that none of the secure messaging protocols could provide every security property. The Signal protocol does not fully support multiple devices, while the Matrix protocol does this well. On the other hand, the Matrix protocol does not fully provide forward and
future secrecy in the protocol, and leave it up to the implementation to support it. 
The Signal protocol is designed to use a server for achieving the asynchronous messaging property. Even if messages are encrypted end-to-end there is still important metadata that is being manipulated and stored on the server. Therefore one would wish to secure better the server side, especially when deployed in a cloud infrastructure or in a country with legislation that disregards privacy. Initial results in this direction have been presented in \cite{kristoffer2017HotSpot} using the recent technology of Intel SGX. This is applicable to Matrix too.

The second conducted analysis was the research experiment of applications which support these secure messaging protocols for their conversations, i.e., in Section~\ref{review-applications}. The applications offer useful usability together with security, and we would strongly recommend general public to adopt one of them (i.e., the one most suited for their needs, after reading through our analysis). We believe that even for common (non-technology people) would be very easy to adopt one of these applications; i.e., it would have the same difficulties (if one could call the such) as any other chat application, whereas the ones related to encryption would not add many difficulties. 
However, there are multiple applications which still could benefit from improvements. We have provided recommendations, in Section~\ref{discussion}, for each application to harden their security around the user's account, but at the same time keep the useful usability properties. The given recommendations are not tested to confirm that it is the best choice for the applications, and it is up to the developers to prioritise if they think it would benefit their application and users.

\newpage

\end{document}